\documentclass[
aps
,groupedaddress
,amsfonts,amsmath,floatfix,eqsecnum
,nofootinbib
,twocolumn
]{revtex4}

\usepackage{hyperref}
\usepackage{xcolor}

\usepackage[utf8]{inputenc}
\usepackage{epsfig,graphicx}
\usepackage{bm}
\usepackage{amsmath,amsfonts}
\usepackage{enumerate}
\usepackage{ytableau}
\usepackage{cancel}
\usepackage{multirow}
\usepackage{mathptmx}      
\usepackage{latexsym}
\usepackage{amstext}
\usepackage{array}
\usepackage[T1]{fontenc} 

\DeclareMathSymbol{\gtrless} {\mathrel}{AMSa}{"3F}

\newcommand{\ignore}[1]{}
\newcommand{\nec}[1]{\eqref{eq:#1}}

\newcommand{\Eq}[1]{Eq. \eqref{eq:#1}}
\newcommand{\Eqs}[1]{Eqs. \eqref{eq:#1}}
\newcommand{\be}{\begin{equation}}
\newcommand{\ee}{\end{equation}}
\def\bes#1\ees{%
  \begin{equation}
    \begin{split}
      #1
    \end{split}
  \end{equation}
}
\def\bs#1\es{%
    \begin{split}
      #1
    \end{split}
}

\newcommand{\R}{\mathbb{R}}

\def\slashchar#1{\setbox0=\hbox{$#1$}
   \dimen0=\wd0 \setbox1=\hbox{/} \dimen1=\wd1
   \ifdim\dimen0>\dimen1 \rlap{\hbox to \dimen0{\hfil/\hfil}} #1
   \else  \rlap{\hbox to \dimen1{\hfil$#1$\hfil}} / \fi}

\newcommand{\ben}{\begin{enumerate}}
\newcommand{\een}{\end{enumerate}}

\newcommand{\ds}{\displaystyle}

\newcommand{\cH}{{\mathcal H }}

\newcommand{\cL}{{\mathcal L }}

\newcommand{\hA}{{\hat{A}}}
\newcommand{\hB}{{\hat{B}}}

\newcommand{\hF}{{\hat{F}}}
\newcommand{\hG}{{\hat{G}}}
\newcommand{\hH}{{\hat{H}}}

\newcommand{\hK}{{\hat{K}}}

\newcommand{\hT}{{\hat{T}}}

\newcommand{\bR}{{\mathbf{R}}}

\newcommand{\esp}[1]{\langle #1 \rangle}

\newcommand{\miem}[1]{\textbf{\textsf{#1}}}

\newcommand{\PM}[1]{ \begin{pmatrix} #1 \end{pmatrix} }

\ignore{$}

\newcommand{\fig}[1]{Fig. \ref{fig:#1}}
\newcommand{\Df}{ D_0 }
\newcommand{\fp}{D}
\newcommand{\hfp}{{\hat{\fp}}}
\newcommand{\fv}{\Gamma}
\newcommand{\fh}{H}
\newcommand{\hfh}{{\hat{\fh}}}
\newcommand{\Hs}{ {H^\prime} }
\newcommand{\mspan}{ \mathrm{span} }

\newcommand{\muRC}{ \mu_\mathrm{RC} }

\renewcommand{\Gamma}{\varGamma}
\renewcommand{\Lambda}{\varLambda}

\renewcommand{\bR}{\bar{R}}
\newcommand{\bO}{O^+}
\newcommand{\rd}{\mathrm{d}}
\newcommand{\rf}{\mathrm{f}}
\newcommand{\ct}{{\rm ct}}
\newcommand{\Tmin}{ T_{\rm min}}
\newcommand{\bd}{{\bar{d}}}

\newcommand{\espp}[1]{\esp{#1}^\phi}
\newcommand{\HC}{\cH_\mathrm{C}}
\newcommand{\HNC}{\cH_\mathrm{NC}}  
\newcommand{\HS}{\cH_\mathrm{S}}
\newcommand{\HNS}{\cH_\mathrm{NS}}

\newcommand{\tg}{\tilde{g}}
\newcommand{\tG}{\tilde{G}}
\newcommand{\tH}{\tilde{H}}

\begin{document}

\title{\textsf{
        Linearized renormalization
  }}

\author{L. L. Salcedo}
\email{salcedo@ugr.es}

\affiliation{Departamento de F\'{\i}sica At\'omica, Molecular y Nuclear and \\
  Instituto Carlos I de F\'{\i}sica Te\'orica y Computacional, \\ Universidad
  de Granada, E-18071 Granada, Spain.  }

\date{\today}

\begin{abstract}  
  
  Using an infinitesimal approach, this work addresses the renormalization
  problem to deal with the ultraviolet divergences arising in quantum field
  theory. Under the assumption that the action has already been renormalized
  to yield an ultraviolet-finite effective action that satisfies a certain set
  of renormalization conditions, we analyze how the action must be adjusted to
  reproduce a first-order change in these renormalization conditions.  The
  analysis then provides the change that is induced on the correlation
  functions of the theory. This program is successfully carried out in the
  case of super-renormalizable theories, namely, a scalar field with cubic
  interaction in four space-time dimensions and with quartic interaction in
  three space-time dimensions. Relying on existing results in the theory of
  perturbative renormalization, we derive explicit renormalized expressions
  for these theories, each of which involves only a finite number of graphs
  constructed with full propagators and full $n$-point vertices.  The
  renormalizable case is analyzed as well; the derived expressions are
  ultraviolet finite as the regulator is removed but cannot be written without
  a regulator. In this sense, the renormalization is not fully explicit in the
  renormalizable case. Nevertheless, a perturbative solution of the equations
  starting from the free theory provides the renormalized Feynman graphs,
  similar to the BPHZ program. For compatibility with the preservation of the
  renormalization conditions, a projective renormalization scheme, as opposed
  to a minimal one, is also introduced. The ideas developed are extended to
  the study of the renormalization of composite operators and the
  Schwinger-Dyson equations.

\end{abstract}


\maketitle
\flushbottom
\setlength{\unitlength}{1mm}

\tableofcontents


\section{\textsf{Introduction and conclusions}}
\label{sec:1}

Quantum field theory is an extremely beautiful and successful tool in
theoretical physics for describing the interactions between particles and in
other areas such as condensed matter \cite{Weinberg:1995mt,Zee:2003mt}. It
accommodates the Standard Model of Particles which currently is being actively
tested \cite{Engel:2013lsa,Alimena:2019zri}. Perturbative calculations using
Feynman diagrams typically display divergences associated with virtual
particles in loops in the asymptotic region of large loop momenta. The same
ultraviolet (UV) divergences appear in the correlation functions for fields at
short distances. Such divergences are not pure mathematical artifacts, rather,
they signal the failure of the model to include the proper degrees of freedom
relevant at very short distances \cite{Wilson:1974mb}. In general, a field
theory is designed to work at scales below a certain momentum cutoff $\Lambda$
and becomes inconsistent above it, developing unphysically large
contributions. Through the process of renormalization, the divergences can be
removed by enlarging the action through the addition of suitable counterterms,
in general involving operators of increasingly larger
dimension. Renormalizable and super-renormalizable theories are those that
only require a finite number of operators in the action to produce finite
correlation functions. While a renormalizable theory such as Quantum
Chromodynamics needs not, and in fact does not, describe the physics at
arbitrarily short scales, it remains an internally consistent theory at all
scales. Quantum electrodynamics is also renormalizable at the perturbative
level, although it develops a non-perturbative Landau pole at extremely high
energies \cite{Landau:1954nau}. Due to the difficult renormalization problem,
only a few quantum field theories currently exist from a rigorous mathematical
point of view, at the non-perturbative level \cite{Glimm:1987ylb}. These
theories exist in lower-dimensional spaces where ultraviolet divergences are
less severe. Certainly, no four-dimensional renormalizable gauge theory has
yet been constructed in a mathematical sense.

The physics of the non-perturbative renormalization and its naturalness were
greatly clarified after the analysis by Wilson and others. Nevertheless, even
if the perturbative analysis does not cover the whole subject of
renormalization, it is still an indispensable and insightful tool in any
theory. The systematics of the renormalization at the perturbative level has
been the subject of intense mathematical work in the past. In particular, the
work of Bogoliubov and Parasiuk \cite{Bogoliubov:1957gp} and Hepp
\cite{Hepp:1966eg} allowed to organize in a systematic way the renormalization
of the momentum space Feynman diagrams of any theory and the identification of
the counterterms. Epstein and Glaser \cite{Epstein:1973gw} developed an
equivalent procedure in space-time space (see \cite{Rejzner:2016} for a modern
presentation). The forest formula by Zimmermann provided an explicit solution
to Bogoliubov's recursion that, as a by-product, allowed to yield manifestly
renormalized ultraviolet-finite Feynman graphs for massive theories
\cite{Zimmermann:1969jj,Collins:1984xc}. The same subtraction rules work
regardless of the regularization and apply to the massless case and gauge
theories, where dimensional regularization becomes the method of choice
\cite{tHooft:1972tcz}.

A field theory is solved when its correlation functions are known. In
perturbation theory, this corresponds to summing the relevant Feynman
graphs. The effective action functional is the classical action, or tree-level
action, that reproduces the correlation functions of a theory, while the full
propagator and full vertices contain all loop effects
\cite{Zinn-Justin:2002ecy}.  In \cite{delaPlata:1996kku}, the following
problem was posed and answered: If a perturbation is added to the action, must
the Feynman graphs be recomputed, including both the old and new vertices, to
obtain the new effective action?  Such procedure corresponds to a scheme
`$S_\text{free} + S_I + \delta S$', where $S = S_\text{free} + S_I$ is the
original action, and $\delta S$ is the perturbation.  In fact, it is not
mandatory to follow such route. The new effective action can also be obtained
using the scheme `$\Gamma + \delta S$', where $\Gamma$ is the effective action
of the original theory. In this scheme, standard Feynman graphs are used with
vertices of $\delta S$, as well as full vertices and full propagators of the
effective action, but include only graphs without unperturbed loops. This
means that every loop must include a vertex of $\delta S$. If the vertices of
$\delta S$ are deleted in one such graph, what remains is a tree graph of
$\Gamma$.

In this work, we explore how the above result can be exploited in the context
of renormalization theory. In a renormalizable theory, every correlation
function remains finite (barring coincident points) when the parameters in the
action are given a suitable dependence on the UV regulator, say $\Lambda$, as
the regulator is removed, i.e., for large $\Lambda$. In this way, the theory
becomes a renormalized one. In a theory that is already renormalized, one can
introduce a first-order perturbation in the action. In the
`$\Gamma + \delta S$' scheme, all the UV divergences from the loops inside
$\Gamma$ have been canceled, but the perturbation $\delta S$ introduces new
loops and thus new divergences that need a regulator. These new divergences
must be canceled through a suitable dependence on the regulator of the
parameters in $\delta S$. As usual, the cancellation of divergences leaves
finite ambiguities in the parameters. Rather than applying a minimal
subtraction scheme, the approach we adopt here is to enforce some
renormalization conditions on the correlation functions by defining some
renormalized parameters. Correspondingly, the subtractions appearing in, e.g.,
the forest formula, follow a projective renormalization scheme. The procedure
then yields the change in the correlation functions induced by the first-order
variation in the renormalized parameters.

This infinitesimal approach, which is implemented through a form of
Schwinger's principle, has an obvious limitation, namely, it does not directly
provide the renormalized effective action, only its variation as the
renormalized parameters change. On the other hand, the renormalization takes
place at the linear level, hence it is structurally simpler and the
divergences are softer. Another virtue is that the expressions involve the
full propagator and vertices, then the equations found are
non-perturbative. This means that each equation involves only a finite number
of graphs constructed with the propagator and vertices of the effective
action, rather than an infinite number of graphs of the action, much as the
Schwinger-Dyson equations do. Of course, a different question is that some of
the theories only exist, in principle, in a perturbative sense, i.e., as a
formal power series in the coupling or in $\hbar$. The fact that the equations
do not directly depend on perturbation theory is an interesting feature, which
could potentially be adapted for a non-perturbative construction of the
theory. This is not attempted in this work.

The linearized approach explored here is applied to super-renormalizable
theories, namely, scalar $\phi^3_4$ and $\phi^4_3$. The fact that in these
theories the UV divergences are rather mild allows us to obtain explicit and
fully renormalized non-perturbative equations for the variation of the
effective with respect to the renormalized parameters. No regulator is
needed. Not unexpectedly, the renormalizable case, e.g. $\phi^3_6$ or
$\phi^4_4$, is not so well behaved since the UV divergences are stronger
there. In the renormalizable case, expressions are found that still involve
only a finite number of graphs constructed with the propagator and vertices of
the effective action, however, the momentum integrals need to be
regulated. The renormalizability of the theory guarantees that there is a
finite limit after removing the regulator, but the limit itself cannot be
expressed in closed form.
In the renormalizable case, the equations display a finite number of explicit
loops, but the divergences involved are not accounted by a naive expectation
from these loops only.
Ultimately, the reason why the limit cannot be taken explicitly is that the
composite operators in the perturbation $\delta S$ activate subdivergences
involving vertices from $\delta S$ and propagator lines (of the action) hidden
within the effective action even if the latter has been renormalized.
Nevertheless, the enforcement of the renormalization conditions, which is
built into the formulation, ensures that when the equations are solved
perturbatively as a formal series in the renormalized coupling, each
contribution is a subtracted and UV-finite Feynman graph so that the regulator
can be removed there.

The subject of perturbing the action to first order is, of course, closely
related to the study of composite operators and their renormalization. In
fact, most of the tools developed in one case immediately apply to the other,
so this topic is also discussed.

The linearized renormalization is also closely related to the subject of
Schwinger-Dyson equations and their renormalization. In both cases, the
equations follow from graphs involving vertices and lines from the effective
action and from the action itself (in the case of the Schwinger-Dyson
equations) or from $\delta S$ in the linearized renormalization approach. The
two sets of equations are related and fulfill consistency conditions between
them.

The paper is organized as follows: Sec. \ref{sec:2} revisits the effective
action machinery. The concept of $\esp{A}^\phi$ and some related diagrammatic
properties are discussed. Our form of the Schwinger principle is also
introduced along with the definition of the coefficients $H$ and
$G$. Sec. \ref{sec:3} details the application of the linearized approach to
the super-renormalizable scalar field theories $\phi^3_4$ and $\phi^4_3$. The
result is an infinite hierarchy of non-perturbative equations that are free
from UV divergences and express the variation of the effective action with
respect to the renormalized parameters.
The renormalizable case is addressed in Sec. \ref{sec:4}. First, the
projective renormalization scheme is introduced.  This is presented within the
framework of a Bogoliubov-Parasiuk-Hepp-Zimmermann (BPHZ) approach, by
specifying the operation $T$ to be employed in the forest formula.  Then, the
linearized equations are constructed. As previously mentioned, these equations
are less explicit than those of the super-renormalizable case since the
regulator cannot be lifted in their non-perturbative form. However, the
regulator can be removed in a perturbative solution of the equations and
this yields subtracted UV-finite Feynman graphs.
Sec. \ref{sec:5} deals with the renormalization of composite operators,
building on the formalism previously developed for the linearized
renormalization. In the super-renormalizable case this approach produces
UV-finite non-perturbative expressions for the matrix elements of the
operators. The renormalization approach is adapted to the Schwinger-Dyson
equations in Sec. \ref{sec:6}. The renormalized Schwinger-Dyson equations
follow the same pattern as the linearized renormalization and the composite
operators; they are manifestly UV-finite in the super-renormalizable case but
not in the renormalizable one.
Some properties of the effective action exposed in Sec. \ref{sec:1}, including
the Schwinger principle, are proven in App. \ref{app:A}.  App.  \ref{app:B}
illustrates properties of the projective renormalization scheme developed in
Sec. \ref{sec:4a}. App. \ref{app:C} elucidates further the concept of
anti-canonical pattern discussed in Sec. \ref{sec:4b} by analyzing a simple
subcase of the scalar theory $\phi^3_6$. It also illustrates the consistency
of other results obtained in the main text. Finally App. \ref{app:D} discusses
a technical aspect of the linearized approach, namely, that the renormalized
results do not depend on the assumption that the effective action has already
been renormalized prior to its perturbation.

\section{\textsf{
The Schwinger action principle
  }}
\label{sec:2}

For simplicity, let us consider a theory $\phi^\kappa_d$, a single scalar
field $\varphi(x)$ with action
\be
S[\varphi] = \int d^dx \left(
  \frac{1}{2} Z (\partial_\mu\varphi)^2
+ \frac{1}{2} m^2 \varphi^2
+ \frac{1}{\kappa!} g \varphi^\kappa
\right)
.
\label{eq:2.1a}
\ee
Here, a Euclidean signature-like notation is used. For convenience, an explicit
wave function renormalization factor $Z$ has also been introduced.

The generating functional $Z[J]$ of the Green or correlation functions is then
\be
Z[J] = \int D\varphi \, e^{\int d^dx ( \cL(x)+J(x)\varphi(x) ) }
:= e^{W[J]}
\ee
so that
\bes
\esp{T \varphi(x_1)\cdots \varphi(x_n) }
&=
\frac{1}{Z[J]}
\frac{\delta}{\delta J(x_1)} \cdots
\frac{\delta}{\delta J(x_n)} Z[J] \Big|_{J=0}
,
\\
\esp{T \varphi(x_1)\cdots \varphi(x_n) }_c
&=
\frac{\delta}{\delta J(x_1)} \cdots
\frac{\delta}{\delta J(x_n)} W[J] \Big|_{J=0}
.
\ees

Here and in what follows, we will use a convention with Boltzmann weight
$e^{+S}$ instead of $e^{-S}$. This implies that $Z$, $m^2$, $g$, $S[\varphi]$,
$W[J]$, $\Gamma[\phi]$, etc, have non-standard signs. In particular, the free
propagator in momentum space becomes
\be
\Df(k) = - (Z k^2 + m^2)^{-1}
.
\ee

While this convention may be inconvenient for performing detailed
calculations, it enjoys the crucial advantage that there are no minus signs
between Green functions contributions and their Feynman graphs, nor in the
Feynman rule of the vertices.

\subsection{\textsf{
The effective action functional
  }}

Using the DeWitt notation $\varphi(x) \to \varphi^i$ (with $\varphi^i$ real
and $i$ including all labels present in the field) a general action takes the
form
\be
S[\varphi] =
c + h_i \varphi^i + \frac{1}{2} m_{ij}\varphi^i\varphi^j
+ \sum_{n\ge 3} \frac{1}{n!} g_{ i_1 \ldots i_n}
\varphi^{i_1} \cdots \varphi^{i_n}
\ee
or simply
\be
S[\varphi] =
\sum_{n\ge 0} \frac{1}{n!} g_{ i_1 \ldots i_n} \varphi^{i_1} \cdots
\varphi^{i_n}
.
\ee
Here $m_{ij}$ and $g_{ i_1 \ldots i_n}$ are completely symmetric covariant
tensors with respect to linear transformations of the $\varphi^i$, which are
contravariant.  The (free connected) propagator is
\be
s^{ij} := (-m^{-1})^{ij},
\qquad
s^{ij}m_{jk} = -\delta^i_k
.
  \ee

In this notation, the generating function is then
\be
Z[J] = e^{W[J]} = \int D\varphi \, e^{S[\varphi]+J_i \varphi^i }
\label{eq:2.7}
\ee
and
\be
\esp{\varphi^{i_1}\cdots \varphi^{i_n}}_c
=
\partial^{i_1} \cdots \partial^{i_n} W[J] \big|_{J=0}
,\qquad
\partial^i := \partial/\partial J_i
\,.
\ee

The effective action $\Gamma[\phi]$ will be used extensively. It is defined as
the Legendre transformation of the connected generating functional,
\bes
\Gamma[\phi] &= W[J] - J_i \phi^i ,
\qquad
\phi^i[J] = \partial^i W[J] = \esp{\varphi^i}^J,
\qquad
\\
J_i[\phi] &= - \partial_i \Gamma[\phi] ,
\qquad
\partial_i := \partial/\partial \phi^i
.
\ees
$\phi^i$ is known as the classical field. The expansion
\bes
\Gamma[\phi] &=
C
+ H_i \phi^i + \frac{1}{2} M_{ij}\phi^i\phi^j
+
\sum_{n\ge 3} \frac{1}{n!}
\fv_{ i_1 \ldots i_n} \phi^{i_1} \cdots \phi^{i_n}
\\ &=
\sum_{n\ge 0} \frac{1}{n!}
\fv_{ i_1 \ldots i_n} \phi^{i_1} \cdots \phi^{i_n}
,
\label{eq:2.2}
\ees
provides the full propagator (the connected two-point function)
\be
\fp^{ij} := (-M^{-1})^{ij} = \esp{\varphi^i \varphi^j }_c,
\qquad
\fp^{ij} M_{jk} = -\delta^i_k,
\label{eq:2.2a}
\ee
and the vertex functions $\fv_{ i_1 \ldots i_n}$, which are the connected,
irreducible amputated graphs of the action $S$.\footnote{Exceptionally, for
  $n=2$ these graphs produce the selfenergy $\Sigma_{ij}$ and
  $M_{ij}=m_{ij} + \Sigma_{ij}$.} In turn, the Green functions of $S$ are
obtained by using the standard Feynman rules with propagator $\fp^{ij}$ and
vertices $\fv_{ i_1 \ldots i_n}$ (for $n\neq 2$) {\em but including only tree
  graphs}. Hence $\Gamma$ is the action that produces classically (i.e. at
tree level) the correlation functions generated by $S$ quantum-mechanically
(i.e. allowing loops),
\be
Z[J] = e^{\Gamma[\phi] + J_i \phi^i}
.
\ee

\subsection{\textsf{
Schwinger's action principle
  }}

If the current $J$ is not set to zero in the functional integral, the
expectation value of an observable $A[\varphi]$ becomes
\be
\esp{A}^J =
\frac{1}{Z[J]} \int D\varphi \, e^{S[\varphi]+J_i\varphi^i} A[\varphi]
.
\ee
Let us introduce the notation
\be
\espp{A} := \esp{A}^{J=J[\phi]}
.
\ee
That is, $\espp{A}$ denotes the expectation value in the presence of the
current $J$ such that $\esp{\varphi}^J = \phi$. In particular
\be
\espp{\varphi^i} = \phi^i
.
\ee

Under a first-order variation of the action $S\to S+\delta S$, the generator
of the connected Green functions is modified as
\be
\delta W[J] = \esp{\delta S }^J
.
\ee
As a consequence, for the effective action one obtains immediately (see
Appendix \ref{app:A}) that
\be
\delta \Gamma[\phi] = \espp{ \delta S }
,
\label{eq:2.1}
\ee
which is likely a version of Schwinger's quantum action principle.
This identity will play a central role in this work.

The first-order variation in \Eq{2.1} is consistent (in the technical sense
$[\delta_1,\delta_2]=0$); therefore, if one wants to compute variations of
higher order it suffices to use the identity recursively. Alternatively, the
theorem proven in \cite{delaPlata:1996kku} applies: If $S$ is perturbed to
$S+S_I$ the new correlation functions follow from using the Feynman rules of
`$\Gamma+S_I$' (namely, the propagator and vertices of $\Gamma$ plus vertices
of $S_I$) but retaining only graphs such that any loop must have at least one
vertex of $S_I$.

\subsection{\textsf{
    The coefficients $H
    $
\label{sec:3.a}
}}

Clearly, when the current $J$ is not removed, the expectation values
$\esp{A}^J$ can be computed with the same Feynman rules of $S$ but using
$h_i+J_i$ as the new 1-point vertex. A related result that will be needed is
as follows:

\miem{Theorem} ~For $n\ge 2$, the expectation values
$\espp{\varphi^{i_1}\cdots \varphi^{i_n}}$ are obtained from the Feynman
rules at the tree level but using the following propagator (line) and
vertices:
\bes
\hfp^{ij}[\phi] &:= ((-\partial^2\Gamma[\phi])^{-1})^{ij}
,
\\
\qquad
\hat{\Gamma}_{i_1\ldots i_n}[\phi] &:=
\left\{ \ \begin{matrix}
\partial_{i_1}\cdots \partial_{i_n}
\Gamma[\phi]
&
n \ge 3
\\
0 & n \le 2
\end{matrix}
\right.
\,.
\label{eq:2.19}
\ees
This statement is proven in Appendix \ref{app:A}. Obviously when $\phi$ is
set to $0$, $\hfp^{ij}[\phi]$ and $\hat{\Gamma}_{i_1\ldots i_n}[\phi]$
become $\fp^{ij}$ and $\Gamma_{i_1\ldots i_n}$ introduced in \nec{2.2a} and
\nec{2.2}.

Applying the Theorem, for $n=1,2,3$ one obtains\footnote{The case $n=1$ is
  included for completeness; it is not derived from the above rules.}
\bes
\espp{\varphi^i} &= \phi^i  ,
\\
\espp{\varphi^i\varphi^j}_c &= \hfp^{ij}[\phi]  ,
\\
\espp{\varphi^i\varphi^j\varphi^k}_c &=
\hfp^{ia}[\phi]
\hfp^{jb}[\phi]
\hfp^{kc}[\phi]
\hat{\Gamma}_{abc}[\phi]  .
\label{eq:20}
\ees
 Schematically,
\bes
\espp{\varphi} &= \phi ,
\\
\espp{\varphi^2}_c &= \hfp ,
\\
\espp{\varphi^3}_c &= \hfp^3\hat{\Gamma}_3 ,
\\
\espp{\varphi^4}_c &= \hfp^4\hat{\Gamma}_4 + 3\times
\hfp^2\hat{\Gamma}_3 \hfp \hat{\Gamma}_3 \hfp^2
.
\ees

The rule is that each $\partial^i= \partial/\partial J_i$ generates a new leg
from either a vertex ($\hat{\Gamma}_n$) or a line ($\hfp$). Likewise
$\partial_j= \partial/\partial \phi^j$ generates an amputated leg because
$\partial^i = \hfp^{ij}[\phi]\partial_j$. So, for instance
\bes
\espp{\varphi^i\varphi^j\varphi^k}_c
&=
\partial^i \espp{\varphi^j\varphi^k}_c
=
\hfp^{ia} \partial_a \hfp^{jk}
\\
&= \hfp^{ia} \hfp^{jb} \hfp^{kc} \hat{\Gamma}_{abc}
.
\ees

Let us introduce the family of functionals $\hH$ from
\be
\espp{ \varphi^{i_1}\cdots \varphi^{i_\ell} }
=
\phi^{i_1}\cdots \phi^{i_\ell}
+ 
\hH^{i_1\ldots i_\ell}[\phi]
.
\label{eq:2.20}
\ee
Note that these expectation values are not connected. Also, for
convenience, the completely disconnected term has been removed from the
definition of $\hH$. Hence, in particular $\hH^{i_1\ldots i_\ell}[\phi]$
vanishes for $\ell=0,1$.  Explicitly for $\ell=2,3$ one has
\bes
\hH^{i_1 i_2 }[\phi] &= \hfp^{i_1 i_2}[\phi]
\\
\hH^{i_1 i_2 i_3}[\phi] &=
\phi^{i_1} \hfp^{i_2 i_3}[\phi]
+
\phi^{i_2} \hfp^{i_1 i_3}[\phi]
+
\phi^{i_3} \hfp^{i_1 i_2}[\phi]
\\&\quad +
\hfp^{i_1 a}[\phi] \hfp^{i_2 b}[\phi] \hfp^{i_3 c}[\phi] \hat\Gamma_{abc}[\phi]
\,.
\ees
Schematically
\bes
\hH^2 &= \hfp ,
\\
\hH^3 &= 3 \times \phi \hfp + \hfp^3\hat{\Gamma}_3 ,
\\
\hH^4 &=
6 \times \phi^2 \hfp
+ 4 \times \phi \hfp^3\hat{\Gamma}_3
+ 3 \times \hfp^2
\\ &\quad
+ 3\times \hfp^2 \hat{\Gamma}_3 \, \hfp \, \hat{\Gamma}_3 \, \hfp^2
+ \hfp^4\hat{\Gamma}_4
,
\ees
and one can define the coefficients $H$ from
\be
\hH^{i_1\ldots i_\ell}[\phi] =
\sum_{n=0}^\infty \frac{1}{n!} H^{i_1\ldots i_\ell}_{j_1\ldots j_n}
\phi^{j_1}\cdots\phi^{j_n}
\,.
\label{eq:2.23}
\ee
These coefficients $H^{i_1\ldots i_\ell}_{j_1\ldots j_n}$ represent tree level
$\Gamma$-graphs (in general disconnected) with legs $i_1, \ldots, i_\ell$
(contravariant) plus the amputated legs $j_1,\ldots, j_n$ (covariant),
constructed with the effective action propagator $\fp^{ab}$ and vertices
$\fv_{a_1\ldots a_k}$ (hence $\Gamma$-graphs). They are fully symmetric
tensors with respect to the covariant and contravariant indices separately.

Since \nec{2.23} represents a Taylor expansion in $\phi$, the coefficients $H$
are easily constructed by applying $\partial_j$ on $\hH[\phi]$ to produce the
amputated legs $j_1, \ldots, j_n$, and then setting $\phi=0$.
Diagrammatically, this corresponds to recursively extracting the amputated
legs in all possible ways from the initial graph of
$\hH^{i_1\ldots i_\ell}[\phi]$. The legs are extracted from explicit $\phi$
(once), from unamputated legs and from vertices, but not from amputated
legs. At the end $\phi$ must be set to zero for each coefficient, but not of
course during the recursion.

For $\ell=2$ one obtains
\bes
H^{i_1i_2} &= \fp^{i_1 i_2}
,
\\
H^{i_1i_2}_{j_1} &= \fp^{i_1 a}\fp^{i_2 b}
\fv_{j_1 a b }
,
\\
H^{i_1i_2}_{j_1 j_2} &=
\fp^{i_1 a}\fp^{i_2 b} \fv_{j_1 j_2 a b }
\\ & \quad
+
\fp^{i_1 a} \fp^{i_2 b} \fp^{cd} \fv_{j_1 ac } \fv_{j_2 db }
+
\fp^{i_1 a} \fp^{i_2 b} \fp^{cd} \fv_{j_2 ac } \fv_{j_1 db }
.
\\
\ees
Furthermore, similarly for higher values of $n$.

Likewise, for $\ell=3$ the extraction of the amputated legs $j_1,\ldots,j_n$
in all possible ways yields
\bes
H^{i_1 i_2 i_3} &= \fp^{i_1 a} \fp^{i_2 b} \fp^{i_3 c} \fv_{abc}
,
\\
H^{i_1 i_2 i_3}_{j_1} &=
\delta^{i_1}_{j_1} \fp^{i_2 i_3}
+
\delta^{i_2}_{j_1} \fp^{i_1 i_3}
+
\delta^{i_3}_{j_1} \fp^{i_1 i_2}
\\
& \quad
+
\fp^{i_1 d}\fv_{j_1 d e}  \fp^{e a} 
\fp^{i_2 b} \fp^{i_3 c} \fv_{abc}
\\
& \quad+
\fp^{i_1 a} \fp^{i_2 d}\fv_{j_1 d e}  \fp^{e b} 
\fp^{i_3 c} \fv_{abc}
\\ & \quad
+
\fp^{i_1 a} \fp^{i_2 b}
\fp^{i_3 d}\fv_{j_1 d e}  \fp^{e c} 
\fv_{abc}
\\
& \quad
+
\fp^{i_1 a} \fp^{i_2 b} \fp^{i_3 c} \fv_{j_1 abc}
.
\\
\ees

For convenience, let us also introduce the notation
\be
\hG^{i_1\ldots i_\ell}[\phi] = \espp{ \varphi^{i_1}\cdots \varphi^{i_\ell} }
,
\ee
that is, as $\hfh^{i_1\ldots i_\ell}[\phi]$ but including the completely
disconnected term. These functionals can be collected into generating
functionals, namely,\footnote{Here $J$ is an independent variable, not
  $J[\phi]$.}
\bes
G[J,\phi] &=
\espp{e^{J_i \varphi^i} }
=
\frac{ Z[J+J[\phi]] } {Z[J[\phi] ]}
\\
&=: e^{J_i \phi^i} + H[J,\phi]
,
\ees
so that
\bes
G[J,\phi]
&=
\sum_{\ell=0}^\infty \frac{1}{\ell!}
J_{i_1} \cdots J_{i_\ell}
\hG^{i_1\ldots i_\ell}[\phi]
\\
&= \sum_{\ell=0}^\infty \sum_{n=0}^\infty
\frac{1}{\ell!} \frac{1}{n!}
G^{i_1 \ldots i_\ell}_{j_1 \ldots j_n}
J_{i_1} \cdots J_{i_\ell}
\phi^{j_1} \cdots \phi^{j_n}
.
\ees

In addition, for an observable $A[\varphi]$,
\be
A[\varphi] = \sum_{\ell=0}^\infty \frac{1}{\ell!}
A_{i_1 \ldots i_\ell}
\varphi^{i_1} \cdots \varphi^{i_\ell}
,
\ee
one can define
\be
G^A[\phi] := A[\phi] + H^A[\phi]
:=
\espp{A[\varphi]}
.
\label{eq:2.33}
\ee
This functional can be expanded as
\be
G^A[\phi] =  
\sum_{n=0}^\infty
\frac{1}{n!}
G^A_{j_1 \ldots j_n}
\phi^{j_1} \cdots \phi^{j_n}
,\ee
where
\be
G^A_{j_1 \ldots j_n} =
\sum_{\ell=0}^\infty \frac{1}{\ell!}
A_{i_1 \ldots i_\ell}
G^{i_1 \ldots i_\ell}_{j_1 \ldots j_n}
.
\ee

From Schwinger's principle, it follows that if $S[\phi]$ is perturbed by adding
a term $\delta S[\phi]=\delta \lambda A[\phi]$,
\be
\delta \Gamma[\phi] = \delta \lambda \espp{ A[\varphi] }  =
\delta \lambda G^A[\phi]
.
\label{eq:2.36}
\ee
Hence, diagrammatically, $G^A_{j_1 \ldots j_n}$ is given by connected and
irreducible graphs with amputated legs $j_1 \ldots j_n$, constructed with the
propagator and the vertices of the action plus exactly one vertex of the
composite operator $A[\phi]$.\footnote{The Feynman rules of these vertices are
  $A_{i_1 \ldots i_\ell}$ and diagrammatically are represented by a single
  point.} We will refer to the $G^A_{j_1 \ldots j_n}$ as the {\em amputated
  matrix elements} of $A$, since the standard matrix elements of $A$,
$\esp{A \varphi^{i_1}\cdots \varphi^{i_m}}$, can be recovered from them, to
wit, by adding tree graphs of $\Gamma$ (in general several trees) in all
possible ways saturating all amputated legs $j_r$ and fields $i_s$, provided
that each tree contains at most one end of type $j_r$ (so that no new loops
are generated).

\subsection{\textsf{
Momentum representation conventions
\label{sec:2.a}
}}

In momentum space we use the following conventions:
\be
\Gamma[\phi] = \sum_{n\ge 0} \frac{1}{n!}\int
\prod_{i=1}^n\frac{d^d p_i}{(2\pi)^d} 
\tilde\Gamma_n(p_1,\ldots,p_n) \tilde\phi_1(p_1)\cdots\tilde\phi_n(p_n)
\label{eq:3.4}
\ee
with $\tilde\phi(p)= \int d^dx \, e^{ipx}\phi(x)$ and
\be
\tilde \Gamma_n(p_1,\ldots,p_n)
=
(2\pi)^d\delta(p_1+\cdots +p_n)
\Gamma_n(p_1,\ldots,p_{n-1})
\label{eq:3.5}
\ee
due to invariance under translations.\footnote{The functions
  $\Gamma_n(p_1,\ldots,p_n)$ or simply $\Gamma_n(p)$, are only defined in the
  subspace $\sum p=0$; therefore, depending on the context, we may write them as
  $\Gamma_n(p_1,\ldots,p_{n-1})$ for convenience. The same notational liberty
  will be taken for similar functions, e.g. $H_n(q;p)$ below.}

Likewise, we introduce the momentum version of the coefficients
$H^{ i_1 \ldots i_\ell }_{ j_1 \ldots j_n }$, denoted
\bes
\tH^\ell_n & (q_1,\ldots,q_\ell;p_1,\ldots,p_n) \\ &= 
(2\pi)^d \delta \big( \sum_{i=1}^\ell q_i
+ \sum_{j=1}^n p_j \big) H^\ell_n(q_1,\ldots,q_\ell;p_1,\ldots,p_n)
\ees
or simply $(2\pi)^d \delta( \sum q + \sum p ) H^\ell_n(q;p)$. Explicitly, for
$n=\ell=2$,
\bes
H^2_2(q_1, & q_2;p_1,p_2) =
\\ &
\fp(q_1) \fp(q_2) \fp(q_1+p_1) \Gamma_3(p_1,q_1,q_1+p_1)
\\ &\quad +
\fp(q_1) \fp(q_2) \fp(q_1+p_2) \Gamma_3(p_2,q_1,q_1+p_2)
\\ &\quad +
\fp(q_1) \fp(q_2) \Gamma_4(p_1,p_2,q_1,q_2)
,
\ees
where
\be
\fp(p) := (-\Gamma_2(p))^{-1}
\ee
denotes the full propagator (connected 2-point function).

Similarly $H^A_{j_1 \ldots j_n}$ corresponds to $\tH^A_n(p)$, or
$H^A_n(p)$ after extracting $(2\pi)^d \delta( \sum p )$ if $A[\phi]$ is
translationally invariant, and the same goes for $G^A_n(p)$.

In what follows, different types of Feynman graphs will appear:
\begin{itemize}
\item[i)] $S$-graphs. They are constructed using the free propagator $\Df(k)$
  and the vertex of the action $g$, equivalently, $s^{ij}$ and
  $g_{ i_1 \ldots i_n}$.
  \item[ii)] $\Gamma$-graphs. They are constructed using the full propagator $\fp(k)$ and the full vertices
    $\Gamma_n(p)$, equivalently, $\fp^{ij}$ and $\Gamma_{ i_1 \ldots i_n}$.
  \item[iii)] $\hat{\Gamma}$-graphs. These are those of the Theorem, using
    $\hfp^{ij}[\phi]$ as line and $\hat{\Gamma}_{i_1\ldots i_n}[\phi]$ as
    vertices and $\phi$ is not set to zero.
\end{itemize}

Here, and in what follows, we use a Euclidean signature.

\section{\textsf{
    Linearized renormalization of super-renormalizable theories
  }}
\label{sec:3}

We want to investigate the idea of using the Schwinger principle \nec{2.1} to
perform the renormalization of a renormalizable theory. In this Section we
start with the more amenable case of a super-renormalizable theory.

A (first-order) infinitesimal variation of the bare action $S$, in DeWitt
notation
\be
\delta S[\phi] = 
\sum_{n\ge 0} \frac{1}{n!} \delta g_{ i_1 \ldots i_n} \phi^{i_1} \cdots
\phi^{i_n}
,
\ee
will induce an infinitesimal variation in the effective action
\bes
\delta \Gamma[\phi]
&= \espp{\delta S[\varphi]}
\\
&= \delta S[\phi] +
\sum_{ \ell \ge 2} \frac{1}{\ell!} \delta g_{ i_1 \ldots i_\ell}
 \hH^{ i_1 \ldots i_\ell } [\phi] 
,
\label{eq:3.22aa}
\ees
and
\be
\delta \Gamma_{j_1 \ldots j_n} =
\delta g_{ j_1 \ldots j_n}
+
\sum_{\ell \ge 2} \frac{1}{\ell!} \delta g_{ i_1 \ldots i_\ell}
H^{ i_1 \ldots i_\ell }_{ j_1 \ldots j_n }
.
\label{eq:3.22a}
\ee

In the spirit of \cite{delaPlata:1996kku}, our point of view is that
$\Gamma[\phi]$ is already renormalized and the correlation functions are free
from ultraviolet (UV) divergences, and the same goes for the functionals $\hH$
and the coefficients $H$, which are finite combinations of the propagators and
vertices of $\Gamma[\phi]$. However, new UV divergences arise in
$\delta \Gamma[\phi]$ since $\esp{\delta S[\phi] }$ requires the expectation
value of the composite local operators present in the action. In \Eq{3.22a}
the divergences are introduced by the sum over the indices
$ i_1 \ldots i_\ell$, or equivalently due to loops in a momentum space
representation. To render $\delta \Gamma[\phi]$ finite, such divergences have
to be canceled by divergences in the bare parameters of the action,
$\delta g_{i_1 \ldots i_\ell}$.

\subsection{\textsf{ The $\phi^3_4$ theory 
   \label{sec:3.b}
  }}

In order to analyze this matter, let us consider the theory $\phi^3_4$ (i.e.,
$\kappa=3$ and $d=4$ in \nec{2.1a}) which is super-renormalizable. In this
theory, the bare coupling and the bare wavefunction renormalization factor are
finite and need not be renormalized; they will be denoted $g$ and $Z$ (rather
than $g_0$ and $Z_0$). Only the bare mass, $m_0$, requires renormalization, so
we only introduce one renormalization condition,
\be
m_R^2 = \Gamma_2(p)\Big|_{p=0} \,.
\label{eq:3.a1}
\ee
The parameters $Z$ and $g$ are those of the action but they can also be
recovered from the effective action in the large momentum limit (\Eq{3.34}
below).

We will use a Euclidean cutoff $\Lambda$ to regularize the UV divergences. In
the $\phi^3_4$ theory, there is a single primitive divergent $S$-graph,
namely, \includegraphics[height=8mm]{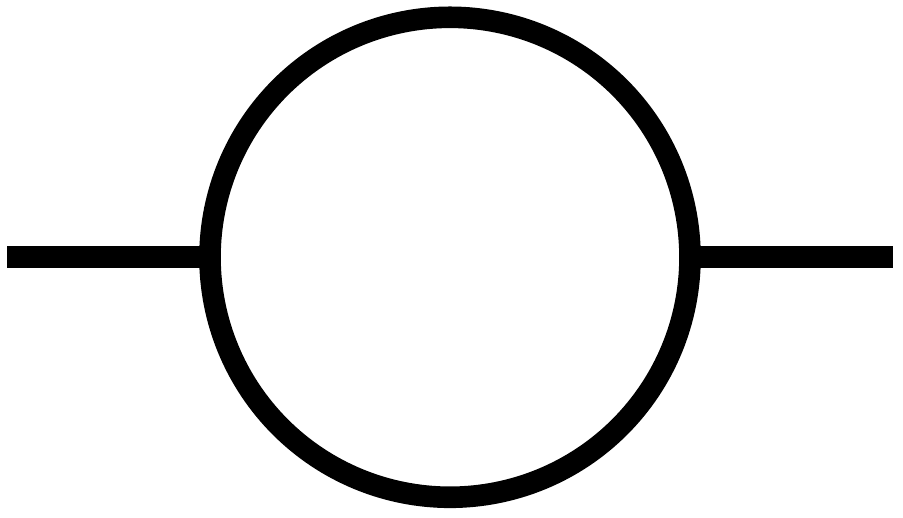}, which is canceled by using
as the bare mass
\be
m_0^2 = m_R^2 + m^2_\ct
\label{eq:3.19}
\ee
with
\be
m^2_\ct(\Lambda) = - \frac{1}{2} \frac{g^2}{Z^2}\Omega_4 L_\Lambda
+ \mathrm{s.d.t.}
,
\label{eq:3.22}
\ee
where ``s.d.t.'' stands for subdominant terms.
Also, we have introduced the notations
\be
L_\Lambda := \log(\Lambda/\mu)
\ee
and
\be
\int \frac{d^d q}{(2\pi)^d} f(q^2)
= \Omega_d \int_0^\infty dt \, t^{d-1} f(t^2)
.
\ee
The value of the scale $\mu$ is not relevant as the renormalization condition
eliminates its dependence. Using the cutoff action with bare mass
$m_0(\Lambda)$ in the Feynman rules and then removing the cutoff produces a
finite $\Gamma[\phi]$. This is the standard approach.

Now, our point of view will be that the effective action $\Gamma[\phi]$ of the
$\phi^3_4$ theory is already renormalized and so it no longer depends on any
regulators. Such an effective action provides finite values for the
correlation functions.\footnote{Of course, new divergences arise as two or
  more fields are located at ever closer points to produce a composite
  operator.}  The functional $\Gamma[\phi]$ is completely determined by the
values of the parameters $m_R^2$, $Z$, and $g$. We consider a first-order
variation in the action
\be
\delta S = \delta m_0^2 \, O^m + \delta Z \, O^Z + \delta g \, O^g
.
\label{eq:3.9}
\ee
Here all three operators include cutoffs through profile factors $F$, which we
assume to be equal for simplicity:
\bes
O^m &:= \int d^d x \frac{1}{2} \hat\phi^2(x) ,
\\
O^Z &:= \int d^d x  \frac{1}{2} (\partial_\mu \hat\phi(x))^2 ,
\\
O^g &:= \int d^d x \frac{1}{\kappa!} \hat\phi^\kappa(x) .
\label{eq:3.2a}
\ees
with $d=4$ and $\kappa=3$, and
\be
\hat{\phi}(x):=
F(-\partial^2/\Lambda^2)\phi(x) 
.
\label{eq:3.11}
\ee

Here $F(x)$ is a real decreasing smooth function rapidly approaching $1$ as
$x \to 0$, and $0$ as $x \to \infty$. Valid convergence conditions are
\be
\forall \alpha \in \R \quad
\lim_{x\to 0^+} x^\alpha ( F(x) -1)  = \lim_{x\to +\infty} x^\alpha F(x)  = 0
.
\ee
For convenience, we assume that $F(x) = 1$ for $0\le x \le 1$ and $F(x) = 0$
for $x\ge 2$; otherwise $F$ is smooth and decreasing. The effect of the
regulator $F$ is to suppress the interaction at momenta well above
$\Lambda$. For regularization in coordinate space see
\cite{Ivanov:2022grx,Ivanov:2024xpr}.

At the level of the Feynman rules in momentum space, the interaction vertex of
$\delta S$ are
\be
(\delta m_0^2 + p^2 \delta Z ) F^2(p^2/\Lambda^2)
\ee
for $\delta m_0^2 \, O^m + \delta Z \, O^Z$
and
\be
\delta g \prod_{j=1}^\kappa F(p_j^2/\Lambda^2)
\ee
for $\delta g \, O^g$, where the $p_j$ denote the momenta at the $\kappa$-point
vertex. In the presence of $\Lambda$, that is, before taking the limit
$\Lambda\to \infty$, all divergences from the loops are regulated in
$\esp{\delta S[\phi]}$.

In \nec{3.9} 2- and 3-point operators are present. Terms corresponding to $0$-
and $1$-point vertices should also be included in $\delta S[\phi]$ to
renormalize the $0$- and $1$-point vertex functions in $\delta \Gamma[\phi]$,
but we will often omit them as they will not be relevant for the discussion
(they do not induce divergences on $m_0^2$).

\begin{figure}[ht]
  \begin{center}
    \includegraphics[height=28mm]{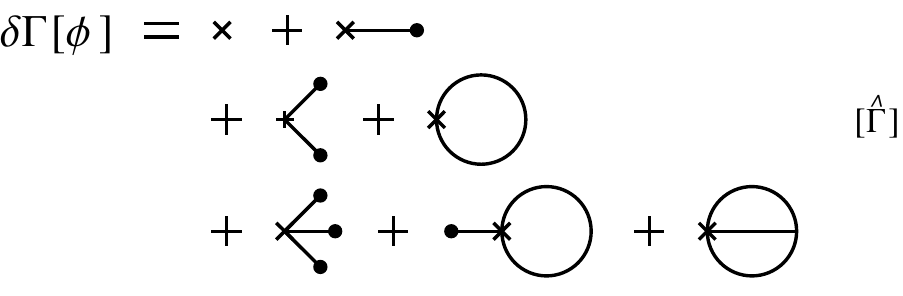}
    \end{center}
    \caption{ Graphs contributing to $\delta \Gamma[\phi]$ (\Eq{3.26}) for a
      variation $\delta S[\phi]$ with two- and three-point vertices
      (\Eq{3.25a}). These are $\hat\Gamma$-graphs (indicated by the tag
      $[\hat{\Gamma}]$ in the figure), that is, the lines and (uncrossed)
      vertices in the graphs are $\hfp^{ij}[\phi]$ and
      $\hat{\Gamma}_{i_1\ldots i_n}[\phi]$ (without setting the field $\phi$
      to zero).  The vertices represented by a cross are those of $\delta S$,
      i.e., $\delta g_{i_1 \ldots i_\ell}$ (for $\ell=2,3$). In the figure,
      the contributions from the vertices of $\delta S$ of $0$- and $1$-points
      --omitted in the formulas-- have been included by completeness.}
    \label{fig:2}
  \end{figure}

For a $\phi^3$ theory
\be
\delta S = \frac{1}{2} \delta m_{ij} \phi^i\phi^j
+ \frac{1}{3!} \delta g_{ ijk} \phi^i \phi^j \phi^k
\label{eq:3.25a}
\ee
\Eq{3.22aa} yields
\be
\delta \Gamma[\phi] = \delta S[\phi] +
\frac{1}{2} \delta m_{ij} \hH^{ij}[\phi] 
+ \frac{1}{3!} \delta g_{ ijk} \hH^{ijk}[\phi] 
.
\label{eq:3.26}
\ee
The graphs are displayed in \fig{2}. A key point here is that the number of
$\hat\Gamma$-graphs involved is finite. Upon expansion
\bes
\delta \Gamma_{j_1\ldots j_n} &=
\delta m_{j_1 j_2} \delta_{n,2}
+
\delta g_{j_1 j_2 j_3} \delta_{n,3}
+
\\ & \quad
\frac{1}{2} \delta m_{ij} H^{ij}_{j_1\ldots j_n}
+ \frac{1}{3!} \delta g_{ijk} H^{ijk}_{j_1\ldots j_n}
.
\label{eq:3.26a}
\ees
The number of $\Gamma$-graphs is also finite for each given $n$. They
correspond to an infinite number of $S$-graphs.

In a momentum representation, \nec{3.25a} becomes
\be
\delta S_n (q) = \big( \delta m_0^2
+
\delta Z q_1^2 \big)  F^2_{q_1}  \delta_{n,2}
+
\delta g F_{q_1} F_{q_2} F_{q_3} \delta_{n,3}
,
\ee
where $F_q := F(q^2/\Lambda^2)$. In turn, \nec{3.26a} becomes
\bes
\delta \Gamma_n (p) &= \delta m_0^2  \big( \delta_{n,2} + \fh^m_n (p) \big)
\\ & \quad +
\delta Z \big( \delta_{n,2} p^2 + \fh^Z_n (p) \big)
\\ & \quad +
\delta g \big( \delta_{n,3}
+ \fh^g_n (p) \big)
.
\label{eq:3.32a}
\ees
Here, we use $p^2$ to refer to $p_1^2$ (or $p_2^2$), but more importantly, in
writing this expression, we have dropped explicit factors $F_p$ (they are
implicit). The form factor $F_q$ is relevant when $q$ can be large, but does
not differ from $1$ for momenta that are never in the UV region, so we adopt
this convention to have simpler expressions.

The contributions from the loops have (relevant) form factors in the internal
lines. Explicitly,
\bes
\fh^m_n (p) &= \int \frac{d^d q}{(2\pi)^d} F_q^2 A_n(q;p)
\\
\fh^Z_n (p) &= \int \frac{d^d q}{(2\pi)^d} F_q^2 q^2 A_n(q;p)
\\
\fh^g_n (p) &= \int \frac{d^d q_1}{(2\pi)^d} \frac{d^d q_2}{(2\pi)^d}
F_{q_1} F_{q_2} F_{q_1+q_2}  B_n(q;p)
\label{eq:3.32b}
\ees
with $d=4$ and
\bes
A_n(q;p) &:= \frac{1}{2} H^2_n(q,-q;p),
\\
B_n(q;p) &:= \frac{1}{3!}  H^3_n(q_1,q_2,-q_1-q_2;p)
.
\label{eq:3.32}
\ees
The functions $\fh^\alpha_n(p)$ (where the label $\alpha$ refers to $m$, $Z$ or
$g$) as well as $A_n$ and $B_n$, are only defined in the subspaces $\sum p=0$
and $\sum q=0$.

The functions $A_n(q;p)$ and $B_n(q;p)$ depend only on the effective action
and not on the regularization details nor the cutoff. They can be represented
diagrammatically as
\bes
A_n(q;p) &= \includegraphics[height=12mm]{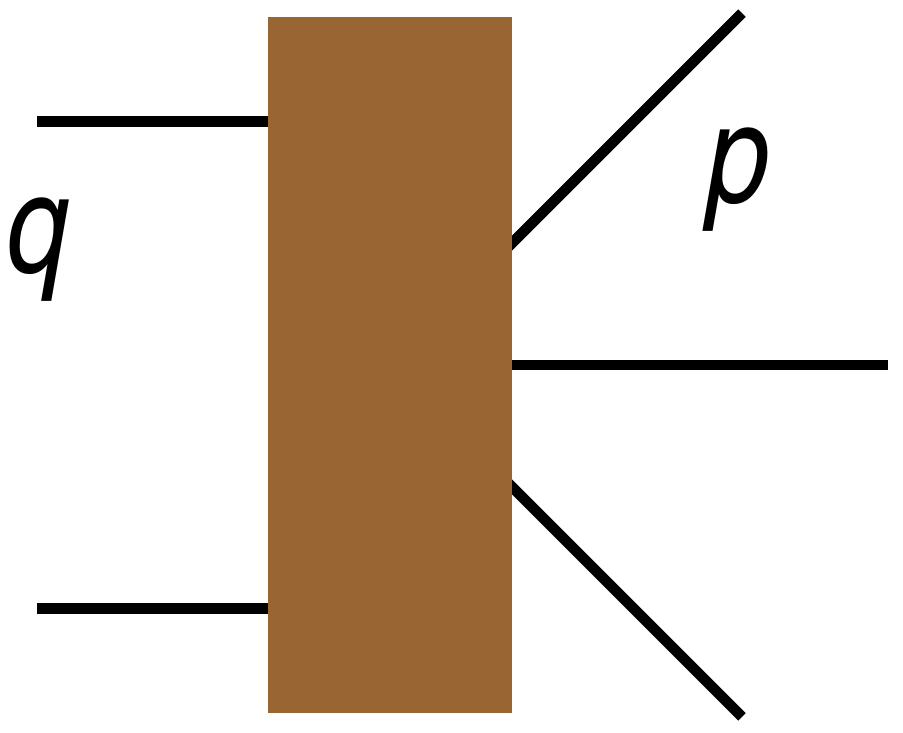}
\\
B_n(q;p) & = \includegraphics[height=12mm]{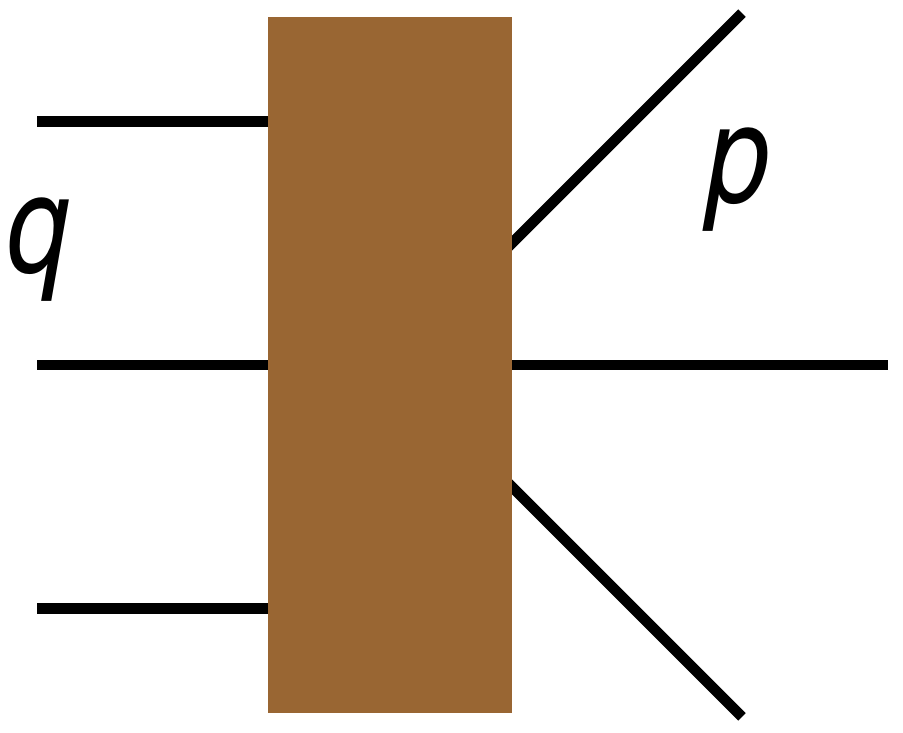}
\ees
It should be recalled that in these functions, the $n$ lines $p$ are
amputated, but not the lines $q$. Likewise $\fh^m_n (p)$, $\fh^Z_n (p)$ and
$\fh^g_n (p)$ can be represented as
\bes
\fh^m_n (p) &=  \includegraphics[height=12mm]{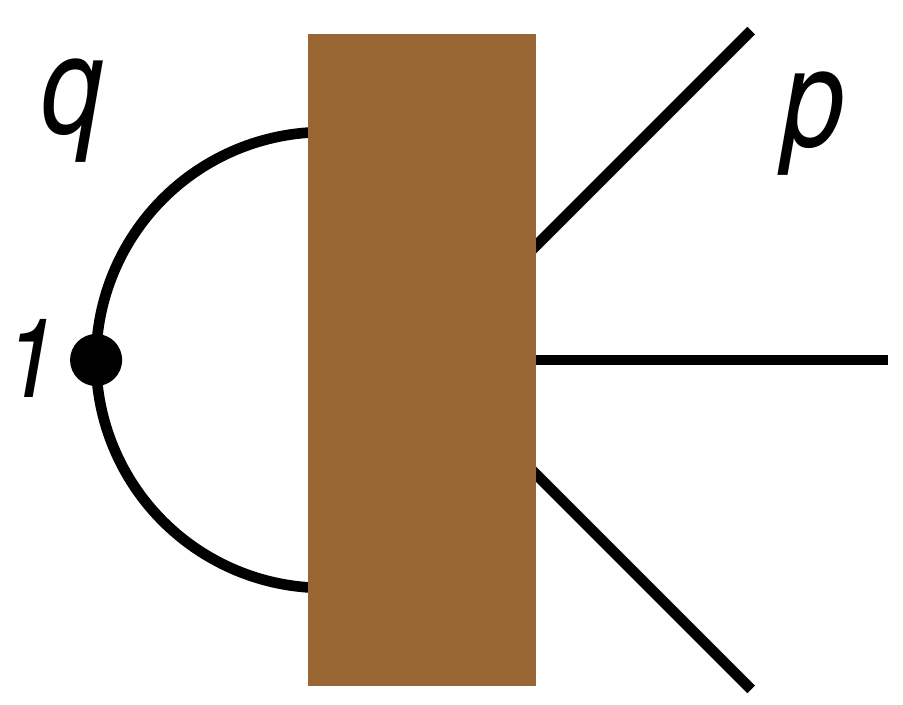}
\\
\fh^Z_n (p) &=   \includegraphics[height=12mm]{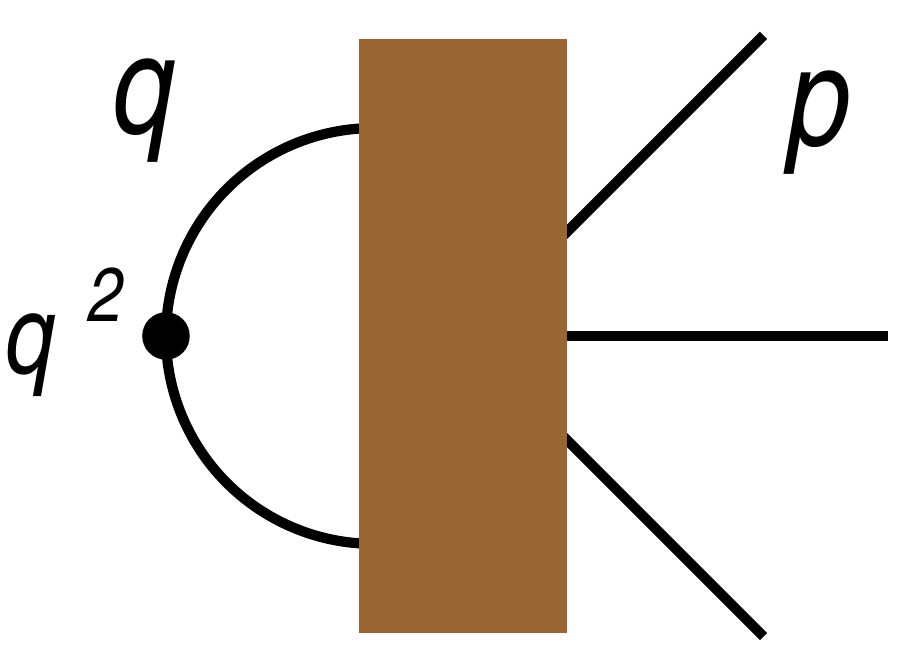}
\\
\fh^g_n (p) &=   \includegraphics[height=12mm]{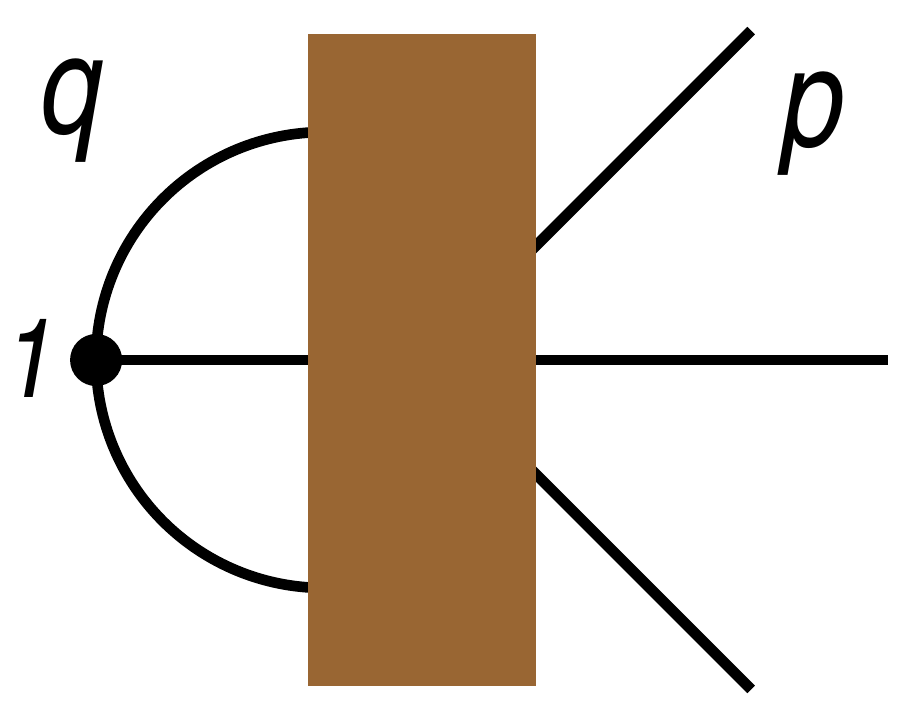}
\label{eq:3.23}
\ees

\subsubsection{\textsf{ Renormalization from $\delta  Z$
  }}

To simplify the discussion, we will start with the case $\delta g=0$.  In the
theory $\phi^3_4$ the mass term $\delta m_0^2$ is divergent but does not
itself introduce divergences; instead the divergences are induced on
$\delta m_0^2$ by $\delta Z$ and $\delta g$, which are finite.\footnote{This
  is already clear from the form of $m^2_\ct$ in \nec{3.22}, which depends on
  $Z$ and $g$ but not on the mass.} Therefore, while the mass term
$\delta m_0^2$ cannot be assumed to vanish if either $\delta Z$ or $\delta g$
are present, the choices $\delta g=0$ or $\delta Z=0$ are consistent. This
is no longer true in the renormalizable case (e.g., the theory $\phi^3_6$).

\begin{figure}[ht]
  \begin{center}
    \includegraphics[height=24mm]{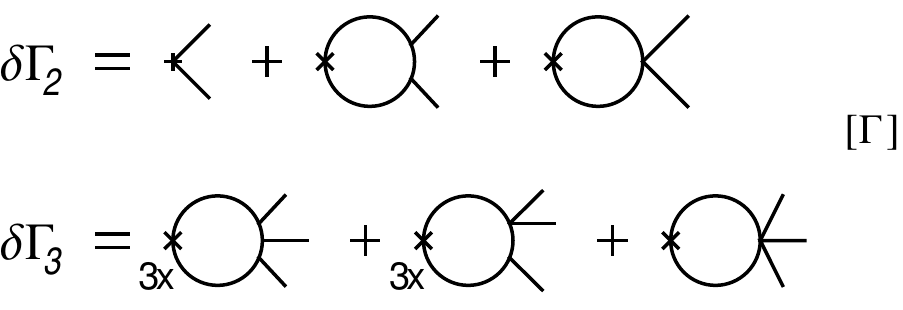}
    \end{center}
    \caption{ For the theory $\phi^3_4$, graphs contributing to
      $\delta \Gamma_2(p)$ and $\delta \Gamma_3(p)$, \Eq{3.32a} for
      $\delta g=0$. The lines and (uncrossed) vertices in the graphs are those
      of $\Gamma[\phi]$, i.e., $\fp(p)$ and $\Gamma_n(p)$, hence these are
      $\Gamma$-graphs. This is indicated by the tag $[\Gamma]$ in the
      figure. The vertices represented by a cross are those of $\delta S$,
      with Feynman rule $F_q^2 (\delta m_0^2 + q^2 \delta Z)$. The $p$-legs
      are distinguishable; therefore, the first two graphs of $\delta
      \Gamma_3(p)$ have three versions.}
\label{fig:3}
\end{figure}

The contributions to $\Gamma_2(p)$
\be
\delta \Gamma_2 (p) = \delta m_0^2 \big( 1 + \fh^m_2 (p) \big)
+ \delta Z \big( p^2 + \fh^Z_2 (p) \big)
\label{eq:3.36a}
\ee
are displayed in the first row of \fig{3}.

The asymptotic UV behavior of $S$-graphs is described by Weinberg's theorem
\cite{Weinberg:1959nj,Collins:1984xc}. In the $\phi^3_4$ theory, the effect of
the quantum fluctuations is suppressed in the asymptotic region; for large
momenta $q$ with fixed $p$ and $k$,\footnote{For a quantity $f(x)$ defined as
  a formal power series, $f(x) = \sum_k f_k(x) g^k$, the statement
  $f(x)=O(x^n)$ should be understood as $f_k(x)=O(x^n) ~\forall k$.}
\bes
\fp(q) &= \Df(q) + O(\log(q^2)/q^4)
\\
\Gamma_3(q,k-q,p) &= g + O(\log(q^2)/q^2)
\\
\Gamma_n(q,k-q,p) &=  O(1/q^2) \quad n \ge 4
,
\label{eq:3.34}
\ees
where
\be
\Df(q) := - \frac{1}{ Z q^2 + m_R^2 }
.
\ee
The first \Eq{3.34} would be equally correct using any other mass scale
instead of $m_R$. The choice adopted might be preferable to connect with
perturbation theory and the BPHZ renormalization scheme \cite{Collins:1984xc}.

Due to of this asymptotic behavior and attending to the graphs in
\fig{3}, $\fh_2^m(p)$ is UV-finite as $\Lambda \to \infty$, while $\fh_2^Z(p)$
has a logarithmic divergence. In \nec{3.36a} the divergence from
$\delta Z \fh_2^Z(p)$ must be canceled by a logarithmically divergent
$\delta m_0^2(\Lambda)$, which cannot depend on $p$. This requires
that the {\em divergent component} of $\fh_2^Z(p)$ should be proportional to
$1 + \fh_2^m(p)$ as $p$ changes. Moreover (from \nec{3.32a}) the same
(divergent) proportionality constant must hold between the divergent component
of $\fh_n^Z(p)$ and $\fh_n^m(p)$ (which is finite) for $n>2$.

To see how this works, we will analyze the asymptotic behavior of $A_n(q;p)$ in
the regime of large $q$ with fixed $p$.  To this end, let us first establish
the relation
\be
\Gamma_{n+2}(q,-q,p)
=
2 g^2 \fp(q) \fh^m_n(p) + \bO(1/q^3)
\quad n \ge 2
\,.
\label{eq:3.28}
\ee
Here, we have introduced the notation $f(x)=\bO(x^n)$ when $x\to \infty$ to
denote $f(x)=O(x^{n+\eta})$ for all $\eta>0$. In particular
$O(x^n \log^k(x)) \subseteq \bO(x^n)$.

Asymptotically the leading terms are those with a minimum number of
propagator-lines traversed by the momentum flow of $q$.  In the $S$-graphs of
$\Gamma_{n+2}(q,-q,p)$ with $n \ge 2 $, the two external lines $q$ and $-q$
must go to two different vertices, because the vertex functions are
one-particle irreducible. The leading terms are those with just one internal
line carrying a large momentum $q+k$ connecting the two vertices (see
\nec{3.27}). The two outgoing lines attached to those two vertices must be
internal (otherwise the graph would be reducible) and recombine to produce the
final $n$ $p$-legs; the recombination involves precisely the amplitude
$A_n(k;p)$. Diagrammatically,
\be
\includegraphics[height=16mm]{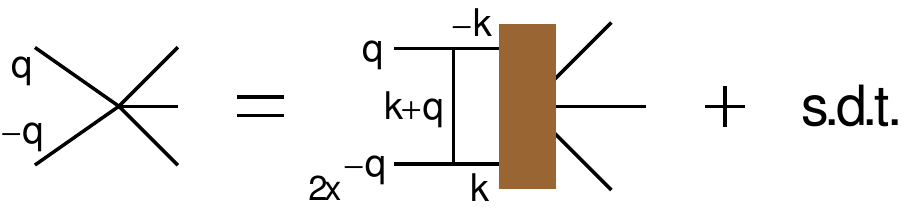}
\label{eq:3.27}
\ee
The left-hand side (LHS) represents $\Gamma_{n+2}(q,-q,p)$; all legs are
amputated. There are two $q$-legs and $n$ $p$-legs. The box is $A_n(k;p)$. The
legs are distinguishable, so there are two versions of the graph.
Furthermore, for large $q$ the internal propagator
$\fp(k+q) = \fp(q) + O(1/q^3)$; in the leading term, that with
$\fp(k+q)\to\fp(q)$, the integration over $k$ can be carried out and it
produces $\fh^m_n(p)$. This proves \nec{3.28}. The factor $2$ compensates for
the symmetry factor $1/2$ present in $\fh^m_n(p)$ but not in
$\Gamma_{n+2}(q,-q,p)$.

Let us now analyze the asymptotic behavior of $A_n(q;p)$.  The cases $n=2$ and
$n\ge 3$ are discussed separately. Let us consider $A_n(q;p)$ for $n=3$,
\be
\includegraphics[height=16mm]{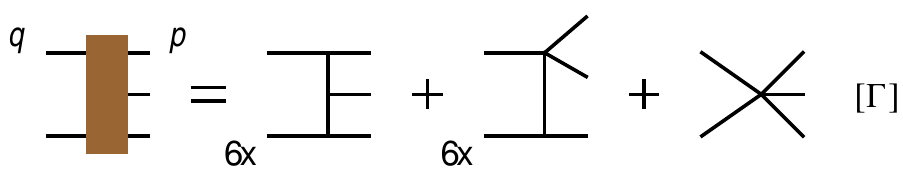}
\ee
Recall that in $A_n(q;p)$ the $p$-lines are amputated and the $q$-lines are
not. The first $\Gamma$-graph in the right-hand side (RHS) is $O(1/q^8)$ from
the four explicit propagators (two external and two internal) carrying the
large momentum $q$.  The second graph is also $O(1/q^8)$ from the three
explicit propagators and the vertex $\Gamma_4=O(1/q^2)$. The third graph, with
two explicit external propagators and a vertex $\Gamma_5 = O(1/q^2)$,
dominates the asymptotic behavior. More generally, for $n\ge 3$ the leading
term of $A_n(q;p)$ comes from the graph with a vertex $\Gamma_{n+2}$ and
\be
A_n(q;p) = \frac{1}{2} \fp^2(q) \Gamma_{n+2}(q,-q,p) + O(q^{-8})
\quad n \ge 3
\,.
\ee
Therefore, using \nec{3.28},
\be
A_n(q;p) = g^2 \fp^3(q) \fh^m_n(p) + \bO(q^{-7})
\quad n \ge 3
\,.
\ee
Diagrammatically, for $A_n(q;p)$ and $n\ge 3$,
\be
\includegraphics[height=28mm]{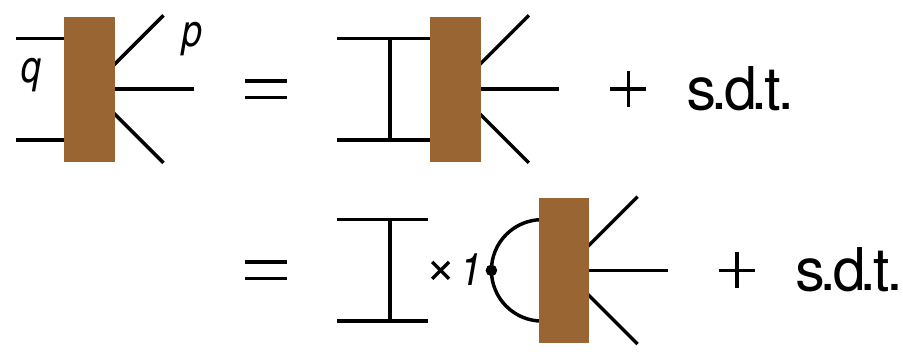}
\ee
Here, we have chosen to use the full propagator $\fp(q)$ in the three lines
carrying the momentum $q$. $\Df(q)$ would be also a valid choice. The choice
affects the precise definition of the ``finite component'' of $\fh^Z_n(p)$ but
this ambiguity will be removed by the renormalization condition.

The case $n=2$ is special because the two $\Gamma$-graphs in $A_2(q;p)$ are of
the same order
\be
\includegraphics[height=13mm]{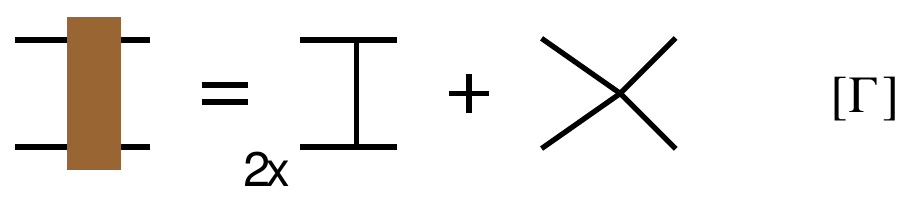}
\ee
The graph with $\Gamma_4$ does not dominate; hence, there is an additional
term in $A_2(q;p)$:
\be
\includegraphics[height=28mm]{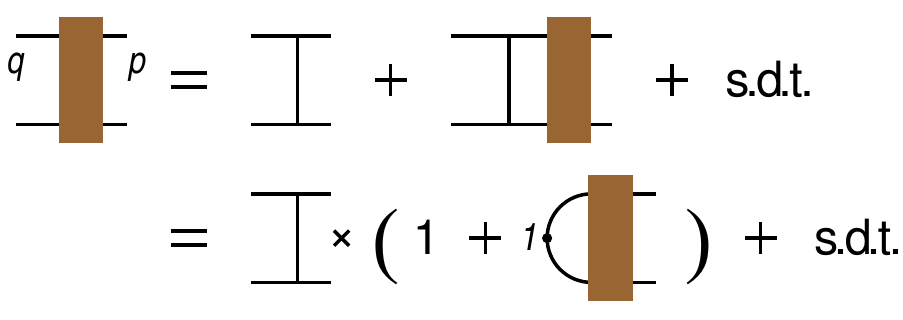}
\ee
In summary,
\bes
A_n(q;p) &=  g^2 \fp^3(q) ( \delta_{n,2} + \fh^m_n(p) ) +
\hA_n(q;p)
,
\qquad
\\
\hA_n(q;p) &= \bO(1/q^7)
.
\label{eq:3.41}
\ees

We can use this result in \nec{3.32b}. Since $A_n(q;p)=O(q^{-6})$, the
integral over $q$ in $\fh^m_n(p)$ is UV-finite, while $\fh^Z_n(p)$ diverges
logarithmically,
\be
\fh^Z_n(p) = C_Z(\Lambda) (\delta_{n,2} + \fh^m_n(p)) + \hfh^Z_n(p)
,
\label{eq:3.42}
\ee
where the remainder $\hfh^Z_n(p)$ is UV-finite and
\be
C_Z(\Lambda) = \int \frac{d^4 q}{(2\pi)^d} F_q^2 q^2 g^2 \fp^3(q)
=
- \Omega_4 \frac{g^2}{Z^3} L_\Lambda
+ \mathrm{s.d.t.}
\ee

The structure of \nec{3.42} introduced in \nec{3.32a} allows to write
\bes
\delta \Gamma_n (p) &= \delta m^2  \left( \delta_{n,2} + \fh^m_n (p) \right)
+
\delta Z \left( \delta_{n,2} p^2 + \hfh^Z_n (p) \right)
\\ &\quad
+
\delta g \left( \delta_{n,3} + \hfh^g_n (p) \right)
\label{eq:3.44}
\ees
where
\be
\delta m^2 := \delta m_0^2 + \delta Z \, C_Z
+ \delta g \, C_g
\,,
\label{eq:3.45a}
\ee
and we have already allowed for the presence of a $\delta g$ contribution,
which will be discussed subsequently.

The crucial observation is that the bare-mass parameter $\delta m_0^2(\Lambda)$
can be chosen so that $\delta m^2$ is finite, and this renders all
$\delta \Gamma_n(p)$ also finite.

In \nec{3.44} it can be assumed that the limit $\Lambda\to\infty$ has been
taken, so the result no longer depends on the value of the cutoff or the
regularization procedure. However $\hfh^Z_n (p) $, $\hfh^g_n (p) $ and
$\delta m^2$ still have a dependence on the conventions adopted regarding how
to split some functions into their leading and subleading components. For
instance, in \nec{3.41} the leading term was chosen with $\fp^3(q)$ instead of
$\fp^2(q)\Df(q)$.  In this sense, these quantities have no direct physical
meaning as they are not determined solely from $\Gamma$ and $\delta\Gamma$.

The ambiguities are fixed by enforcing the renormalization condition in
\nec{3.a1},
\bes
\delta m_R^2 &= 
\delta \Gamma_2 (0)
\\ &= \delta m^2  ( 1 + \fh^m_2 (0) )
+
\delta Z \, \hfh^Z_2 (0) + \delta g \, \hfh^g_2 (0)
.
\label{eq:3.45}
\ees
Using this relation to eliminate $\delta m^2$ in favor of $\delta m_R^2$ and
substituting in \nec{3.44}, yields
\bes
\delta \Gamma_n (p) &= \delta m_R^2  ( \delta_{n,2} + \fh^m_{R,n} (p) )
\\ & \quad +
\delta Z (\delta_{n,2} p^2 +  \fh^Z_{R,n} (p) )
\\ & \quad +
\delta g (\delta_{n,3} +  \fh^g_{R,n} (p) )
\label{eq:3.47}
\ees
with
\bes
\fh^m_{R,n}(p) &= \frac{ \fh^m_n (p) -  \delta_{n,2} \fh^m_2 (0) }{ 1 + \fh^m_2 (0) }
,
\\
\fh^Z_{R,n}(p) &= \hfh^Z_n (p) -
\frac{ \delta_{n,2} + \fh^m_n (p) }{ 1 + \fh^m_2 (0) } \hfh^Z_2 (0)
,
\\
\fh^g_{R,n}(p) &= \hfh^g_n (p) -
\frac{ \delta_{n,2} + \fh^m_n (p) }{ 1 + \fh^m_2 (0) } \hfh^g_2 (0)
.
\label{eq:3.48}
\ees
By construction, $\fh^m_{R,2}(0) = \fh^Z_{R,2}(0) = \fh^g_{R,2}(0) = 0$, so that
$\delta m_R^2 = \delta\Gamma_2(0)$, regardless of the values of $\delta Z$
and $\delta g$.

Note that the same Eqs. \nec{3.48} hold using $\fh^Z_n$ and $\fh^g_n$ instead of
$\hfh^Z_n$ and $\hfh^g_n$,\footnote{Such equations are obtained directly from
  \nec{3.32a} by eliminating $\delta m_0^2$ in favor of $\delta m_R^2$.}
although in that case each term would be divergent separately.

After removing the cutoff, the function $\fh^m_{R,n}(p)$ is well-defined and
is determined solely and completely from the effective action. The same
statement holds for $\fh^Z_{R,n}(p)$. In fact, whatever the concrete choice of
the leading term of $A_n(q;p)$ used in the definition of the remainder
$\hA_n(q;p)$, it is erased in $\fh^Z_{R,n}(p)$, as this function can be
expressed as
\bes
\fh^Z_{R,n} (p) &= \int \frac{d^4 q}{(2\pi)^4} q^2 A_{R,n}(q;p)
,
\\
A_{R,n}(q;p) &:= A_n(q;p) - \frac{ \delta_{n,2} + \fh^m_n (p) }{ 1 + \fh^m_2 (0) }
A_2(q;0)
.
\label{eq:3.54}
\ees
By construction, $A_{R,2}(q;0)=0$. The integrand $A_{R,n}(q;p) = \bO(q^{-7})$,
hence no regulator is required.

As a technical aside, a function $F(q;p)$ {\em factorizes for large $q$ and
  fixed $p$} when there exist functions $A(q)$ and $B(p)$ such that
$\lim_{q\to\infty} F(q,p)/A(q) = B(p)$. Then $F(q;p)= A(q) B(p) + \hF(q;p)$
where $\hF(q;p)$ is a subleading component. For a given $F(q;p)$ there is an
ambiguity in $A(q)$ since it is always possible to use instead
$A'(q)= \lambda A(q)+R(q)$, where $\lambda \neq 0$, and $R(q)$ is
subleading. However, the subtracted function
\bes
F_R(q;p) &:= F(q;p) - \frac{B(p)}{B(p_0)}F(q;p_0)
\\ &
= \hF(q;p) - \frac{B(p)}{B(p_0)}\hF(q;p_0)
\ees
is free from such ambiguity and is more convergent than the original $F$.
In turn $F_R(q;p)$ depends on the choice of $p_0$ such that $F_R(q;p_0)=0$.

\begin{figure}[ht]
  \begin{center}
\includegraphics[height=35mm]{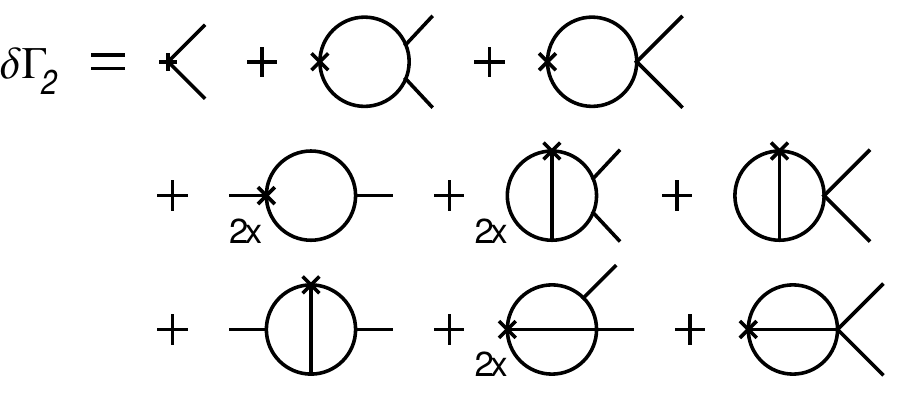}
\end{center}
\caption{ For the $\phi^3_4$ theory, $\Gamma$-graphs contributing to
  $\delta \Gamma_2(p)$. The lines and (uncrossed) vertices in the graphs are
  $\fp(p)$ and $\Gamma_n(p)$. The vertices represented by a cross are those of
  $\delta S$, with Feynman rule $F_q^2 ( \delta m_0^2 + q^2 \delta Z )$ for
  the two-point vertex and $+F_{q_1}F_{q_2}F_{q_1+q_2} \delta g$ for the
  three-point vertex.}
\label{fig:4}
\end{figure}

\subsubsection{\textsf{ Renormalization from $\delta  g$
  }}

Let us include the effect of a non-vanishing $\delta g$. Since the
infinitesimal contributions are additive, we can set $\delta Z=0$ in this
discussion. We start with the case $n=2$.
\be
\delta \Gamma_2 (p) = \delta m_0^2 \big( 1 + \fh^m_2 (p) \big)
+ \delta g \, \fh^g_2 (p)
.
\label{eq:3.33a}
\ee
The corresponding $\Gamma$-graphs are displayed in \fig{4}. The graphs may
contain divergences coming from i) loops and ii) from parameters (namely,
$\delta m_0^2$). Because $\delta Z=0$, the divergences from the loops are
present exclusively in the three graphs in the second row, and the divergences
from the parameters are present exclusively in the three graphs of the first
row. The three graphs in the third row are fully finite when the regulator in
$\delta S$ is removed.

Clearly, the logarithmic loop-divergence in \fig{4}d =
\includegraphics[height=8mm]{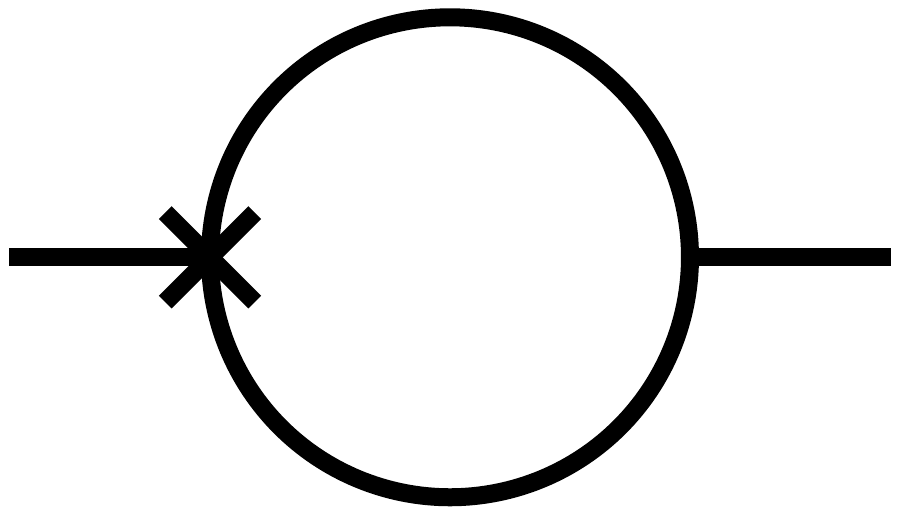} can be can removed by a mass
counterterm in \fig{4}a = \includegraphics[height=8mm]{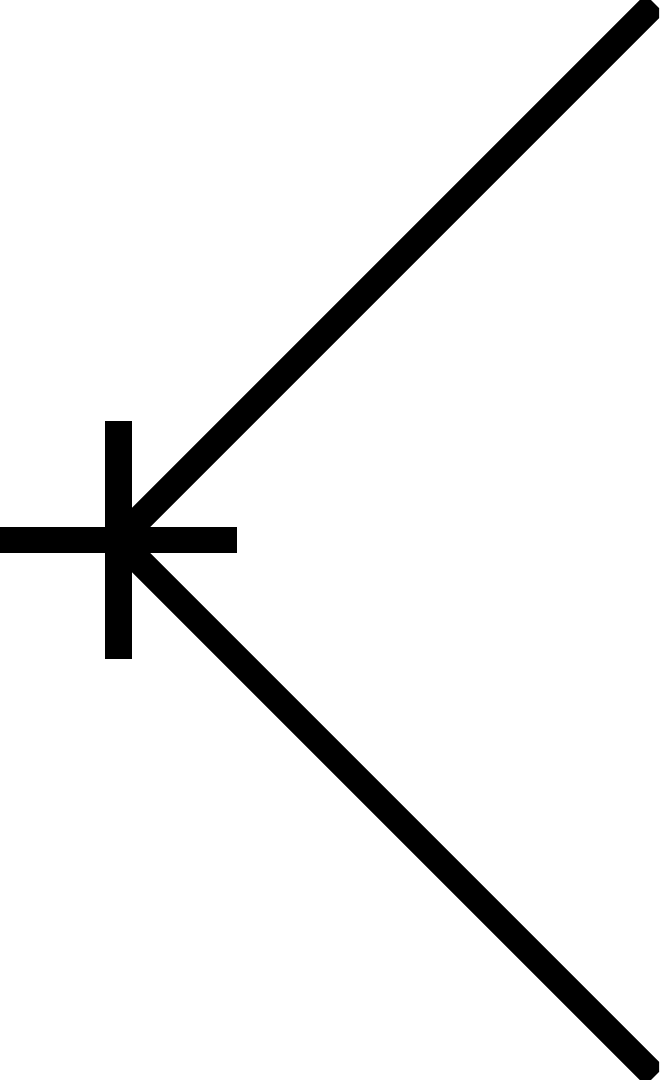}. In turn,
since \fig{4}d appears as subgraph in \fig{4}e =
\includegraphics[height=8mm]{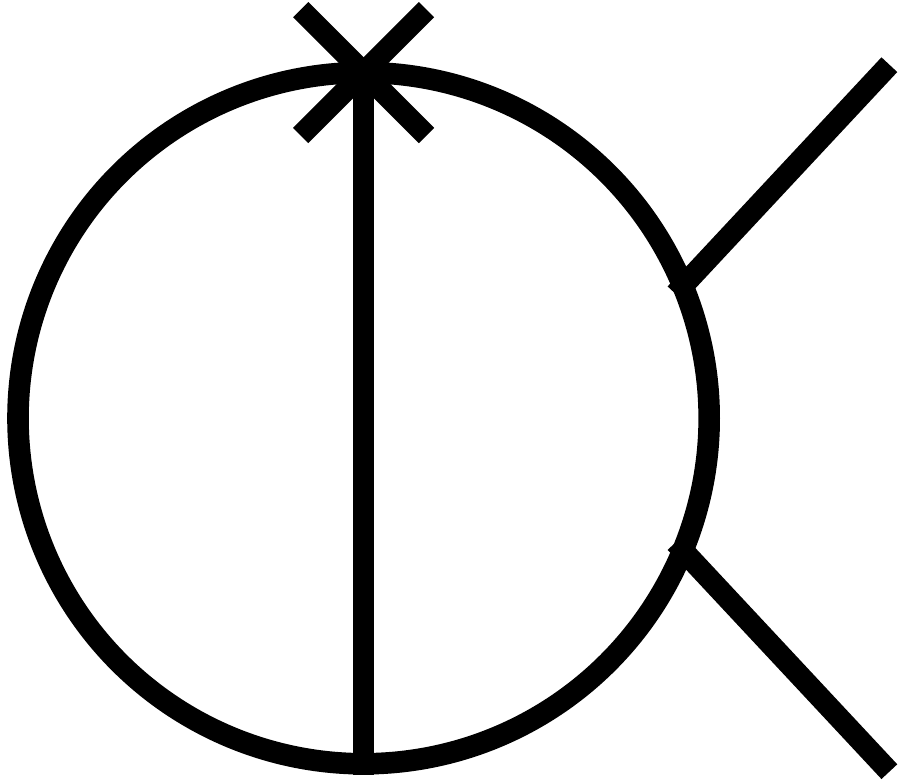} and \fig{4}f =
\includegraphics[height=8mm]{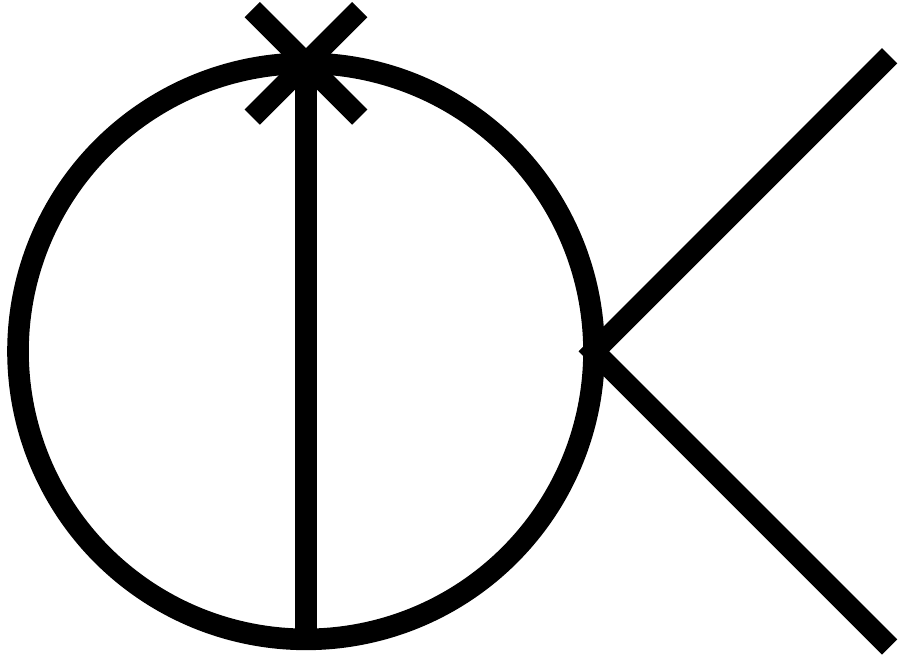} , the same counterterm acting in
\fig{4}b = \includegraphics[height=8mm]{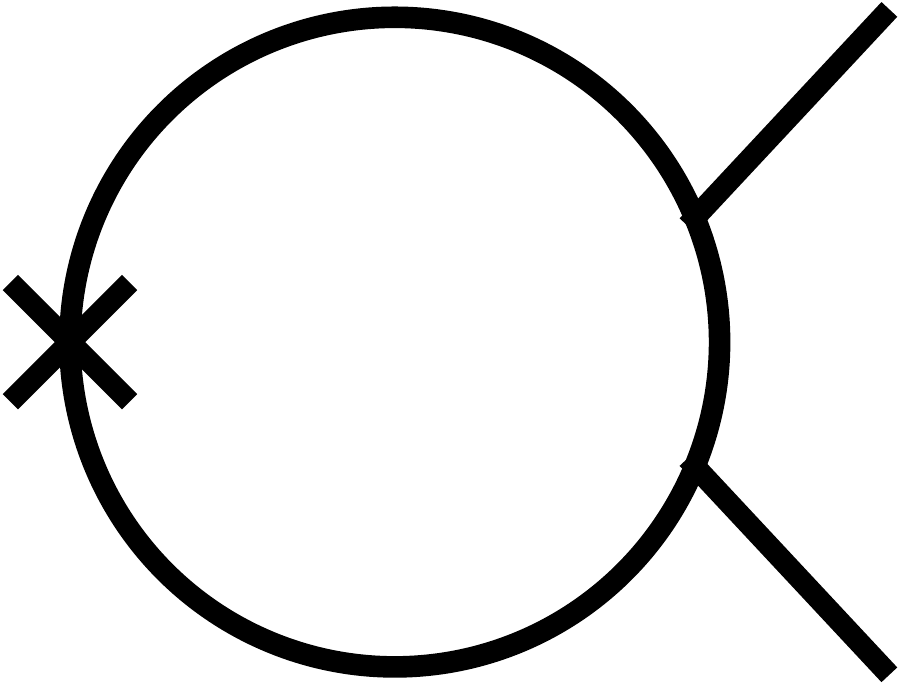} and \fig{4}c =
\includegraphics[height=8mm]{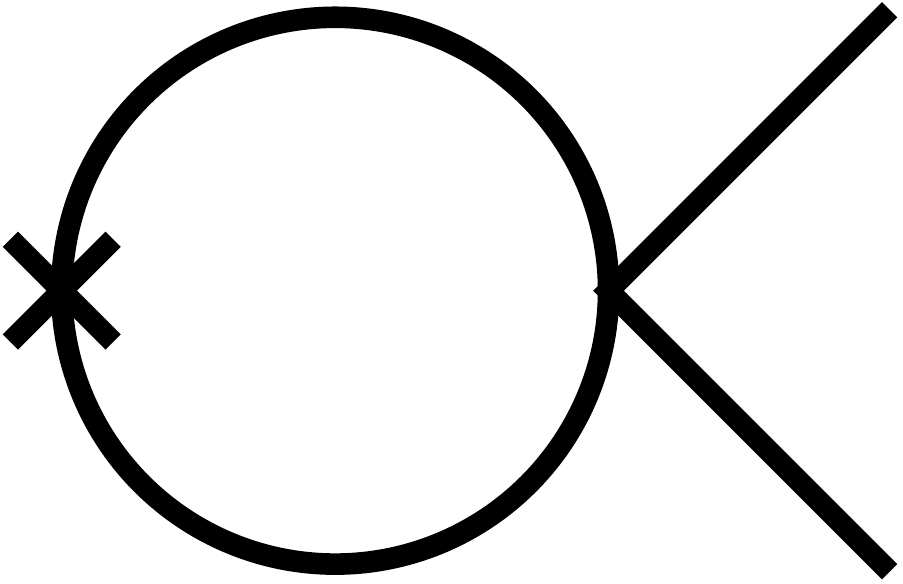} will cancel the subdivergences in
\fig{4}e and \fig{4}f, respectively.

The three graphs in the first row of \fig{4} correspond to the
$\delta m_0^2 \big( 1 + \fh^m_2 (p) \big)$ terms depending on $A_2(q;p)$ already
discussed. The remaining six graphs correspond to $\delta g \, \fh^g_2 (p)$, as
defined  in \nec{3.32b}.  It will be convenient to
split this function into two components (d) and (f) (for {\em divergent} and
{\em finite}) corresponding respectively to the second and third
row graphs of \fig{4}:
\be
\fh^g_2(p) = \fh_2^{g(\rd)}(p) + \fh_2^{g(\rf)}(p)
.
\ee
$\fh_2^{g(\rf)}(p)$ is already UV-finite, while $\fh_2^{g(\rd)}(p)$ contains the
UV-divergent graph \fig{4}d, and the same graph as a subgraph in \fig{4}e and
\fig{4}f. When \includegraphics[height=8mm]{GF9.pdf} is contracted to a
point, the three graphs of $\fh_2^{g(\rd)}(p)$ turn into the three graphs of
$\fh_2^m(p)$. Algebraically, let
\bes
K(q) &:= \int \frac{d^4 k}{(2\pi)^4} F_k F_{k+q} B(k;q)
\\
B(k;q) &:= \fp(k) \fp(k+q) \Gamma_3(k,q)
\label{eq:3.59}
\ees
so that
\be
q\includegraphics[height=8mm]{GF9.pdf}
=
\frac{1}{2} \delta g \, F_q K(q)
\ee
and
\be
\fh_2^{g(\rd)}(p) =
  K(p)
 +
 \int \frac{d^4 q}{(2\pi)^4} F_q K(q) \, A_2(q;p)
 .
 \label{eq:3.56}
\ee
If $K$ is replaced by $1$ (and the regulator is removed), this expression
becomes $1+\fh^m_2(p)$.

$K(p)$ is logarithmically divergent. To disentangle the (sub)divergence, let
us extract the leading term. A possible choice is
\be
B(q;p) = g \fp^2(q) + \hB(q;p),
\quad
\hB(q;p) = \bO(q^{-5})
,
\label{eq:3.56b}
\ee
so that
\be
K(p) = C_g + \hK(p)
\ee
with $\hK(p)$ UV-finite and\footnote{Actually the factor in \nec{3.52} would
  be $F_q F_{q+p}$ instead of $F_q^2$, however, in the UV limit, the
  difference yields a constant term which is absorbed in subleading terms.}
\be
C_g(\Lambda) = \int \frac{d^4 q}{(2\pi)^4} F_q^2 g \fp^2(q)
=
\Omega_4 \frac{g}{Z^2} L_\Lambda
+ \mathrm{s.d.t.}
\label{eq:3.52}
\ee

Correspondingly, we can define
\be
\hfh_2^{g(\rd)}(p) :=
  \hK(p)
 +
 \int \frac{d^4 q}{(2\pi)^4} \hK(q) \, A_2(q;p)
 ,
 \label{eq:3.56a}
\ee
which is UV-finite since $\hK(q) = O(\log(q^2))$ and $A_2(q;p)=O(q^{-6})$.

As a consequence
\be
\fh_2^g(p) = \hfh_2^g(p) +
 C_g \left( 1 +  
 \int \frac{d^4 q}{(2\pi)^4} F_q \, A_2(q;p)
\right)
,
\label{eq:3.60}
\ee
with
\be
\hfh_2^g(p) := \hfh_2^{g(\rd)}(p) + \fh_2^{g(\rf)}(p)
,
\label{eq:3.66}
\ee
which is also UV-finite.

The integral in the second term of \nec{3.60} is very similar to $\fh_2^m(p)$,
but not identical: In \nec{3.32b} the form factor appears as $F_q^2$ while in
\nec{3.60} it appears as $F_q$. The difference $F_q(1-F_q)$ vanishes outside
the range $\Lambda^2 < q^2 < 2 \Lambda^2$, hence the integral in \nec{3.60}
differs from $\fh_2^m(p)$ by an $O(\Lambda^{-2})$ contribution, which vanishes
as the cutoff is removed. Note that the form factor of the $3$-point vertex
needs not coincide with that of the $2$-point vertex (or even the form factors
of $\delta m^2_0$ and $\delta Z$ could be different). The $O(\Lambda^{-2})$
dependence does not rely on the details of the form factors, provided they are
sufficiently well-behaved.

It follows that (up to terms irrelevant after removing the cutoff)
\be
\fh_2^g(p) = \hfh_2^g(p) + C_g (\Lambda) ( 1 + \fh_2^m(p) )
.
\label{eq:3.60a}
\ee
When this expression is introduced in \nec{3.33a}, the Eqs. \nec{3.44} (for
$n=2$) and \nec{3.45a} are verified. Not surprisingly, the values of $C_Z$ and
$C_g$ are such that $\delta m^2 - \delta m_0^2 = \delta Z C_Z + \delta g C_g$
matches $\delta (-m^2_\ct )$, with $m_\ct^2$ introduced in \nec{3.19},
\be
- \delta Z \Omega_4 \frac{g^2}{Z^3} L_\Lambda
+ \delta g \Omega_4 \frac{g}{Z^2} L_\Lambda
=
\delta \left(
  \frac{1}{2} \frac{g^2}{Z^2}\Omega_4 L_\Lambda
\right)
.
\ee

As before, $\delta m^2$ can be traded for $\delta m_R^2$ by imposing the
renormalization condition $\delta m_R^2=\delta \Gamma_2(0)$, and in this case
\nec{3.47} (for $n=2$) and \nec{3.48} are also fulfilled. The concrete choice
of the leading and subleading components in $B(q;p)$ in \nec{3.56b} is not
relevant. In fact $\fh_{R,2}^g(p)$ can be rewritten as
\be
\fh^g_{R,2}(p) = \fh_{R,2}^{g(\rd)}(p) + \fh_{R,2}^{g(\rf)}(p)
\label{eq:3.69}
\ee
with
\bes
 \fh_{R,2}^{g(\rd)}(p)  &=
   K_R(p)
 +
 \int \frac{d^4 q}{(2\pi)^4} K_R(q) \, A_{R,2}(q;p)
 ,
 \\
 \fh_{R,2}^{g(\rf)}(p)  &= \fh_2^{g(\rf)} (p) -
 \frac{ 1 + \fh^m_2 (p) }{ 1 + \fh^m_2 (0) } \fh_2^{g(\rf)} (0),
 \label{eq:3.70a}
\ees
and
\be
K_R(p) := \int \frac{d^4 q}{(2\pi)^4} ( B(q;p) - B(q;0) )
.
\ee
$\fh_{R,2}^{g(\rd)}(0) =  \fh_{R,2}^{g(\rf)}(0) = 0$, so that the renormalization
condition is preserved.

So $\fh_{R,2}^g(p)$ is expressed entirely in terms of the propagator and
vertices of $\Gamma[\phi]$ with UV convergent integrals without a regulator,
just like $\fh_{R,n}^m(p)$ and $\fh_{R,n}^Z(p)$.

\begin{figure}[ht]
  \begin{center}
\includegraphics[height=80mm]{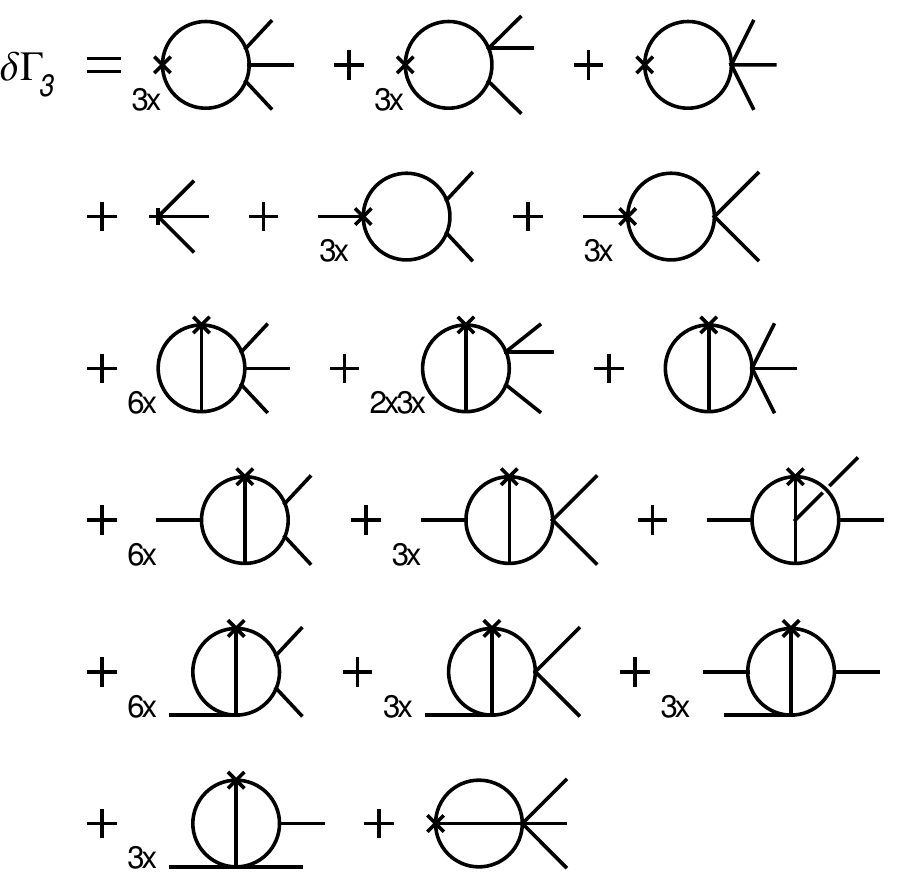}
\end{center}
\caption{$\Gamma$-graphs for $\delta\Gamma_3(p)$ in the $\phi^3_4$ theory.}
\label{fig:11}
\end{figure}

The previous formulas can be extended to $n > 2$. The $\Gamma$-graphs in
\fig{4}, for $n=2$, were obtained from the $\hat\Gamma$-graphs in \fig{2} by
extracting two (amputated) $p$-legs in all possible ways, either from the
lines or the vertices. The same procedure applies for $n\ge 2$. The result for
$n=3$ is displayed in \fig{11}. Algebraically
\be
\delta \Gamma_n (p) = \delta m_0^2 \big( \delta_{n,2} + \fh^m_n (p) \big)
+ \delta g \big( \delta_{n,3} + \fh^g_n (p) \big)
.
\label{eq:3.33b}
\ee

In \nec{3.33b} the term $g\delta_{n,3}$ of course comes from
\includegraphics[height=8mm]{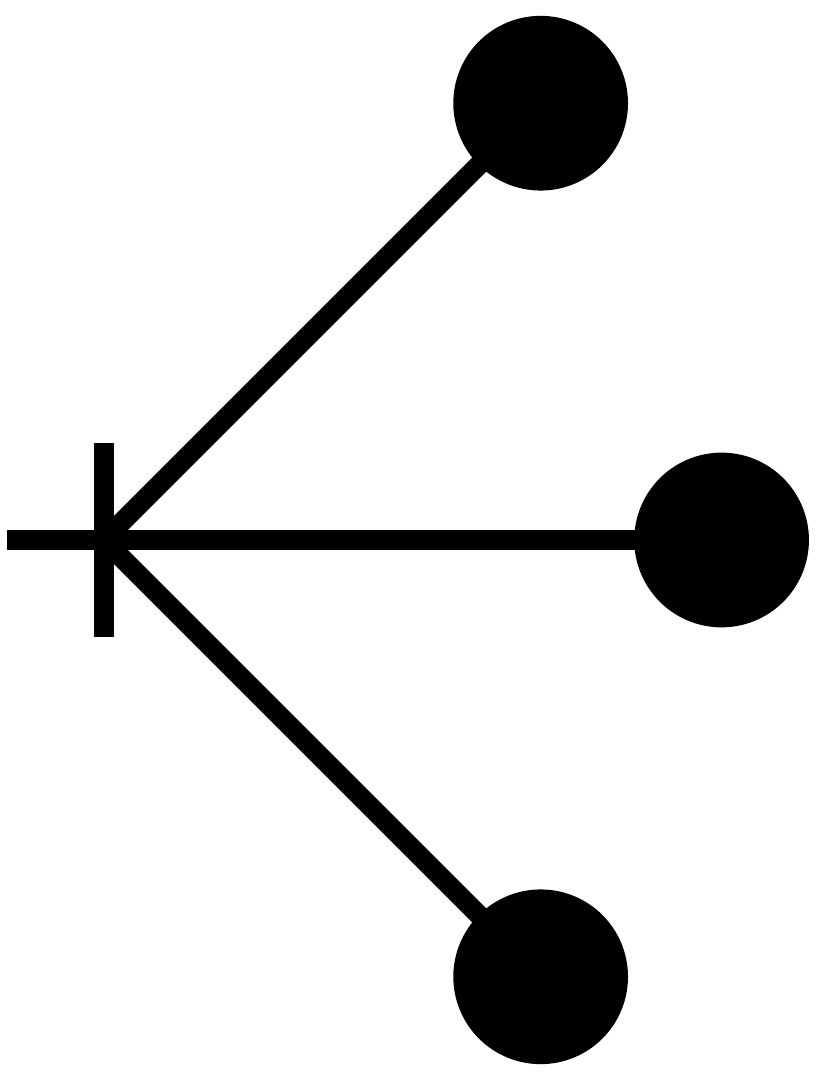} of \fig{2}. The term $g \fh^g_n (p)$
comes from extracting $n$ $p$-legs from the graphs with loops in the third
line of \fig{2}. The different contributions are eventually classified into
two types
\be
\fh^g_n (p) =
\fh^{g(\rd)}_n (p) 
+
\fh^{g(\rf)}_n (p) 
,
\ee
which are divergent and finite, respectively. Let us analyze these
contributions.

In the $\hat\Gamma$-graph
\includegraphics[height=8mm]{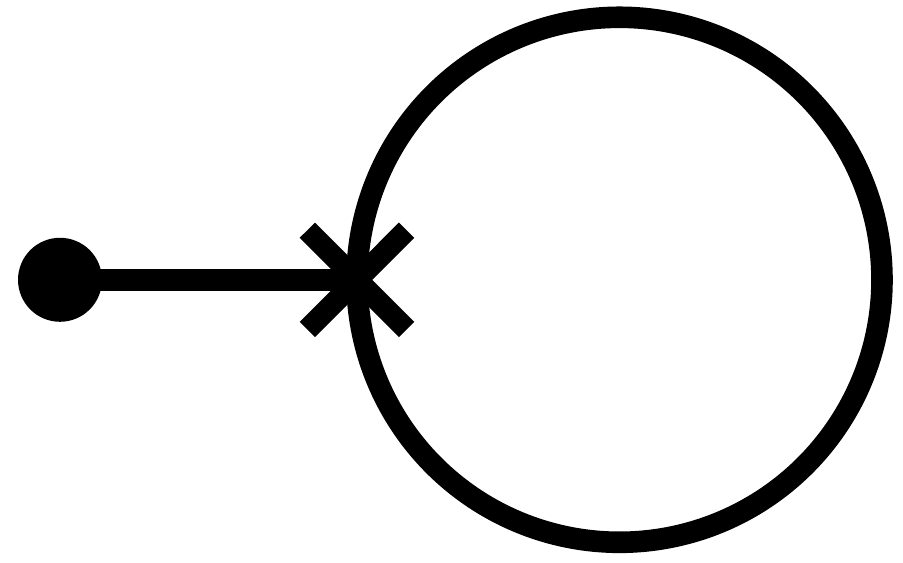}
, necessarily one of the $p$-legs is extracted from the crossed vertex, and
$n-1$ legs are extracted from the line. For instance, for $n=3$, the
$\Gamma$-graphs produced are
\be
\includegraphics[height=10mm]{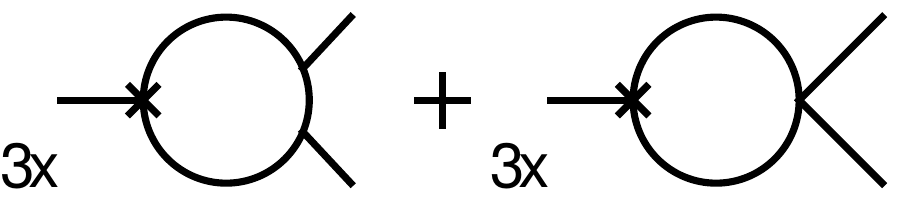}
\,.
\ee
This type of terms yield a contribution
\be
\sum_{j=1}^n
\int \frac{d^4 q}{(2\pi)^4} F_q^2 A_{n-1}(q;p)
=
\sum_{j=1}^n \fh^m_{n-1} (p_1,\ldots,\widehat{p}_j,\ldots,p_n)
,
\label{eq:3.70}
\ee
which is convergent without a regulator for $n \ge 3$ and adds to
$\fh^{g(\rf)}_n (p)$. For $n=2$ such a contribution is
\includegraphics[height=8mm]{GF9.pdf}
; it is just $K(p)$, which is logarithmically divergent; therefore, it was
included in $\fh^{g(\rd)}_2 (p)$ in \nec{3.56}.

In the $\hat{\Gamma}$-graph
\includegraphics[height=8mm]{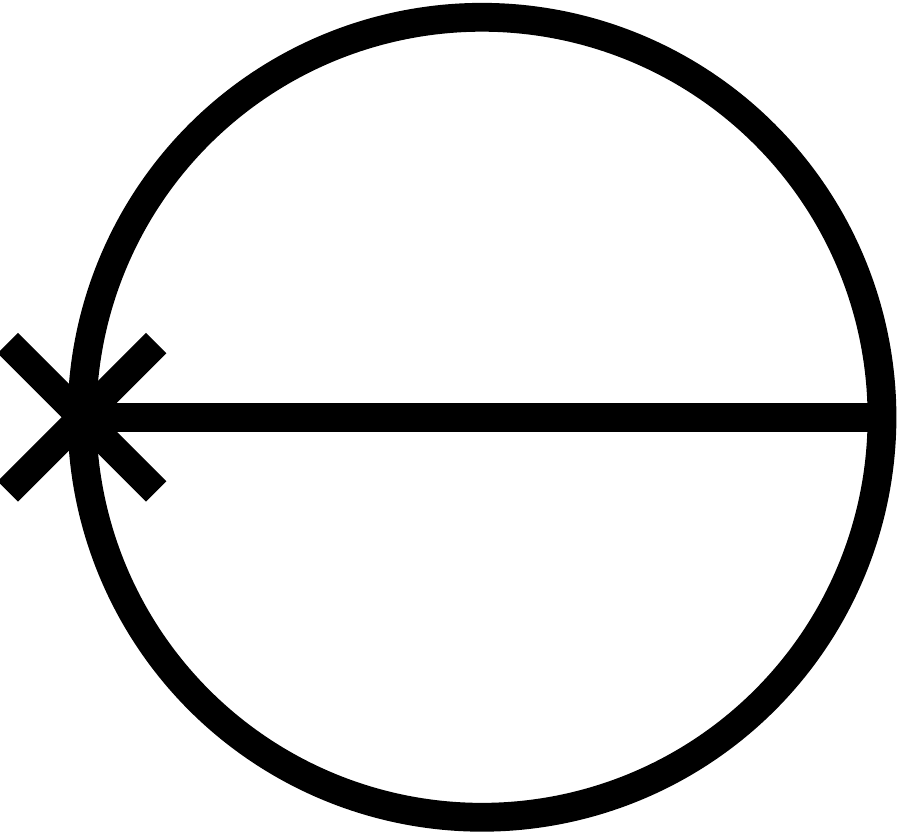}
all $p$-legs are to be extracted from the three lines and the uncrossed
vertex. Two types of $\Gamma$-graphs may be distinguished:
\begin{itemize}
\item[(i)] Those where all $n$ legs are extracted from the same line.
\item[(ii)] Otherwise.
\end{itemize}
    
Type (ii) graphs are UV-finite and so they go into $\fh^{g(\rf)}_n (p)$.

Type (i) graphs contain the subgraph
\includegraphics[height=8mm]{GF9.pdf}
and in fact, this is the best way to characterize type (i) graphs. Hence, they
are divergent and go into $\fh^{g(\rd)}_n (p)$. Explicitly, for $n = 3$, these
are
\be
\includegraphics[height=11mm]{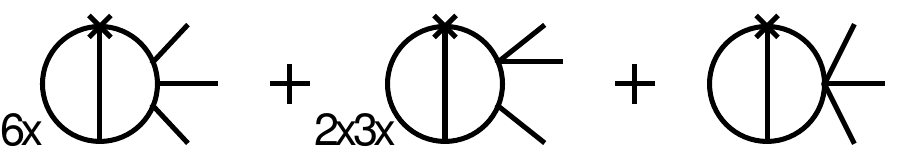}
\ee
The assembling of the $n$ legs is such that they are organized to form the
amplitude $A_n(q;p)$, thus (including now the $n=2$ exceptional term)
\be \fh^{g(\rd)}_n (p) = \delta_{n,2} K(p) + \int \frac{d^4 q}{(2\pi)^4} F_q
K(q) A_n(q;p)
.
\ee
Once again, this leads to
\be
\fh_n^g(p) = \hfh_n^g(p) + C_g ( \delta_{n,2} + \fh_n^m(p) )
,
\ee
that introduced in \nec{3.33b} comply with \nec{3.44} and \nec{3.45a}.
Elimination of $\delta m^2$ in favor of the renormalization condition then
produces \nec{3.48}. Finally, $\fh^g_{R,n}(p)$ can be arranged as
\be
\fh^g_{R,n}(p) = \fh_{R,n}^{g(\rd)}(p) + \fh_{R,n}^{g(\rf)}(p)
\label{eq:3.78}
\ee
with
\bes
 \fh_{R,n}^{g(\rd)}(p)  &=
   \delta_{n,2} K_R(p)
 +
 \int \frac{d^4 q}{(2\pi)^4} K_R(q) \, A_{R,n}(q;p)
 ,
 \\
 \fh_{R,n}^{g(\rf)}(p)  &= \fh_n^{g(\rf)} (p) -
 \frac{ \delta_{n,2} + \fh^m_n (p) }{ 1 + \fh^m_2 (0) } \fh_2^{g(\rf)} (0)
 ,
\label{eq:3.79}
\ees
where everything is convergent without the UV regulator. Unlike the case of
$\fh^Z_{R,n}(p)$, $\fh^g_{R,n}(p)$ requires two subtractions (one in
$A_{R,n}(q;p)$ and another in $K_R(q)$); a consequence of the two-loop
structure of the $\hat{\Gamma}$-graph
\includegraphics[height=8mm]{MG7.pdf}.

\subsection{\textsf{ The $\phi^4_3$ theory 
   \label{sec:3.c}
  }}

The theory $\phi^4_3$ is also super-renormalizable, and in fact, it is one of
the few quantum field theories for which there exists a mathematically
rigorous non-perturbative treatment. Unlike $\phi^3_4$, the Hamiltonian of
$\phi^4_3$ is bounded from below and the renormalized correlation functions
exist beyond perturbation theory \cite{Guerra:1973gd,Glimm:1987ylb}.

Despite that, both theories have substantial similarities and most of the
treatment just developed for the renormalization of $\phi^3_4$ carries over to
$\phi^4_3$. There are only two primitive divergent Feynman $S$-graphs:
\includegraphics[height=8mm]{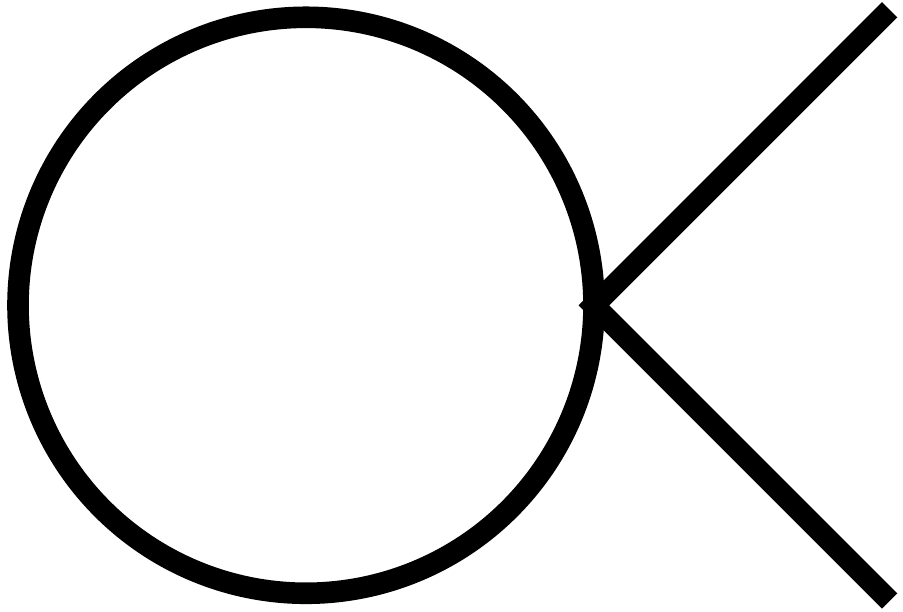}
and
\includegraphics[height=8mm]{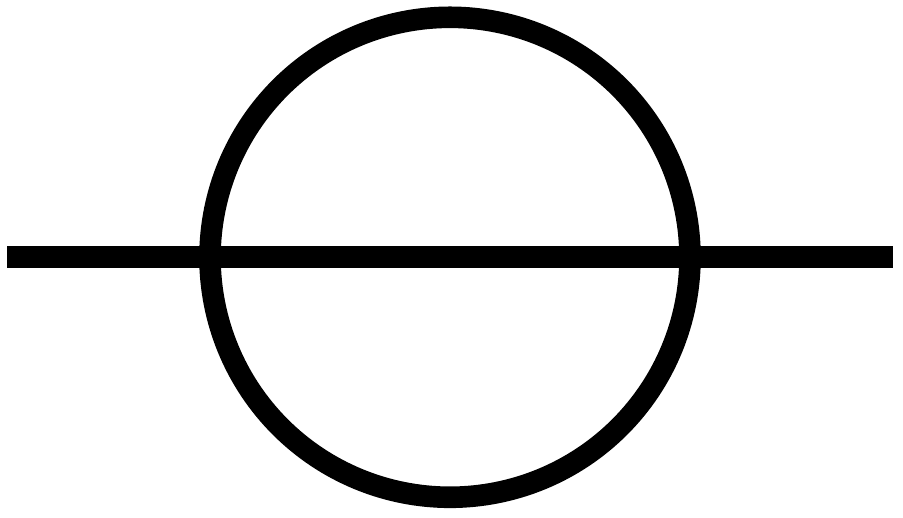}
.
The first one diverges linearly and the second one logarithmically. They are
canceled by a counterterm mass term
\be
m_\ct^2 = \frac{1}{2} \frac{g}{Z} \Omega_3 \Lambda
+
\frac{1}{3!} \frac{g^2}{Z^3} \frac{1}{8}\Omega_3 L_\Lambda + \mathrm{s.d.t.}
\ee
Once again, the parameters $Z$ and $g$ need no renormalization.

In this theory there is symmetry under $\phi \to -\phi$, and we will assume
that all vertex functions of odd order vanish.

\begin{figure}[ht]
  \begin{center}
\includegraphics[height=36mm]{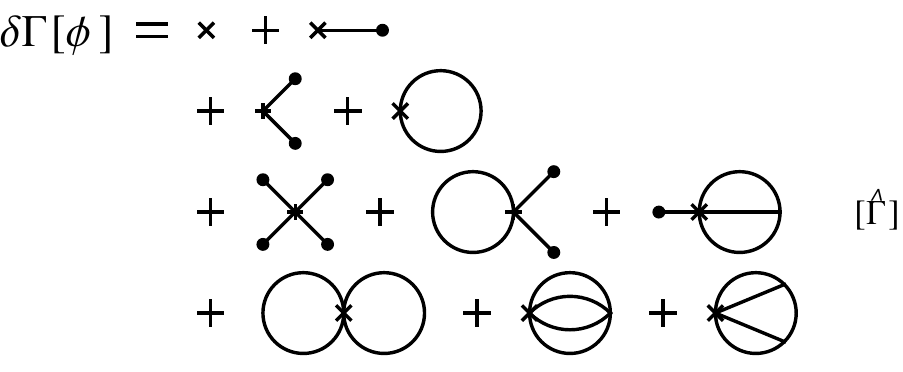}
\end{center}
\caption{ $\hat\Gamma$-graphs contributing to $\delta \Gamma[\phi]$ for a
  variation $\delta S[\phi]$ with two- and four-point vertices. The
  contributions from the vertices of $\delta S$ of $0$- and $1$-points have
  been included by completeness.}
\label{fig:5}
\end{figure}

\begin{figure}[ht]
  \begin{center}
\includegraphics[height=48mm]{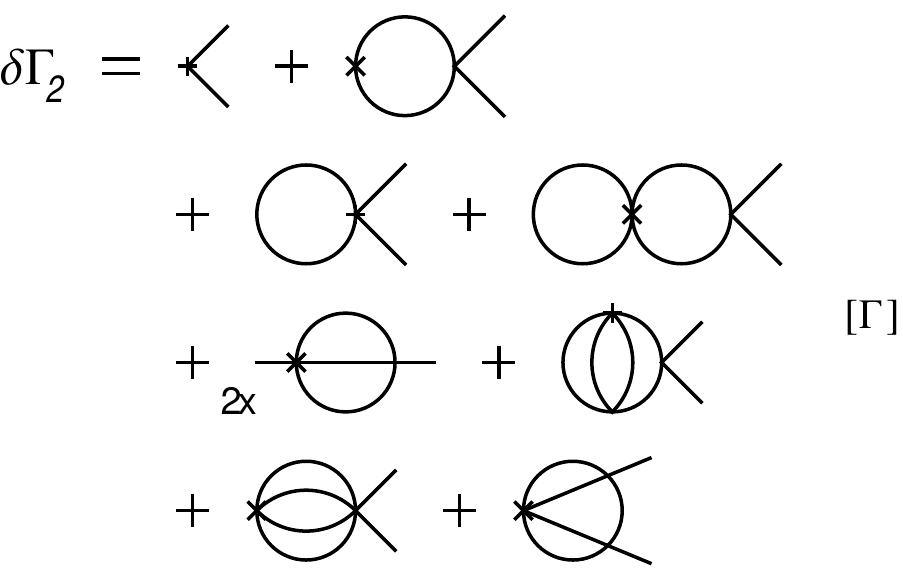}
\end{center}
\caption{ $\Gamma$-graphs for the $\phi^4_3$ theory contributing to
  $\delta \Gamma_2(p)$, assuming $\Gamma_n(p)=0$ for odd $n$. The vertices
  represented by a cross are those of $\delta S$, with Feynman rule
  $F_q^2 (\delta m_0^2 + q^2 \delta Z)$ for the two-point vertex and
  $+F_{q_1}F_{q_2}F_{q_3}F_{q_1+q_2+q_3} \delta g$ for the four-point vertex.}
\label{fig:6}
\end{figure}

\fig{5} displays the contributions to $\delta \Gamma[\phi]$ induced by
$\delta S$ with two- and four-point vertices by applying Schwinger's
principle.\footnote{Note that $\Gamma_n(p)$ vanishes for odd $n$ but not so
  $\hat\Gamma_n[\phi]$.}  The various vertex functions $\delta\Gamma_n(p)$ are
then obtained by extracting the $n$ $p$-legs in all possible ways. The
$\Gamma$-graphs corresponding to the contributions to $\delta \Gamma_2(p)$ are
displayed in \fig{6}.

In this theory
\be
\delta S_n (q) = \big( \delta m_0^2
+
\delta Z q^2 \big)  F^2_{q}  \delta_{n,2}
+
\delta g F_{q_1} F_{q_2} F_{q_3} F_{q_4} \delta_{n,4}
,
\ee
and
\bes
\delta \Gamma_n (p) &= \delta m_0^2  \big( \delta_{n,2} + \fh^m_n (p) \big)
\\ & \quad +
\delta Z \big( \delta_{n,2} p^2 + \fh^Z_n (p) \big)
\\ & \quad +
\delta g \big( \delta_{n,4}
+ \fh^g_n (p) \big)
.
\label{eq:3.32a4}
\ees
The expressions of $\fh^m_n (p)$ and $\fh^Z_n (p)$, as well as $A_n(q;p)$, are as in
\nec{3.32b} and \nec{3.32}, while $\fh^g_n (p)$ and $B_n(q;p)$ become
\bes
\fh^g_n (p) &= \int \frac{d^d q_1}{(2\pi)^d} \frac{d^d q_2}{(2\pi)^d}
\frac{d^d q_3}{(2\pi)^d}
F_{q_1} F_{q_2} F_{q_3} F_{q_1+q_2+q_3}  B_n(q;p)
\\
B_n(q;p) &:= \frac{1}{4!}  H^4_n(q_1,q_2,q_3,-q_1-q_2-q_3;p),
\label{eq:3.324}
\ees
with $d=3$.

As happened with $\phi^3_4$, here a finite variation of the mass does not
introduce UV divergences while $\delta Z$ and $\delta g$ do induce divergences
to be absorbed by the mass counterterm.

\subsubsection{\textsf{ Renormalization from $\delta  Z$  }}

Let us start by analyzing the effect of $\delta Z$, with $\delta g=0$. Our aim
is to establish a relationship as in \nec{3.42}. Let us consider the case $n=2$.
$\delta \Gamma_2(p)$ only receives contributions from the first two graphs in
the RHS of \fig{6}. Explicitly,
\be
A_2(q;p) = \frac{1}{2} \fp^2(q) \Gamma_4(q,-q,p)
.
\ee
Since, in this theory,
\be
\fp(q) = O(1/q^2), \qquad
\Gamma_n(q,-q,p) = O(1) \quad n \ge 4
,
\ee
$\fh^m_2(p)$ is UV-finite while $\fh^Z_2(p)$ has a linear divergence.

In more detail, when the integration over $q$ in \nec{3.32b} is carried out to
yield $\fh^Z_2(p)$, a linear divergence is produced from terms $O(1)$ in
$\Gamma_4(q,-q,p)$ and a logarithmic divergence is produced from terms
$O(1/q)$; more convergent terms do not produce UV divergences. An analysis
similar to that performed for $\phi^3_4$ can be carried out for $\phi^4_3$. By
inspection of the $S$-graphs of the theory, it can be seen that the $O(1)$
terms are in fact $q$-independent: they come from graphs where the two
external lines $q$ and $-q$ go to the same vertex (with Feynman rule
$+g$). The two outgoing lines may not interact again or else they may interact
producing the amplitude $A_2(k;p)$. Diagrammatically, the $q$-independent
terms of $\Gamma_4(q,-q,p)$ are
\be
\includegraphics[height=12mm]{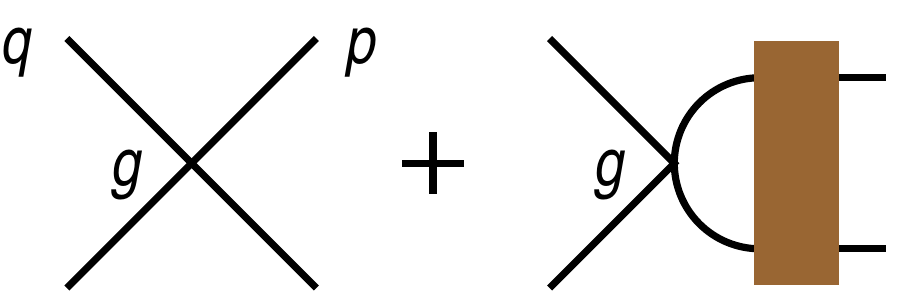}
\ee
and algebraically
\be
\Gamma_4(q,-q,p) \Big|_{\mbox{$q$-indep.}} = g ( 1  + \fh^m_2(p) )
.
\ee
On the other hand, the terms $O(1/q)$ in $\Gamma_4(q,-q,p)$ are produced when
the large momentum $q$ flows through the graph
\includegraphics[height=7mm]{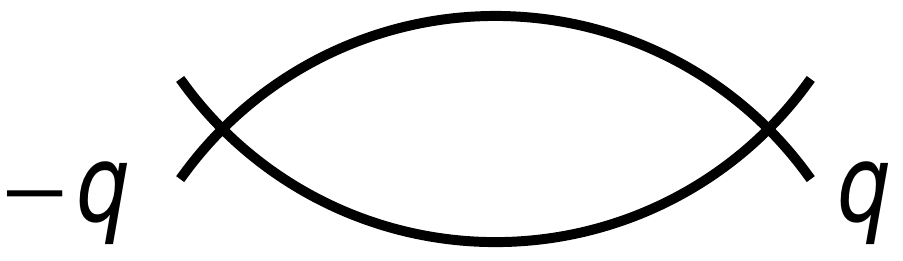}
which behaves as a $\ds \frac{1}{2}\frac{g^2}{Z^2}\frac{1}{8}\frac{1}{q}
+ \bO(1/q^2)$. Once again, the two outgoing legs may or may not interact
again. Diagrammatically, the $O(1/q)$ contributions to
$\Gamma_4(q,-q,p)$ are
\be
\includegraphics[height=28mm]{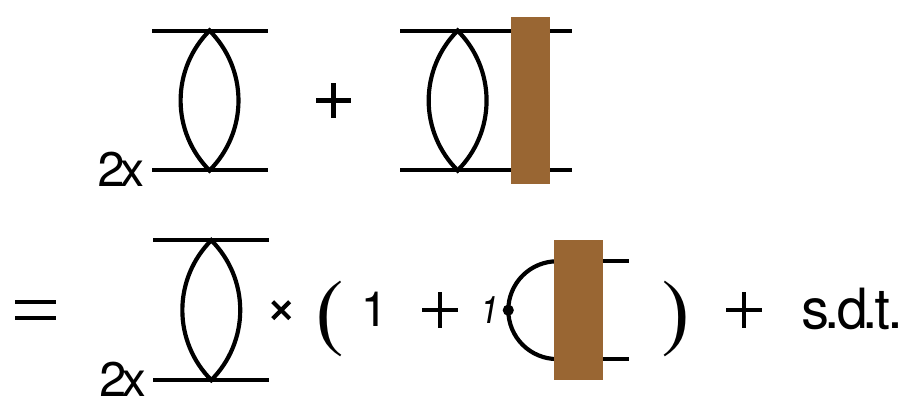}
\ee
corresponding to
\be
\Gamma_4(q,-q,p) \Big|_{O(1/q)} =\frac{g^2}{Z^2}\frac{1}{8}\frac{1}{q}
( 1 + \fh^m_2(p)) + \bO(1/q^2)
.
\ee
Combining both results
\be
\Gamma_4(q,-q,p) = \tg(q) ( 1 + \fh^m_2(p)) + \bO(1/q^2)
\label{eq:3.97}
\ee
where
\be
\tg(q) := g + \frac{g^2}{Z^2}\frac{1}{8}\frac{1}{q}
.
\ee
Thus
\bes
A_2(q;p) &=
\frac{1}{2} 
\tg(q) \fp^2(q) ( 1 + \fh^m_2(p)) + \hA_2(q;p)
,
\\
\hA_2(q;p) &= \bO(1/q^6)
.
\ees
Then \nec{3.42} is reproduced for $n=2$ with
\be
C_Z(\Lambda) =
\frac{1}{2} \Omega_3 \left(
\frac{g}{Z^2} \Lambda  + \frac{1}{8} \frac{g^2}{Z^4}  L_\Lambda 
+ \mathrm{s.d.t.}
\right)
=
- \frac{\partial m_\ct^2(\Lambda)} {\partial Z}
.
\ee
This guarantees that a suitable $\delta m_0^2(\Lambda)$ cancels the
divergences induced by $\delta Z$ on $\delta \Gamma_2(p)$.

The analysis for $n\ge 4$ and still $\delta g=0$ is similar to that done for
$\phi^3_4$. For $\delta \Gamma_4(p)$, the $\Gamma$-graphs are
\be
\includegraphics[height=13mm]{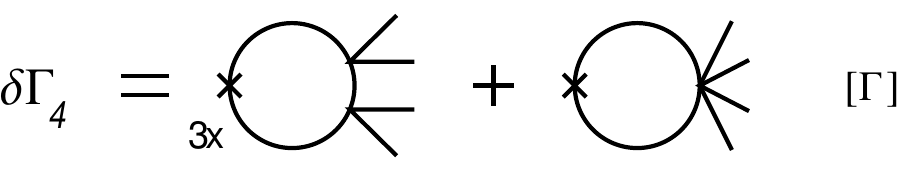}
\ee
The divergence comes only from the second graph, with a vertex $\Gamma_6$, as
the first one has three explicit internal propagators with large momentum $q$.
More generally, for $n\ge 4$, the leading terms in $\delta \Gamma_n(p)$ come
from $\Gamma_{n+2}(q,-q,p)$, that fulfills, using the arguments already
presented for $\Gamma_4(q,-q,p)$,
\be
\Gamma_{n+2}(q,-q,p) = \tg(q) ( \delta_{n,2} + \fh^m_n(p)) + \bO(1/q^2)
\quad
n \ge 2
.
\ee
Therefore
\bes
A_n(q;p) &=
\frac{1}{2} 
\tg(q) \fp^2(q) ( \delta_{n,2} + \fh^m_n(p)) + \hA_n(q;p)
,
\\
\hA_n(q;p) &= \bO(1/q^6)
.
\label{eq:3.41a}
\ees

From this point on, the equations already derived for the theory $\phi^3_4$
regarding $\delta Z$ follow (some with obvious modifications) including
\nec{3.42}, \nec{3.44}, \nec{3.45a}, \nec{3.45}, \nec{3.47}, \nec{3.48} and
\nec{3.54}. Thus $\fh^Z_n(p)$ is expressed through integrals which are UV-finite
without a regulator.

\subsubsection{\textsf{ Renormalization from $\delta  g$  }}

Let us now study the renormalization of $\fh^g_n(p)$, thus we assume
$\delta Z=0$. Let us first consider the case $n=2$. The $\Gamma$-graphs
corresponding to $\delta \Gamma_2(p)$ are displayed in \fig{6}. The two graphs
in the first row are $\delta m_0^2 ( 1 + \fh^m_2(p) )$ already discussed and
$\fh^m_2(p)$ is finite.  The remaining graphs are $\delta g \fh^g_2(p)$. Again this
function can be split into UV-divergent and UV-finite components:
\be
\fh^g_2(p) = \fh_2^{g(\rd)}(p) + \fh_2^{g(\rf)}(p)
.
\ee
$\fh_2^{g(\rf)}(p)$ corresponds to the two graphs in the last row, and 
$\fh_2^{g(\rd)}(p)$ to the four graphs in the second and third rows of \fig{6}.

The graph \fig{6}c = \includegraphics[height=8mm]{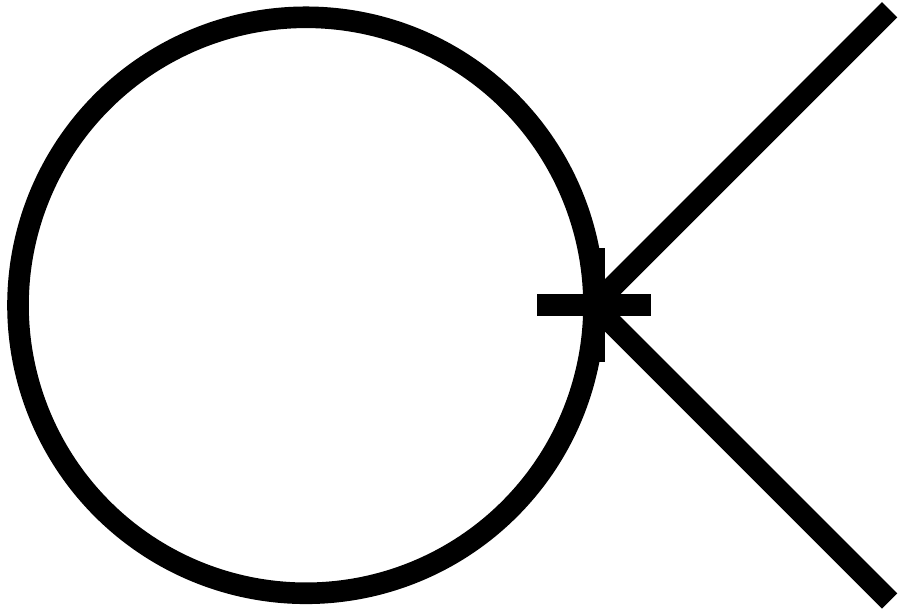} is linearly
divergent, and the divergence can be canceled by a mass-counterterm in \fig{6}a
=\includegraphics[height=8mm]{GF10.pdf}\,.  ~When such a counterterm acts
in \fig{6}b =\includegraphics[height=8mm]{GF12.pdf} it cancels
the divergence in \fig{6}d =\includegraphics[height=8mm]{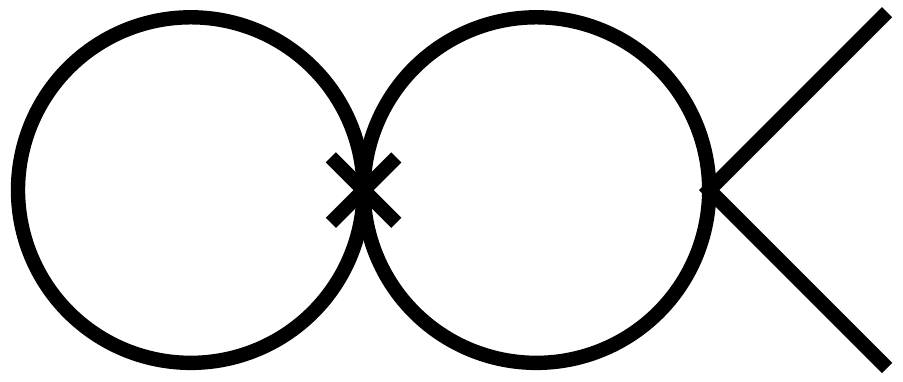}.
Likewise the graph \fig{6}e = \includegraphics[height=8mm]{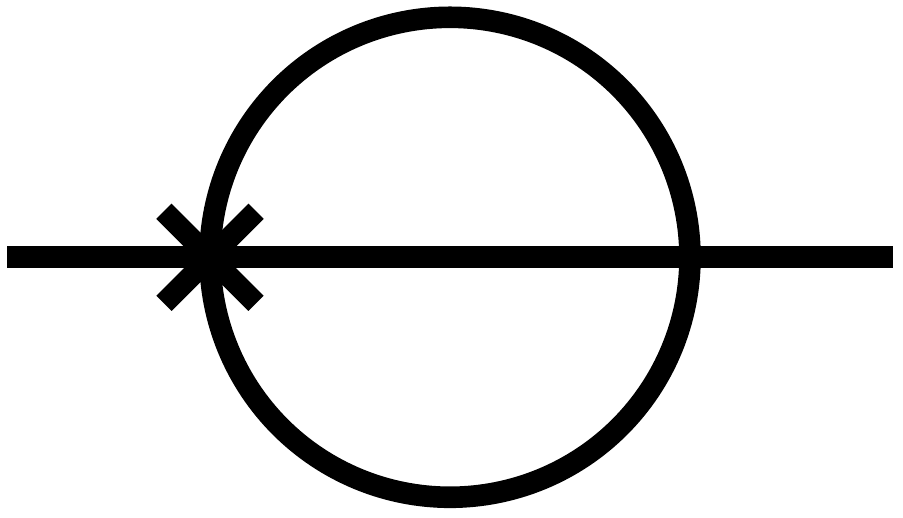}
diverges logarithmically to be canceled by a mass-counterterm in
\fig{6}a. When such counterterm is introduced in \fig{6}b it cancels the
divergence in \fig{6}f = \includegraphics[height=8mm]{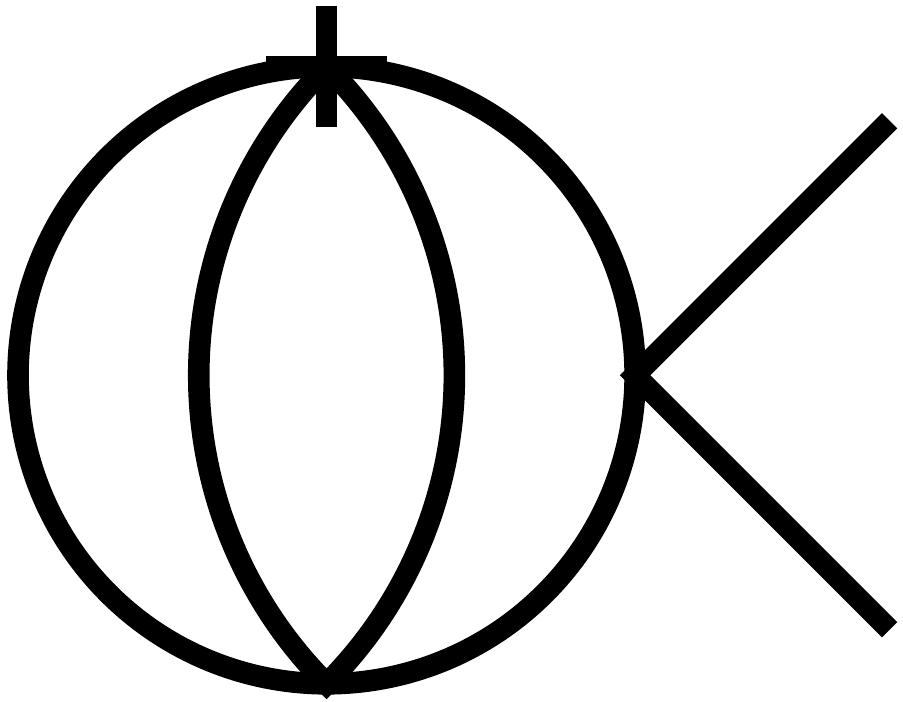}\,.

When \includegraphics[height=8mm]{MG43.pdf} is contracted to a point,
\fig{6}c and \fig{6}d become to \fig{6}a and \fig{6}b, respectively. Likewise
when \includegraphics[height=8mm]{MG45.pdf} is contracted to a point
\fig{6}e and \fig{6}f become \fig{6}a and \fig{6}b. Algebraically, let
\be
C_{g,a}(\Lambda) :=
\frac{1}{2} \int \frac{d^3 k}{(2\pi)^3} F_k^2 \fp(k)
\ee
so that
\be
\includegraphics[height=8mm]{MG43.pdf}
=
\delta g \, F_q^2 C_{g,a}
,
\ee
and
\bes
K(q) &:=
 \int \frac{d^3 k_1}{(2\pi)^3} \frac{d^3 k_2}{(2\pi)^3}
F_{k_1} F_{k_2} F_{k_1+k_2+q} B(k_1,k_2;q)
\\
B(k_1,k_2;q) &:= \frac{1}{3} \fp(k_1) \fp(k_2) \fp(k_1+k_2+q) \Gamma_4(k_1,k_2,q)
\ees
so that
\be
q \includegraphics[height=8mm]{MG45.pdf}
=
\frac{1}{2} \delta g \, F_q K(q)
.
\ee
Then, the four divergent contributions in \fig{6}c-f, can be expressed as
\bes
\fh_2^{g(\rd)}(p) &= 
  C_{g,a} + K(p)
  \\ &\quad
  +
 \int \frac{d^3 q}{(2\pi)^3} \left (
   F_q^2 C_{g,a} + F_q K(q) \right)  A_2(q;p)
 .
 \label{eq:3.109}
\ees
If $C_{g,a}+K$ is replaced by $1$ (and the regulator is removed), this
expression becomes $1+\fh^m_2(p)$.

$C_{g,a}$ is a linearly divergent constant,
\be
C_{g,a} = -\frac{1}{2} \frac{1}{Z} \Omega_3 \Lambda + \mathrm{s.d.t.}
\ee
On the other hand, $K(q)$ has a logarithmic divergence. To isolate it, let us
extract a leading term in $B$,
\bes
B(q_1,q_2;p) &=
\frac{1}{3} g \fp(q_1) \fp(q_2) \fp(q_1+q_2)
\\ &\quad
+ \hB(q_1,q_2;p)
.
\ees
The vertex $\Gamma_4(q_1,q_2,q_3)$ is $O(1)$ when $q_1$ and $q_2$ are large
and more specifically it is $g + O(1/q)$ when $q_3$ is also large. Therefore
\be
\hB(q_1,q_2;p) = \bO(q^{-7})
\ee
except in the zero-measure case $q_1 + q_2 =$ finite. (Such exceptional case
is precisely that in \nec{3.97}, where $q_1=-q_2=q$.) We then obtain
\be
K(p) = C_{g,b}(\Lambda) + \hK(p)
\ee
with $\hK(p)$ UV-finite and
\bes
C_{g,b}(\Lambda) &=
\int \frac{d^3 q_1}{(2\pi)^3} \frac{d^3 q_2}{(2\pi)^3}
F_{q_1} F_{q_2} F_{ q_1 + q_2 } 
\\ & \quad \times
\frac{1}{3} g \fp(q_1) \fp(q_2) \fp(q_1+q_2)
\\ &=
- \frac{1}{3} \frac{g}{Z^3} \frac{1}{8} \Omega_3 L_\Lambda
+ \mathrm{s.d.t.}
\ees

We can now define (similar to \nec{3.56a})
\be
\hfh_2^{g(\rd)}(p) := 
  \hK(p)
  +
 \int \frac{d^3 q}{(2\pi)^3} \hK(q) A_2(q;p)
 ,
 \label{eq:3.109a}
\ee
which is UV-finite since $\hK(q) = O(\log^2(q^2))$ and $A_2(q;p)=O(q^{-4})$,
and also
\be
\hfh_2^g(p) := \hfh_2^{g(\rd)}(p) + \fh_2^{g(\rf)}(p)
.
\ee
As a consequence,
\bes
\fh_2^g(p) &= \hfh_2^g(p)
+
C_{g,a} + C_{g,b}
\\ &\quad +
\int \frac{d^3 q}{(2\pi)^3} ( C_{g,a} F^2_q + C_{g,b} F_q ) A_2(q;p)
.
\ees
By the same argument given after \nec{3.66}, $F_q$ is equivalent to $F_q^2$
within the integral, up to terms vanishing when the cutoff is removed, and we
finally arrive at
\be
\fh_2^g(p) = \hfh_2^g(p) + C_g (\Lambda) ( 1 + \fh_2^m(p) )
,
\label{eq:3.60b}
\ee
with
\be
 C_g := C_{g,a} + C_{g,b} = 
- \frac{\partial m_\ct^2(\Lambda)} {\partial g}
.
\ee

\Eq{3.60b} is in fact identical to \nec{3.60a} of the theory $\phi^3_4$. The
relations found there regarding $\fh^g_2(p)$ apply also here (with $d=3$), in
particular, after imposing the renormalization condition, \Eqs{3.69} and
\nec{3.70a} also hold here with
\be
K_R(p) := \int \frac{d^3 q_1}{(2\pi)^3} \frac{d^3 q_2}{(2\pi)^3}
( B(q_1,q_2;p) - B(q_1,q_2;0) )
.
\label{eq:3.122}
\ee

Let us analyze now $\fh^g_n(p)$ for $n \ge 4$. The discussion is similar to
that for $\phi^3_4$.  The contributions follow by extracting $n$ $p$-legs in
all possible ways from the five $\hat\Gamma$-graphs with loops and $\delta g$
in \fig{5}. The $\Gamma$-graphs so obtained are of two types, divergent and
finite,
\be
\fh^g_n(p) := \fh_n^{g(\rd)}(p) + \fh_n^{g(\rf)}(p)
.
\ee
Divergent are precisely those graphs containing either
\includegraphics[height=8mm]{MG43.pdf} or
\includegraphics[height=8mm]{MG45.pdf} as a subgraph. All other graphs are
finite. Explicitly, for
$n=4$, the divergent graphs are
\be
\includegraphics[height=22mm]{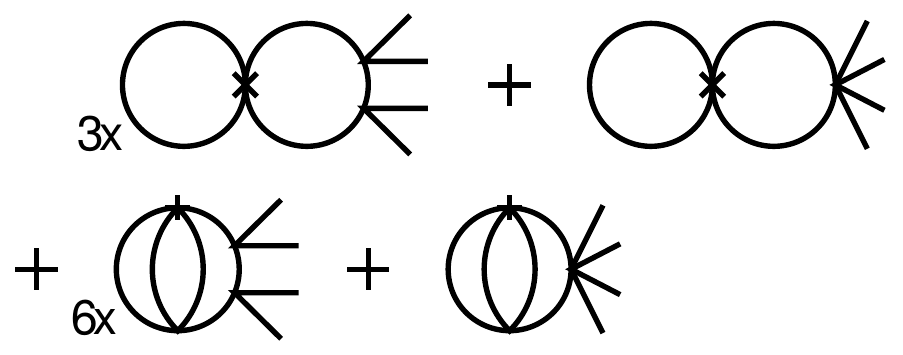}
\ee
Note that a graph like \includegraphics[height=8mm]{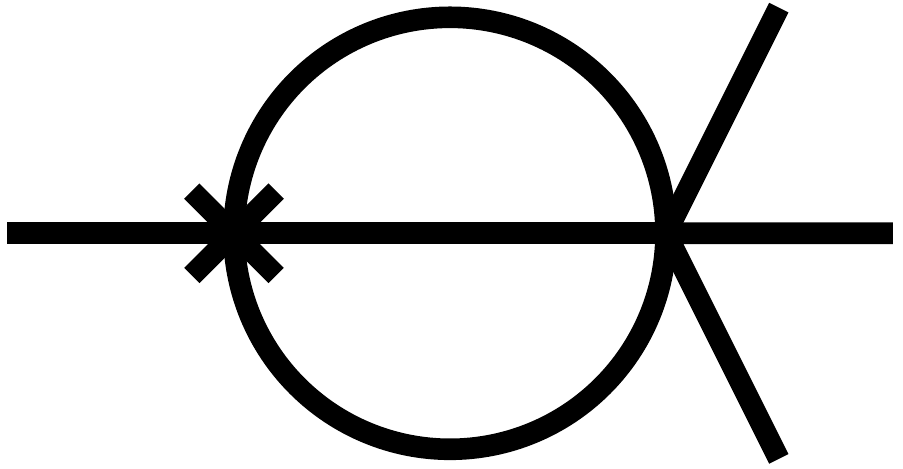}
is UV-finite; a divergence would require the three internal lines with large
momentum, but in that case $\Gamma_6(q)=O(1/q)$. More generally,
$\Gamma_n(q)=O(1/q)$ when $n > 4$ and more than two lines carry a large
momentum. This is not in conflict with the fact that $\Gamma_n(q)$ is indeed
$O(1)$ for $n \ge 4$ in a kinematic configuration $\Gamma_n(q,-q+k,p)$ with
$q$ large and $k$, $p$ finite.

As already happened for $\phi^3_4$, the structure of the divergent graphs is
such that the amplitude $A_n(q;p)$ is produced. Thus
\bes
\fh^{g(\rd)}_n (p) &= \delta_{n,2} ( C_{g,a} + K(p) )
\\ &\quad
+ \int \frac{d^4 q}{(2\pi)^4} \left( F_q^2 C_{g,a} + 
  F_q K(q) \right) A_n(q;p)
.
\ees
This leads to
\be
\fh_n^g(p) = \hfh_n^g(p) + C_g ( \delta_{n,2} + \fh_n^m(p) )
,
\ee
where $\hfh_n^g(p)$ is UV-finite. This expression ensures that a single
$p$-independent parameter $\delta m_0^2(\Lambda)$ renormalizes all divergences
induced by $\delta g$.

After imposing the renormalization conditions, \nec{3.47} (with $\delta_{n,4}$
instead of $\delta_{n,3}$) holds along with \nec{3.78} and \nec{3.79}.  There,
$K_R(p)$ is defined in \nec{3.122} and $A_{R,n}(q;p)$ is defined in
\nec{3.54}. In these expressions all integrals are convergent without a
regulator.

\subsection{\textsf{ The effective action in the manifold $(m_R^2,Z,g)$
 \label{sec:3.d} }}
  
The previous Sec. \ref{sec:3.b} and \ref{sec:3.c} have dealt with the
super-renormalizable theories $\phi^\kappa_d$ for $(\kappa,d) = (3,4)$ and
$(4,3)$. In both cases, the effective action functional is determined by the
three parameters $(m_R^2,Z,g)$,\footnote{Actually two further parameters come
  from the renormalization conditions on $\Gamma_n(p)$ for $n=0$ and $n=1$
  which can be disregarded in this discussion.} and
\bes
\frac{ \partial \Gamma_n(p)}{\partial m_R^2 } &= \delta_{n,2} + \fh^m_{R,n} (p),
\\
\frac{ \partial \Gamma_n(p)}{\partial Z } &= \delta_{n,2} p^2 + \fh^Z_{R,n} (p),
\\
\frac{ \partial \Gamma_n(p)}{\partial g } &= \delta_{n,\kappa} + \fh^g_{R,n} (p)
.
\ees

Explicit expressions were obtained for the $\fh^\alpha_{R,n}(p)$ in terms
of $\Gamma$-graphs (for $n\ge 2)$. Since previously the results themselves
necessarily appear mixed with their derivation and with intermediate
definitions, for convenience we summarize here the main formulas:
\bes
\fh^m_{R,n}(p) &= \frac{ \fh^m_n (p) -  \delta_{n,2} \fh^m_2 (0) }{ 1 + \fh^m_2 (0) }
\\
\fh^Z_{R,n} (p) &= \int \frac{d^d q}{(2\pi)^d} q^2 A_{R,n}(q;p)
\\
\fh^g_{R,n}(p) &= \fh_{R,n}^{g(\rd)}(p) + \fh_{R,n}^{g(\rf)}(p)
\ees
\bes
A_{R,n}(q;p) &= A_n(q;p) - \frac{ \delta_{n,2} + \fh^m_n (p) }{ 1 + \fh^m_2 (0) }
A_2(q;0)
\\
\fh_{R,n}^{g(\rd)}(p)  &=
\delta_{n,2} K_R(p)
+
\int \frac{d^d q}{(2\pi)^d} K_R(q) \, A_{R,n}(q;p)
\\
\fh_{R,n}^{g(\rf)}(p)  &= \fh_n^{g(\rf)} (p) -
\frac{ \delta_{n,2} + \fh^m_n (p) }{ 1 + \fh^m_2 (0) } \fh_2^{g(\rf)} (0)
\label{eq:3.108}
\ees
\bes
K_R(p) &= \int \frac{d^4 q}{(2\pi)^4} ( B(q;p) - B(q;0) ) \quad [\phi^3_4]
\\
K_R(p) &= \int \frac{d^3 q_1}{(2\pi)^3} \frac{d^3 q_2}{(2\pi)^3}
( B(q_1,q_2;p) - B(q_1,q_2;0) )
 \quad [\phi^4_3]
\\
B(q;p) &= \fp(q) \fp(q+p) \Gamma_3(q,p)
\\
B(q_1,q_2;p) &= \frac{1}{3} \fp(q_1) \fp(q_2) \fp(q_1+q_2+p) \Gamma_4(q_1,q_2,p)
\ees
Finally,
\bes
A_n(q;p) &= \includegraphics[height=12mm]{GF2.pdf}
\\
\fh^m_n (p) &= \includegraphics[height=12mm]{GF4.pdf}
\\
\fh^g_n (p) &= \includegraphics[height=12mm]{GF6.pdf} \qquad [\phi^3_4]
\\
\fh^g_n (p) &= \includegraphics[height=12mm]{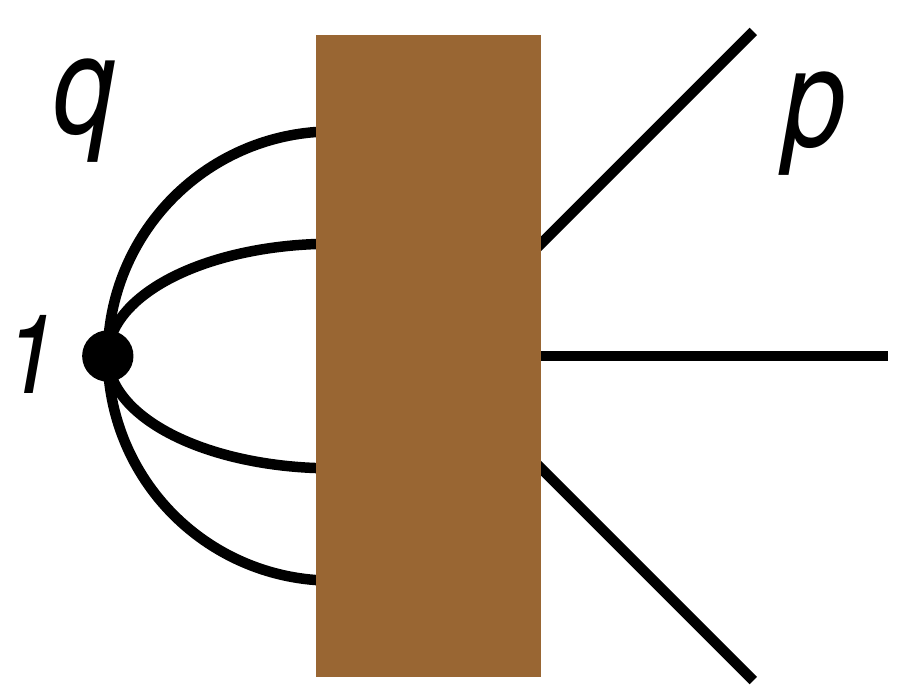} \qquad [\phi^4_3]
\ees
but $\fh_n^{g(\rf)}(p)$ only includes the graphs of $\fh_n^g(p)$ without
\includegraphics[height=8mm]{GF9.pdf} as a subgraph for $\phi^3_4$, and
without subgraphs \includegraphics[height=8mm]{MG43.pdf} or
\includegraphics[height=8mm]{MG45.pdf} for $\phi^4_3$.

\medskip
The expressions enjoy several interesting properties:

\begin{itemize}

\item[i)] They are UV-finite and renormalized without resorting to UV
  regulators. In this sense they are ``manifestly'' or ``explicitly''
  renormalized.

\item[ii)] The expressions are non-perturbative. They are exact and do not
  rely on the analysis of the behavior of an infinite number of graphs of $S$.

\item[iii)] They define an autonomous set of equations. For each value of $n$,
  the three derivatives are expressed using a finite number of
  $\Gamma$-graphs, that is, constructed with the full propagator and vertices
  of the effective action. At most $\kappa-1$ explicit loops are present.

\item[iv)] They allow to move on the manifold $(m_R^2,Z,g)$. In particular,
  starting from the free theory $(m_R^2,Z,0)$, for which
  $\Gamma[\phi]=S[\phi]$, the equation of
  $\partial \Gamma[\phi; g]/ \partial g$ can be solved in powers of $g$. Due
  to the hierarchical structure of the equations, this procedure delivers
  systematically the perturbative series already renormalized within a
  BPHZ-like scheme.

\item[v)] The equations are consistent, that is, successive derivatives
  commute; hence,
  \be
  \frac{\partial \fh^\beta_{R,n}(p)}{ \partial g_{R,\alpha}}
  =
  \frac{\partial \fh^\alpha_{R,n}(p)}{ \partial g_{R,\beta}}
  ,
  \ee
  where $g_{R,\alpha}$ refers to the coordinates $(m_R^2,Z,g)$. The
  consistency properties provide identities among the functions $\Gamma_n(p)$.

\end{itemize}

The constructions performed for $\phi^3_4$ and $\phi^4_3$ show that, at
least in the super-renormalizable case, the renormalization can be achieved by
perturbing an already renormalized and finite $\Gamma[\phi]$. The divergences
introduced by the perturbation through the presence of loops can be canceled
by taking suitable divergent parameters in the perturbing action.

It can be noted that $\delta m_R^2$ is of course an exact differential in the
manifold $(m_R^2,Z,g)$, as are $\delta Z$, $\delta g$ and
$\delta \Gamma[\phi]$, while the auxiliary quantity $\delta m^2$ (\Eq{3.45a})
is UV-finite but may not be an exact differential. Nevertheless, the
coefficients of $\delta m^2$, $\delta Z$ and $\delta g$ in \nec{3.44} (and the
analogous equation for $\kappa=4$) have the nice property of having a polynomial
dependence with respect to the full propagator and vertices, while
$\fh_{R,n}^\alpha(p)$ are rational functions of them.

Another observation is that $\fh^m_n(p)$ is finite without renormalization
(i.e., without requiring any subtractions) as is also
$\fh^m_{R,n}(p)$. Therefore, no (UV) renormalization is required when moving
between different values of $m_R^2$ with $(Z,g)$ fixed.

The derivatives $\fh^\alpha_{R,n}(p)$ are consistent with the symmetries of
the theory. In particular, under a rescaling of the field $\phi(x)$, the
action $S[\lambda \phi]$ has $\Gamma[\lambda \phi] + c(\lambda)$ as effective
action. This can be implemented through a rescaling of the parameters of the
action,
\be
\Gamma[\lambda \phi; m_R^2, Z,g]
=
\Gamma[\phi; \lambda^2 m_R^2, \lambda^2 Z , \lambda^\kappa g ]
\ee
(up to a $\phi$-independent term) or equivalently,
\be
\lambda^n \Gamma_n(p; m_R^2, Z,g)
=
\Gamma_n(p; \lambda^2 m_R^2, \lambda^2 Z , \lambda^\kappa g )
\qquad n > 0
.
\ee
This leads to the identities, such as
\bes
n \Gamma_n(p) &=
2 m_R^2 \big( \delta_{n,2} + \fh^m_{R,n}(p) \big)
  +
2 Z \big( \delta_{n,2} p^2 + \fh^Z_{R,n}(p) \big)
\\ &\quad +
\kappa g \big( \delta_{n,\kappa} +  \fh^g_{R,n}(p) \big)
\qquad  n > 0 \,.
\ees
This formula allows us to express $\fh^g_{R,n}(p)$ in terms of $\Gamma_n(p)$
and the derivatives $\fh^m_{R,n}(p)$ and $\fh^Z_{R,n}(p)$, which are
considerably simpler, as they have just one explicit loop. Naturally, and
regrettably, such a simplified form of $\fh^g_{R,n}(p)$ cannot be used to
solve the equation for $\partial \Gamma[\phi; g]/\partial g$ perturbatively,
as it is singular at $g=0$.

Likewise, if $\hbar$ is introduced explicitly in the functional integral
through
$S[\phi] \to S[\phi]/\hbar$ this is equivalent to a rescaling of the
parameters, that is,
\be
\frac{1}{\hbar} \Gamma[\phi; m_R^2, Z,g;\hbar]
=
\Gamma[\phi; \frac{m_R^2}{\hbar}, \frac{Z}{\hbar},\frac{g}{\hbar};1]
,
\ee
which leads to the identity
\be
\left( -1
  +
  \hbar \frac{\partial}{\partial \hbar }
  +
  m_R^2 \frac{\partial}{\partial m_R^2 }
  +
  Z \frac{\partial}{\partial Z }
  +
  g \frac{\partial}{\partial g }
\right) \Gamma [ \phi; \hbar]
= 0
.
\label{eq:3.116}
\ee
Again, the singularity at $\hbar=0$ prevents to reconstruct $\Gamma[\phi]$ with
loops starting from the tree level effective action $S[\phi]$. \Eq{3.116}
simply states that $\Gamma^{(L)}[\phi]$ (the contributions with $L$ loops) is
a homogeneous function of the variables $(m_R^2,Z,g)$ of degree $1-L$.

The theory  $\phi^4_3$ is stable and admits a mathematically rigorous
construction. Perhaps the closed expressions derived for $\partial
\Gamma[\phi;g]/\partial g$ allow to provide an alternative construction.
The analysis of this theory can be repeated without assuming
$\Gamma_n$ to vanish for odd $n$. The evolution (with respect to $g$) starting
from the free theory should presumably arrive at the same conclusion without
introducing it as an assumption.


\section{ \textsf{ Linearized renormalization of renormalizable theories }
  \label{sec:4} }

\subsection{ \textsf{ Projective renormalization scheme }
  \label{sec:4a} }

The discussion here will concern the {\em perturbative} renormalization of
renormalizable theories of the type $\phi^\kappa_d$, although it may also
apply to the super-renormalizable case. Actually, a theory such as $\phi^4_4$
is afflicted by the problem of triviality; unless $g=0$ it is thought to be
non-renormalizable beyond perturbation theory.\footnote{See however
  \cite{Romatschke:2023sce}.}  Likewise $\phi^3_d$ has a problem of stability
(even for $d=1$) since $\phi^{\rm odd}$ is not semi-bounded.\footnote{Although
  the functional integral could be rendered convergent by taking a suitable
  path in the complex plane of $\phi$ \cite{Pehlevan:2007eq}.} Here, we refer
to the correlation functions at a perturbative level, i.e., as a formal power
series in the coupling constant or in $\hbar$. Hence in that mathematical
sense, such correlation functions do exist and are well-defined.

Of course for, e.g. $\phi^3_6$, the action in \Eq{2.1a}, which contains fully
local operators, would not yield finite and well-defined values for the
Feynman graphs due to UV divergences, and a regularization scheme is required.
One such scheme is dimensional regularization (DR)
\cite{tHooft:1972tcz,Collins:1984xc} where the fields propagate in
$\bd=d-2\epsilon$ space-time dimensions, being $d$ the physical dimension.
Another is a Euclidean cutoff as in Sec. \ref{sec:3}. In what follows, we will
adopt the latter scheme for definiteness and briefly comment on DR below.

To be more specific, let us denote by $S_0[\phi]$ the action of the
theory. This is the {\em true action} that (upon removal of the regulator)
produces all the physical UV-finite correlation functions of $\phi(x)$. We
will call it the {\em action} or the {\em bare action}, but warn that in the
literature it is often denominated the renormalized action
\cite{Collins:1984xc}. This functional takes the form
\be
S_0 = m_0^2 \, O^m + Z_0 \, O^Z + g_0 \, O^g
.
\label{eq:3.9a}
\ee
The operators $O^\alpha$ are those in \nec{3.2a}, with a profile factor $F$
and a cutoff $\Lambda$. Equivalently,
\be
S_0[\phi] = g_{0,\alpha} O^\alpha[\phi]
\label{eq:3.9b}
.
\ee
A sum over $\alpha$ is implicit, $\alpha \in \{m,Z,g\}$ (here $m,Z,g$ are
merely labels, not variables). The bare couplings $g_{0,\alpha}$ are $m^2_0$,
$Z_0$, and $g_0$ and have a dependence on $\Lambda$. This action yields the
regulated effective action $\Gamma[\phi;\Lambda]$.

As is well-known, the theory will be perturbatively renormalizable
(super-renormalizable) when the coupling constant has zero (positive) mass
dimension. The (regulated) effective action itself depends on $\Lambda$ and is
UV-finite. This means that it is possible to choose the $\Lambda$ dependence
in the bare parameters of the action, $g_{0,\alpha}(\Lambda)$, in such a way
that the limit $\Lambda\to \infty $ of $\Gamma[\phi;\Lambda]$, denoted
$\Gamma[\phi]$, exists and is finite for sufficiently regular configurations
$\phi$.

In the renormalization program developed by Bogoliubov and collaborators
\cite{Bogolyubov:1959bfo,Collins:1984xc} the dependence of the bare parameters on
the regulator is obtained through counterterms. Let the functional $S_b$,
\be
S_b[\phi] = m^2 \, O^m + Z \, O^Z + g \, O^g
= g_{b,\alpha} O^\alpha[\phi]
\ee
be (what we will call) the {\em basic action}. Here $g_{b,\alpha}$ (that is,
$m^2$, $Z$, $g$) are {\em finite} ($\Lambda$-independent) parameters that
will serve as coordinates in the manifold of effective actions of the theory
and need not have a particular physical meaning. The basic action yields,
using the Feynman rules, an effective action $\Gamma_b[\phi;\Lambda]$ which
is UV divergent. To remove the divergences and obtain $\Gamma[\phi]$,
counterterms $S_\ct$ are added to the basic action,
\be
S_0 = S_b + S_\ct
\,.
\ee
The graphs in $\Gamma$ and $\Gamma_b$ are identical and only differ by the
couplings, $g_{0,\alpha}$ and $g_{b,\alpha}$, respectively. The basic
parameters are finite and $\Gamma_b$ diverges from the loops. $\Gamma$ is
finite after the divergences of the loops are compensated by the divergences
in the bare parameters.

The counterterms can be determined recursively within a perturbative or loop
expansion. At each new order, the new Feynman graphs of that order have the
{\em subdivergences} automatically canceled by lower order counterterms, while
the remaining superficial divergences are primitive and give a new
contribution to the counterterm action. For a graph with all subdivergences
subtracted, some prescription, implemented by a linear operation $T$, extracts
the divergent component while $(1-T)$ gives a purely finite component. The
operation $T$ can be chosen in many ways, but it must fulfill three essential
requirements, i) $T^2=T$ (idempotent), ii) the component $(1-T)$ must be
finite, hence any possible divergences must be isolated by $T$, and iii) $T$
must always act in the same form when acting on the same graph or subgraph.

Denoting, as usual, $\bR$ the operation of recursively extracting all
subdivergences (but not the overall divergence), the counterterms can be
summarized in the following formula:
\be
S_\ct = - T \bR (\Gamma_b - S_b )
\ee
so
\be
S_0 = S_b - T \bR ( \Gamma_b - S_b )
.
\label{eq:3.25}
\ee
$\Gamma_b - S_b $ is the sum of the graphs of $\Gamma_b$ with loops.
Explicitly \cite{Zimmermann:1969jj,Itzykson:1980rh} on each graph of
$\Gamma_b$,
\be
\bR = \prod_\gamma{}^\prime (1 - T_\gamma)
,
\ee
where $\gamma$ denotes every possible subgraph with loops of effective-action
type, that is, amputated, connected, and irreducible. The full graph itself is
excluded from the product. The factor $1 - T_\gamma$ removes any possible
divergent component of the subgraph $\gamma$. The subtractions are to be
applied orderly, with smaller subgraphs first. The prime indicates that, upon
expansion, monomials with factors of the type $T_{\gamma_1} T_{\gamma_2}$ are
to be removed when the subgraphs $\gamma_1$ and $\gamma_2$ are overlapping,
that is, they share at least one line but neither is a subgraph of the other.

Another standard definition is
\be
R := (1- T ) \bR
.
\ee
The operation $R$ represents the removal of all subdivergences, including the
overall divergence. The effective action is obtained after applying the
operation $R$ to the graphs of $\Gamma_b$ with loops; therefore,
\be
\Gamma = S_b + R (\Gamma_b - S_b )
.
\label{eq:3.30}
\ee
The meaning of this relation is that if one adds all (regulated)
effective-action-like graphs produced by $S_0$, upon removal of the regulator
the functional $\Gamma[\phi]$ is obtained on the LHS, and the same result is
obtained on the RHS by using instead $S_b$ plus systematic subtraction of the
divergences in the graphs with loops.

As said, there is much freedom in the choice of $T$. The most usual choices are
of the minimal type, that is, $\Tmin$ vanishes if the graph or subgraph has no
divergences. This is the case of minimal subtraction (MS) in DR, where $\Tmin$
selects the principal part in the power series in $\epsilon$, or in BPHZ,
where $\Tmin$ extracts the component of the Taylor series in powers of the
external momenta up to a degree equal to the degree of divergence (or nothing if
there is no divergence). In minimal schemes
\be
\Tmin \Gamma = \Tmin S_b = 0
,\qquad
(1 - \Tmin ) S_0 = S_b
\,.
\label{eq:4.10}
\ee
The first equations follow from the fact that $\Gamma$ and $S_b$ are finite,
the latter equation indicates that $S_b$ is the finite component of $S_0$,
because the counterterms are strictly divergent and so $\Tmin S_\ct= S_\ct$.

Here we will discuss a different, non-minimal, scheme, which will be called
{\em projective renormalization scheme}. As already noted in
Sec. \ref{sec:3.d}, a perturbative solution of the equation for
$\partial \Gamma_n(p)/\partial g$ provides something similar to the BPHZ
expansion, where each graph is renormalized before integration. However, in
BPHZ the subtraction is minimal, while here even finite graphs are
subtracted since what is enforced are the {\em renormalization conditions}
(RC).

To see this in more detail, let us consider the renormalizable case, where in
principle all the three bare parameters need renormalization. The effective
action manifold is coordinated by three renormalized finite parameters
$g_{R,\alpha}$, namely, $m_R^2$, $Z_R$ and $g_R$, defined through some RC. A
typical choice (for the massive case) is that of renormalization conditions at
zero momentum, or more generally at a scale $\muRC \ge 0$,
\bes
\Gamma_2(p)\Big|_{\muRC}
- \muRC^2 \frac{\partial\Gamma_2(p)}{\partial(p_1^2)}\Big|_{\muRC}
&= m_R^2 ,
\\
\frac{\partial\Gamma_2(p)}{\partial(p_1^2)}\Big|_{\muRC}  &= Z_R,
\\
\Gamma_\kappa(p)\Big|_{\muRC} &= g_R
,
\label{eq:3.3}
\ees
where $m_R^2$, $Z_R$ and $g_R$ are three finite $\Lambda$-independent
parameters. Here $\Gamma_n(p)$ refers to the $n$-point vertex function, and
the kinematic condition $\big|_{\muRC}$ refers to
\be
p_i p_j = \frac{\muRC^2}{n-1} ( n \delta_{ij} - 1 )
.
\label{eq:4.12}
\ee

If the theory is massless, there are only two parameters: $Z_R$ and $g_R$.  In
the {\em renormalizable} case the theory has no scale and DR does not
introduce one, so one can adopt as renormalization conditions
\bes
\frac{1}{\muRC^2}\Gamma_2(p)\Big|_{\muRC} &= Z_R ,
\\
\Gamma_\kappa(p)\Big|_{\muRC} &= g_R
,
\label{eq:3.6}
\ees
for $\muRC>0$, and the operator $O^m$ is not present in the
actions.\footnote{$O^m$ is needed also in the massless case when the regulator
  is a cutoff.}

In a general renormalization scheme, the basic couplings $g_{b,\alpha}$
together with some subtraction prescription $T$ produce the effective action
functional (\Eq{3.30}). These basic couplings, which define a coordinate
system in the manifold, are then adjusted to reproduce some RC. The idea of
the projective renormalization scheme is to adopt $g_{b,\alpha}=g_{R,\alpha}$
from the beginning, hence,
\be
S_b = S_R
\ee
where
\be
S_R[\phi] :=
m_R^2 \, O^m + Z_R \, O^Z + g_R \, O^g
=
g_{R,\alpha} O^\alpha[\phi]
.
\ee
In addition, within this scheme, the corresponding effective action $\Gamma_b$
will be denoted $\Gamma_R$.

In order to provide proper definitions, let us introduce the linear space
$\HC$ spanned by the functionals $\Gamma[\phi]$ which are translationally
invariant, with arbitrary (but sufficiently regular) coefficients
$\Gamma_n(p)$.\footnote{Here we are using the symbol $\Gamma$ to denote both a
  generic effective action-like functional and the actual effective action
  corresponding to the action $S_0$. We will let the two uses be
  distinguished by the context rather than introducing a different new
  notation.} Then the renormalization conditions in \Eq{3.3} or \Eq{3.6} can
be cast in the form
\be
\hT_m \Gamma = m_R^2,
\qquad
\hT_Z \Gamma = Z_R,
\qquad
\hT_g \Gamma = g_R,
\ee
or
\be
\hT_\alpha \Gamma[\phi] = g_{R,\alpha}
\qquad
\alpha \in \{ m,Z,g\}
,
\label{eq:3.8}
\ee
with $g_{R,\alpha}=m^2_R, Z_R, g_R$ (or just $Z_R, g_R$ in the massless case).
Here the mappings $\hT_\alpha$ are linear forms on $\HC$ (i.e.,
$\hT_\alpha \in \HC^*$). We can further define the subspace $\HS$ as
\be
\HS := \mspan(\{ O^\alpha \} ) \subseteq \HC
,
\ee
where $O^\alpha[\phi]$ are the operators appearing in the action. The 1-forms
$\hT_\alpha$ have been defined in such a way that {\em when restricted to}
$\HS$ they are the dual basis of the $O^\alpha$, i.e. they fulfill the
relation
\be
\hT_\alpha O^\beta = \delta^\beta_\alpha
\,,
\label{eq:3.10}
\ee
and hence also
\be
\hT_\alpha S_0 = g_{0,\alpha}
.
\ee
This is readily verified: for the functional $S_0$ in \nec{3.9a} and using the
definitions in \nec{3.4} and \nec{3.5}, one obtains in momentum
space (dropping irrelevant form factors)
\be
S_{0,2}(p) = m_0^2 + Z_0 p^2,
\quad
S_{0,3}(p)=g_0,
\quad
S_{0,n \ge 4}(p) = 0
,
\ee
hence $\hT_m S_0 = m_0^2$, $\hT_Z S_0 = Z_0$, and $\hT_g S_0 = g_0$ for the 
RC in \nec{3.3}, or in \nec{3.6} when $m_0^2=0$.

The dual basis property of the $\hT_\alpha$ allows us to define the
projector operator $T$ as
\be
T := O^\alpha \hT_\alpha
,
\qquad
T^2 = T
.
\label{eq:4.22}
\ee
Also $\hT_\alpha T = \hT_\alpha$.

$T$ acts on the space $\HC$ and projects onto $\HS$. Hence in particular
\be
TS_0 = S_0
.
\ee
Note that the conditions \nec{3.10} restrict, but do not completely determine,
the $1$-forms $\hT_\alpha$ (e.g., they depend on the scale $\muRC$). By the same
token, while the subspace $\HS$ is determined from (the operators in) $S_0$,
the projector $T$ itself depends on how the complementary subspace is chosen
in the direct sum decomposition $\HC = \HS \oplus \HS^\perp$, i.e., on
the concrete choice of the renormalization conditions. Here $\HS^\perp$ is
just a name for the space $(1-T)\HC$, no scalar product is introduced in
$\HC$ to define the orthogonality.

The basic action functional $S_b=S_R$ is also in $\HS$, hence
\be
T S_R = S_R
.
\ee
From its definition, this action fulfills the identity
\be
T \Gamma = S_R
\,,
\label{eq:4.25}
\ee
which is an alternative form of the RC \nec{3.8}, and
\be
\hT_\alpha S_R = g_{R,\alpha}
.
\ee
By its very definition
\be
\bR S_R = S_R
,
\ee
since the graphs of $S_R$ contain no subgraphs (with loops).
It follows that
\be
R S_R = (1 - T) \bR S_R = (1 - T) S_R = 0
.
\label{eq:4.28}
\ee

Every (effective-action-like) graph defines a functional of $\phi$, and the
action of $T$ only depends on such a functional. This is unlike the operations
$\bR$ and $R$, which depend on the detailed structure of the graph. The
projector $T$ in \nec{4.22} characterizes this renormalization
scheme. $T$ is not minimal and acts as an idempotent operator on the space of
(translationally invariant) functionals of $\phi$, so its action depends on
the subgraph and not on where the subgraph is. In this sense $T$ has a
geometric nature. In a renormalizable (or super-renormalizable) theory, the
superficial divergences have the same form as the action itself, hence $1-T$
selects finite components only.
When a cutoff is used, the operators $O^\alpha$ depend on the profile function
$F$ and the regulator. Both $\HS$ and $T$ inherit such dependence since the
$O^\alpha$ spans $\HS$.
For the theory $\phi^3_6$, or for $\phi^4_4$ assuming no breaking of the
symmetry $\phi\to-\phi$, the three operators in \nec{3.2a} suffice to span
$\HS$. In general all operators required by counterterms need to be included,
and the theory is deemed renormalizable when the dimension of $\HS$ is finite.

The operator $T$ is not of the minimal-subtraction type: the relations
\nec{4.10} are very different from those in \nec{4.25} or
\nec{4.28}. Nevertheless, the two general relations
\be
S_0 = S_R - T \bR ( \Gamma_R - S_R )
\label{eq:3.25b1}
\ee
and
\be
\Gamma = S_R + R (\Gamma_R - S_R )
,
\label{eq:3.25b2}
\ee
are fulfilled, where $\Gamma_R$ denotes the (divergent by loops) effective
action produced by $S_R$. Subtraction of the two equations (noting that
$R + T\bR= \bR$) yields
\be
\Gamma - S_0  = \bR ( \Gamma_R - S_R )
.
\label{eq:3.31}
\ee
This formula is remarkable because it contains the other two: \nec{3.25b1}
follows immediately from \nec{3.31} by applying $T$ on both sides, using
$TS_0=S_0$ and $T\Gamma= S_R$. Likewise, \nec{3.25b2} follows by applying
$1-T$.\footnote{Also a relation $\Gamma - S_0 = \bR ( \Gamma_b - S_b )$ holds
  in any scheme, after subtracting \nec{3.25} from \nec{3.30}, but in
  general neither contains the other two.}

Note that the relations \nec{3.25b1} and \nec{3.25b2} are not specifically
related to UV divergences or even to quantum field theory. These identities
hold equally well in models where the range of the index $i$ in $\phi^i$ is
discrete or even finite. They are diagrammatic identities that follow from
the form of the partition function $Z[J]$ as a sum or integral over the
Boltzmann weight in \nec{2.7}. Given the action functional $S_0[\phi]$, the
functional $\Gamma[\phi]$ is fully determined through the Feynman rules as a
formal sum of graphs constructed with the parameters in $S_0$. The choice of
projector $T$ then completely fixes $S_R = T\Gamma$ in terms of
$S_0$. Everything is expressed in terms of $S_0$ or equivalently in terms of
the parameters $g_{0,\alpha}$.  One can then decide using instead the
variables $g_{R,\alpha}$.  In this view, \nec{3.25b1} is just the change of
variables expressing the parameters $g_{0,\alpha}$ in terms of the parameters
$g_{R,\alpha}$, that is, $S_0$ as a function of $S_R$. Likewise \nec{3.25b2}
expresses $\Gamma$ in terms of $S_R$. The $\bR$ and $R$ operations are
automatically implemented when the inversion $S_R(S_0) \to S_0(S_R)$ is
performed perturbatively. This point is illustrated in Appendix \ref{app:B}.

\Eq{3.31} is noteworthy because, regarding $S_0$ as the independent variable,
the LHS is independent of the RC, that is, the concrete choice of $T$ or
equivalently the choice of the subspace $\HS^\perp$, hence the equation
implies that the dependence on $T$ must also cancel on the RHS (when
everything is expressed in terms of $S_0$). An infinitesimal change in $T$
(e.g. in the scale $\muRC$) leads to the Callan-Symanzik equations
\cite{Callan:1970yg,Symanzik:1970rt,Collins:1984xc,Zinn-Justin:2002ecy} which
we do not analyze further here.

Until now we have assumed that the image of the projector $T$ fills up the
space $\HS$ spanned by the action. More generally, one can consider a subspace
$\cH_T \subseteq \HS$, such that still $T^2=T$ but $T \HC = \cH_T$. Now, the
operation $1-T$ in $\bR$ and $R = (1-T) \bR$ removes only the component
parallel to $\cH_T$ leaving the rest of $\HS$ in place. Accordingly, the
definition of the basic action $S_R$ is generalized to a partial projection
\be
S_R = T \Gamma + (1-T) S_0
.
\label{eq:3.36}
\ee
Consider for instance a theory
\be
S_0 = Z_0 O^Z + g_0 O^g
,
\ee
with $T = O^g \hT_g$, in this case
\be
S_R = Z_0 O^Z + g_R O^g
.
\ee
This amounts to saying that instead of the full change of variables
$(Z_0,g_0) \to (Z_R,g_R)$, a partial change of variables
$(Z_0,g_0) \to (Z_0,g_R)$ is performed, leaving $Z_0$ unchanged because the
removal operation $(1-T)$ only acts on three-point subgraphs.

Since $1-T$ still acts recursively, the three relations
\bes
\Gamma - S_0  &= \bR ( \Gamma_R - S_R )
,
\\
S_0 &= S_R - T \bR ( \Gamma_R - S_R )
,
\\
\Gamma &= S_R + R (\Gamma_R - S_R )
,
\label{eq:3.31a}
\ees
still hold and are consistent with \nec{3.36}.

This partial renormalization may be particularly useful for
super-renormalizable theories, such as $\phi_4^3$. There only $m_0^2$ needs to
be renormalized as $Z_0$ and $g_0$ are UV-finite. There are no UV divergences
in the sector spanned by $O^Z$ and $O^g$, so the action of $T$ can be
restricted to the sector spanned by $O^m$.\footnote{The formalism allows to
  consider non-standard partial projections, mixing subgraphs with different
  number of legs, such as
$
T = (O^Z + O^g) \frac{1}{2} ( \hT_Z + \hT_g ) := O \,\hT_O
.
\nonumber
$
It is not clear whether such a possibility could find any useful applications.}

The projective scheme has an obvious disadvantage in concrete calculations: it
is not minimal, every graph of effective action type with a component along
$\HS$ (i.e., with $2$ or $\kappa$ legs in a $\phi^\kappa_d$ theory) gives a
contribution to the counterterms, even if it is fully UV-finite. On the other
hand, virtues are its geometric and systematic nature, and that by
construction $T(\Gamma-S_R)=0$; the tree level $S_R$ already saturates the RC
and they are preserved at every order hence the effective action is expressed
in terms of physical coordinates $g_{R,\alpha}$ automatically, rather than in
terms of intermediate coordinates $g_{b,\alpha}$. The perturbative solution of
$\partial \Gamma_n(p) / \partial g_R$ yields the graphs renormalized under the
projective scheme. Let us also remark that the use of the projective scheme is
compatible with DR.

\subsection{ \textsf{ Linearized renormalization }
  \label{sec:4b} }

The UV divergences in the renormalizable theories are more severe than those
of super-renormalizable ones, correspondingly, the results that we have
obtained are also less conclusive.

According to Schwinger's principle, for a theory with bare action \nec{3.9b}
\be
\delta \Gamma[\phi] = \delta g_{0,\alpha} \espp{ O^\alpha }
\label{eq:4.b1}
.
\ee
Our point of is be that the unperturbed effective action is already
renormalized and only the new divergences need regularization. To do this a
cutoff will be assumed. The possibility of using DR is briefly discussed
below.

We will use the notation $G^\alpha[\phi] := \espp{ O^\alpha }$
already introduced in \nec{2.33}
, that is,
\be
G^\alpha[\phi] := \frac{\partial \Gamma[\phi] } { \partial g_{0,\alpha} }
,
\ee
or equivalently
\be
G^\alpha_n(p) := \frac{\partial \Gamma_n(p) } { \partial g_{0,\alpha} }
,
\ee

and correspondingly,
\bes
\delta \Gamma_n (p) &=
\delta g_{0,\alpha}  G^\alpha_n (p)
\\
&=
\delta m_0^2 \, G^m_n (p) 
+
\delta Z_0 \, G^Z_n (p) 
+
\delta g_0 \, G^g_n (p) 
.
\label{eq:3.32c}
\ees

We remark that here we are using the notation based on the functions
$G^\alpha_n(p)$ while $H^\alpha_n(p)$ were used in Sec. \ref{sec:3} (e.g., in
\nec{3.32a}). They are related by
$G^\alpha[\phi] = O^\alpha[\phi] + H^\alpha[\phi]$, that is,
\bes
G^m_n(p) &= \delta_{n,2} + H^m_n(p)
,
\\
G^Z_n(p) &= \delta_{n,2} p^2 + H^Z_n(p)
,
\\
G^g_n(p) &= \delta_{n,\kappa}  + H^g_n(p)
.
\ees
The functions $H^\alpha_n(p)$ only include the loop terms while
$G^\alpha_n(p)$ contain also tree graphs. Apart from this difference, the
formulas given for $H^\alpha_n(p)$ in the super-renormalizable case (e.g.
\nec{3.32b}) in terms of the propagators and effective vertices also apply in
the renormalizable case up to the different dimension in the momentum
integrals.

$G^\alpha[\phi]$, or equivalently $G^\alpha_n(p)$, are the quantities that we
can obtain explicitly from $\Gamma[\phi]$, and they are divergent by loops.
The quantities that are completely finite are, instead,
\be
G_R^\beta[\phi] := \frac{\partial \Gamma[\phi] } { \partial g_{R,\beta} }
,
\ee
so that
\bes
\delta \Gamma_n (p) &=
\delta g_{R,\alpha}  G^\alpha_{R,n} (p)
\\
&=
\delta m_R^2  G^m_{R,n} (p) 
+
\delta Z_R  G^Z_{R,n} (p) 
+
\delta g_R  G^g_{R,n} (p) 
.
\ees

In order to relate both sets of quantities, let us introduce the matrix
\be
W^\alpha{}_\beta := \frac{ \partial g_{R,\beta} }{\partial g_{0,\alpha} }
,
\label{eq:4.b2.5}
\ee
which depends on the regulator and is divergent. In terms of it
\be
G_R^\beta[\phi] =
\frac{\partial \Gamma[\phi] } { \partial g_{R,\beta} }
=
\frac{ \partial g_{0,\alpha} }{ \partial g_{R,\beta} }
\frac{ \partial \Gamma[\phi] }{ \partial g_{0,\alpha} }
,
\ee
therefore
\be
G_R^\beta[\phi] = (W^{-1})^\beta{}_\alpha \, G^\alpha[\phi]
,
\label{eq:4.b3}
\ee
or equivalently
\be
G_{R,n}^\beta(p) = (W^{-1})^\beta{}_\alpha \, G^\alpha_n(p)
.
\label{eq:4.b3a}
\ee

The matrix $W$ can be computed from $G^\alpha[\phi]$ by using the RC in
\nec{3.8},
\be
W^\alpha{}_\beta = \frac{ \partial g_{R,\beta} }{\partial g_{0,\alpha} }
= \frac{ \partial \hT_\beta \Gamma[\phi] }{\partial g_{0,\alpha} }
= \hT_\beta \frac{\partial \Gamma[\phi] }{\partial g_{0,\alpha} }
\ee
hence
\be
W^\alpha{}_\beta = \hT_\beta G^\alpha[\phi]
.
\label{eq:4.b4}
\ee
The consistency condition
\be
\hT_\alpha G_R^\beta[\phi] = \delta^\beta_\alpha
,
\ee
which follows from the definition of $G_R^\beta$ and the RC, is also derived from
the relations \nec{4.b3} and \nec{4.b4}.

To be more explicit, for the RC in \nec{3.3} with $\muRC=0$ the matrix $W$ takes
the form
\be
W =
\begin{pmatrix}
  G^m_2(0) & \partial_{p^2}  G^m_2(0) & G^m_\kappa(0)
  \\
  G^Z_2(0) & \partial_{p^2}  G^Z_2(0) & G^Z_\kappa(0)
  \\
  G^g_2(0) & \partial_{p^2}  G^g_2(0) & G^g_\kappa(0) 
\end{pmatrix}
,
\ee
while for the massless case RC in \nec{3.6}
\be
W =
\begin{pmatrix}
  \muRC^{-2} G^Z_2(\muRC) & & G^Z_\kappa(\muRC)
  \\
  \muRC^{-2} G^g_2(\muRC) & & G^g_\kappa(\muRC)
\end{pmatrix}
.
\ee

In the DeWitt notation, from
\be
O^\alpha[\phi] = 
\sum_{\ell \ge 0} \frac{1}{\ell!} O^\alpha_{ i_1 \ldots i_\ell}
\phi^{i_1} \cdots \phi^{i_\ell}
,
\ee
it follows that
\be
G^\alpha[\phi] = O^\alpha[\phi] + 
\sum_{\ell \ge 2} \frac{1}{\ell!} O^\alpha_{ i_1 \ldots i_\ell}
\hH^{i_1\ldots i_\ell}[\phi]
.
\label{eq:4.b0}
\ee

The main formulas are then i) \nec{4.b0} (or its momentum space version
\nec{4.b0a}) expressing $G^\alpha[\phi]$ in terms of the effective action, ii)
\nec{4.b4} which provides the matrix $W^\alpha{}_\beta$, and iii) \nec{4.b3}
that yields $G_R^\beta[\phi]$.

The divergent functions $G^\alpha_n(p)$ enjoy two distinct properties, namely,
their divergences are {\em explicit} and {\em anti-canonical}.

Let us explain the meaning of {\em explicit} in this context. In \nec{4.b0},
both the operator coefficients $O^\alpha_{ i_1 \ldots i_n}$ and the
functionals $\hH^{i_1\ldots i_\ell}[\phi]$ are free from divergences (for
regular field configurations). The divergence arises from the sum over the
indices. In momentum space,\footnote{$O^\alpha_n(p)$ is only defined in the
  subspace $\sum_{i=1}^n p_i=0$. In the integral
  $\sum_{i=1}^\ell q_i= \sum_{j=1}^n p_j= 0$. The form factors are included in
  $O^\alpha_\ell(q)$.}
\be
G^\alpha_n(p) = O^\alpha_n(p) + 
\sum_{\ell \ge 2} \frac{1}{\ell!} \int^\Lambda
\prod_i^{\ell-1}\frac{d^dq_i}{(2\pi)^d}
O^\alpha_\ell(q) H^\ell_n(q;p)
,
\label{eq:4.b0a}
\ee
and $G^\alpha_n(p)$ is expressed as a momentum integral (identical to those
found in the super-renormalizable case) where the integrand itself is
completely finite and independent (or essentially independent) of the
cutoff. The divergence emerges only as the cutoff grows and the integration
region increases, and in this sense the divergence is {\em explicit}, there
are no hidden divergences in $G^\alpha_n(p)$ and so they are also absent in
$W^\alpha{}_\beta$, due to $W=\hT G$.

The fact that the divergences are explicit in $G^\alpha_n(p)$ or
$W^\alpha{}_\beta$ is not to say that they are {\em superficial} in a
technical sense. Let us clarify this point. As is well-known, when an
$S$-graph having only an overall divergence of degree $\gamma \ge 0$ (all
subdivergences removed) is differentiated more than $\gamma$ times with
respect to (external) momenta or masses, it becomes convergent; consequently,
the divergent component of such graph is a polynomial of degree $\gamma$ of
the momenta and the masses \cite{Collins:1984xc}. On the other hand, the
corresponding (regulated) momentum integral of the graph (that is, excluding
coupling constants) has a mass dimension $\gamma$. As a mathematical
consequence, the dependence of the integral on the cutoff $\Lambda$ follows a
{\em canonical pattern}:
\be
\sum_{k=1}^\gamma P_k \Lambda^k +
P_0 \log(\Lambda) + f(\Lambda)
,
\ee
where the $P_k$ are $\Lambda$-independent polynomials of degree $\gamma-k$ in
momenta and masses, and $f(\Lambda)$ has a finite limit as the cutoff is
removed. The presence of subdivergences introduces non-canonical divergent
terms, such as $\Lambda^n \log^m(\Lambda)$. Of course, in the renormalizable
case, the dependence on the momenta in the canonical divergent terms conforms
to the operators in the action, and the divergence in the bare parameters
$g_{0\alpha}(\Lambda)$ is canonical because the counterterms collect only
contributions from superficial divergences.

\begin{figure}[ht]
  \begin{center}
\includegraphics[height=25mm]{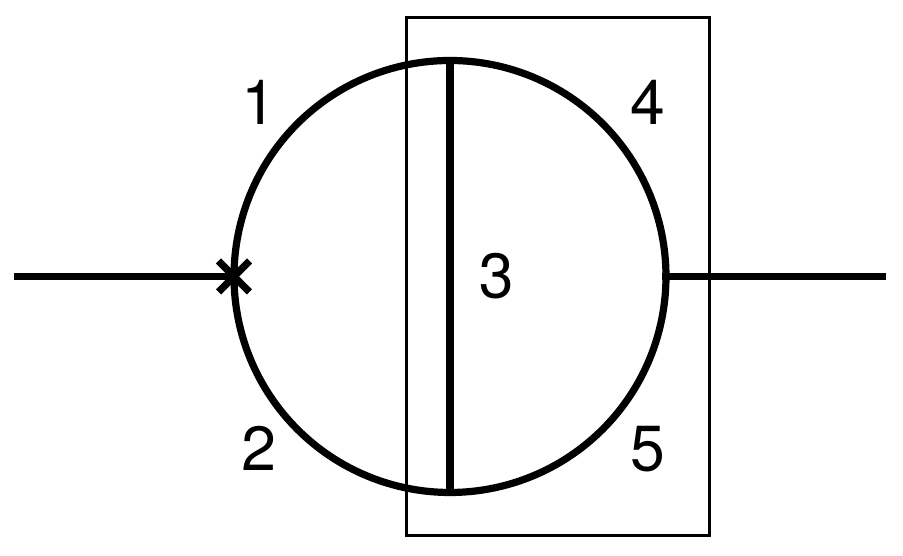}
\end{center}
\caption{An $S_R$-graph of the $\phi^3_6$ theory for $G^g_2(p)$. The graph
  belongs to the $\Gamma$-graph class
  \includegraphics[height=8mm]{GF9.pdf}. The divergence of the subgraph
  $(345)$ has been subtracted, and this is indicated by the thin-line box.}
\label{fig:9}
\end{figure}
The dependence on $\Lambda$ in $G^\alpha_n(p)$ or the matrix $W$ is explicit
but not necessarily canonical due to the presence of subdivergences: even if
the functions $H_n^\ell(q;p)$ are finite, new divergences and subdivergences
are produced by attaching the vertex $O^\alpha_\ell(q)$ and integrating over
$q$.  This is illustrated in \fig{9}, which represents an $S_R$-graph of the
theory $\phi^3_6$ contributing to $G^g_2(p)$. The subgraph with lines $(345)$
comes from $\Gamma[\phi]$ and is already subtracted. When the crossed vertex
is added, two new divergences arise, one is the subgraph $(123)$,
logarithmically divergent, and the other is the full graph $(12345)$, which
diverges quadratically. If they are not subtracted, this is a contribution to
$G^g_2(p)$. Since the subgraph $(123)$ is not subtracted, the divergence in the
full graph is not only superficial and hence $G^g_2(p;\Lambda)$ will contain
non-canonical terms.

\begin{figure}[ht]
  \begin{center}
\includegraphics[height=25mm]{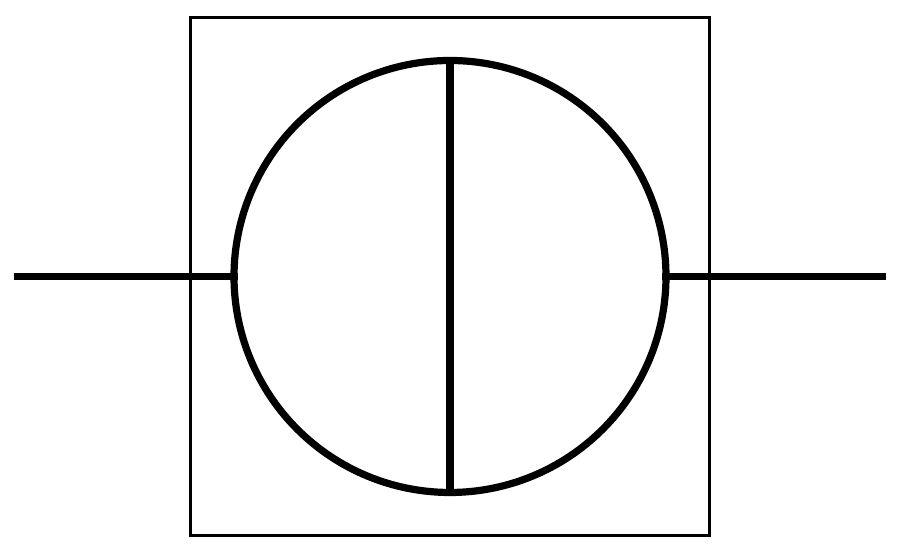}
\end{center}
\caption{An $S_R$-graph of the $\phi^3_6$ theory for $\Gamma_2(p)$. The box
  indicates that all the divergences are subtracted.}
\label{fig:10}
\end{figure}

Using this example, it is interesting to emphasize that if the two new
divergences are also subtracted, what is obtained is a contribution to
$G^g_{R,2}(p)$. Let us discuss this point in more detail. As already noted
after \nec{3.30}, \Eq{3.25b2} states that the same functional $\Gamma$ is
obtained from the $S_0$-graphs without subtractions or from the $S_R$-graphs
with subtractions. This observation can be applied to $G^\alpha[\phi]$ and
$G_R^\alpha[\phi]$. The coefficients $\hH^{i_1\ldots i_\ell}$ or
$H^\ell_n(q,p)$ can be (finitely) expressed through $\Gamma$-graphs. In turn,
they can be expanded in terms of $S_0$-graphs or in terms of {\em subtracted}
$S_R$-graphs. In the latter representation \nec{4.b0} expresses
$G^\alpha[\phi]$ by means of graphs of the theory
`$S_R+O^\alpha$'\footnote{That is, graphs constructed with lines and vertices
  of $S_R$ plus the vertex corresponding to $O^\alpha$.} but including only
graphs with exactly one vertex $O^\alpha$ and with subtraction of all
subgraphs not involving the vertex $O^\alpha$.  The only remaining divergences
are those induced by the composite operator $O^\alpha$. If one proceeds to
subtract these remaining divergences, $G_{R,n}^\alpha(p)$ is obtained instead
of $G_n^\alpha(p)$. To see this consider, for instance, $\Gamma_2(p)$ expanded
as a sum of subtracted $S_R$-graphs, as required by \nec{3.25b2}. \fig{10}
shows one such $S_R$-graph. These contributions depend on the variables
$g_{R,\alpha}$. If a variation $\delta g_R$ is applied, the corresponding
derivative is $G^g_{R,2}(p)$. The variation of $\delta g_R$ in the graph of
\fig{10} produces (among other) the graph in \fig{9} with all divergences
subtracted, including the subgraph $(123)$ and the full graph $(12345)$.

Hence $G^\alpha[\phi]$ and $G_R^\alpha[\phi]$ share the same graphs, with the
partial and full removal of the divergences, respectively. In this view,
\Eq{4.b3} is nothing but the linear version of \nec{3.25b2}. Bogoliubov's $R$
and $\bR$ operators act by eliminating divergences on every subgraph; in the
linearized version, the divergences irradiate from a single composite operator
and their removal is obtained by a simple inversion of the matrix $W$; $G$ and
$W$ share the same divergences and they cancel in the combination
$W^{-1}G=G_R$. Of course, the price to pay for the linear simplicity is that
only the derivatives of the effective action are obtained rather than the
effective action itself.

However, the structure of the divergences in $W$ is by no means arbitrary, on
the contrary: {\em the matrix $W^{-1}$ has only canonical divergences}. To see
this, let us introduce the matrix $V^\beta{}_\alpha$ through
\be
(W^{-1})^\beta{}_\alpha  = \delta^\beta_\alpha - V^\beta{}_\alpha
.
\ee
Using \nec{4.b2.5}, this definition implies the relation
\be
\delta g_{0,\alpha} =
\delta g_{R,\alpha} - \delta g_{R,\beta} V^\beta{}_\alpha
.
\label{eq:4.b6}
\ee
This is the linear version of \nec{3.25b1}. The matrix $V$ collects the
contributions of the counterterms to the (variation of the) bare
couplings. Such contributions come from the superficially divergent component
of the primitively divergent graphs after all subdivergences have been
subtracted, and thus, as $g_{0,\alpha}(\Lambda)$ itself, they have only a
canonical dependence on $\Lambda$.

\begin{figure}[ht]
  \begin{center}
\includegraphics[height=30mm]{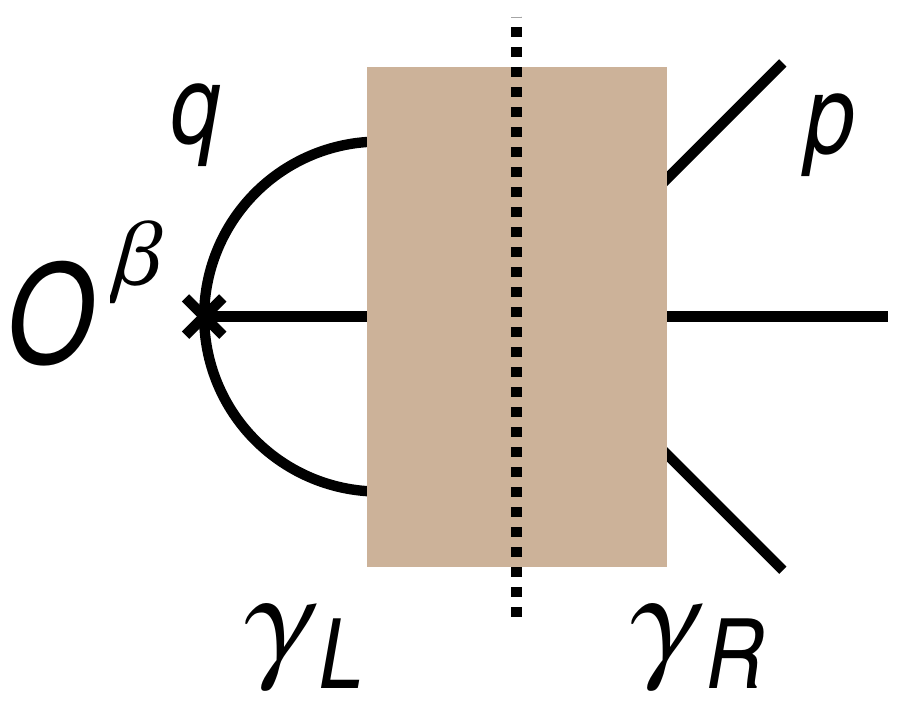}
\end{center}
\caption{Schematics of the graphs in $G^\beta_n(p)$, to expose
  \Eq{4.b5}. The vertical dotted line separates the left-hand subgraphs
  containing the vertex $O^\beta$ from right-hand subgraphs, containing the
  dependence on $p$.}
\label{fig:7}
\end{figure}

Additionally, \nec{4.b3} also implies the relation
\be
G^\beta[\phi]
 =
G_R^\beta[\phi]
+
V^\beta{}_\alpha G^\alpha[\phi]
.
\label{eq:4.b5}
\ee
The interpretation of this formula is as follows. We consider the typical
$S_R$-graphs contributing to $G^\beta_n(p)$, in \fig{7}. They start at the
vertex $O^\beta$, located on the left in the figure, and propagate to the
right, where $n$ (amputated) legs emerge carrying momentum $p$. All
divergences from subgraphs not involving the operator $O^\beta$ have been
subtracted, but there remain the new divergences irradiating from the
composite operator. We can consider all possible cuts of each graph,
represented by the vertical dotted line in the figure, separating the graph
into two subgraphs, $\gamma_L$ and $\gamma_R$, where the former contains the
source $O^\beta$ and the latter does not. To each such cut, we can associate a
decomposition $1 = T_{\gamma_L} + (1-T_{\gamma_L})$. For a renormalizable
theory $\phi^\kappa$ only cuts cutting at most $\kappa$ lines need be
considered, since otherwise $T_{\gamma_L}$ vanishes. As the cut is moved from
left to right, factors $(1-T_{\gamma_L})$ accumulate, hence subtracting the
subdivergences. The product of the $(1-T_{\gamma_L})$ factors produces the
fully subtracted graphs, and this is $G_R^\beta[\phi]$ on the RHS in
\nec{4.b5}. On the other hand, when a factor $T_{\gamma_ L}$ acts for the
first time, and therefore $\gamma_L$ has all subdivergences subtracted and
only displays an overall divergence, the projector onto $\HS$ selects
operators $O^\alpha$, with divergent coefficients that are counterterms of the
canonical type. These add to $V^\beta{}_\alpha$. Once the operator $O^\alpha$
is formed and all possible graphs $\gamma_R$ attached to it are added, they
reproduce the original structure, this time as $G^\alpha[\phi]$.

Summarizing, in all cases, the divergences present in $\Gamma[\phi]$ have been
renormalized (subtracted if expressed in terms of $S_R$-graphs).
$G^\alpha[\phi]$ describes the effect induced by a source $O^\alpha$, and
corresponds to graphs with unsubtracted divergences generated by that
source. $W^\alpha{}_\beta$ is the component of $G^\alpha[\phi]$ along
$O^\beta$.  $V^\beta{}_\alpha$ corresponds to graphs with source $O^\beta$ and
effect along $O^\alpha$ with subtractions of all {\em subdivergences}, but not
the overall divergence. $G_R^\beta[\phi]$ is the renormalized effect of a
source $O^\beta$, it corresponds to graphs with all divergences subtracted.

It is interesting to note that $G^\alpha[\phi]$ does not know about the RC,
hence $T$ or $\hT^\beta$ only appears once in the matrix $W=\hT G$. Upon
matrix inversion, $W^{-1}=1-V$ will produce instances of $\hT$ in several
places. At a perturbative level, it can be checked that such instances of $T$
precisely combine to produce the operators $\hT \bR$, so that $V$ contains
only superficially divergent graphs, without subdivergences. An explicit
perturbative calculation illustrates this point in Appendix \ref{app:C}.

Once again, the consistency conditions
\be
\frac{\partial G^\beta_R[\phi]}{ \partial g_{R,\alpha}}
=
\frac{\partial G^\alpha_R[\phi]}{ \partial g_{R,\beta}}
,
\label{eq:4.59}
\ee
or equivalently
\be
\frac{\partial G^\beta[\phi]}{ \partial g_{0,\alpha}}
=
\frac{\partial G^\alpha[\phi]}{ \partial g_{0,\beta}}
,
\ee
apply and lead to multiple identities among the correlation functions.

The formulas derived above apply to the super-renormalizable case as well. To
this end, for the theories $\phi^3_4$ or $\phi^4_3$ we define\footnote{ $Z_0$
  and $g_0$ were denoted $Z$ and $g$ in Sec. \ref{sec:3}.}
\be
m_R^2 = \Gamma_2(0)
,
\qquad
Z_R := Z_0,
\qquad
g_R := g_0
.
\ee
Besides convenience, these are proper RC, since $Z_0$ and $g_0$ can be
extracted from $\Gamma_2(p)$ and $\Gamma_\kappa(p)$ (with $\kappa=3$ or $4$)
in the large momentum limit. Specifically, these RC correspond to \nec{3.3}
with a suitable choice of the scale $\muRC$ in each case,
\bes
m_R^2 &= \hT_m \Gamma \qquad\mbox{with $\muRC = 0$},
\\
Z_R^2 &= \hT_Z \Gamma \qquad\mbox{with $\muRC = \infty$},
\\
g_R^2 &= \hT_g \Gamma \qquad\mbox{with $\muRC = \infty$}
.
\label{eq:4.62}
\ees
Therefore, in the super-renormalizable case, the matrix $W$ takes the form
\be
W =
\begin{pmatrix}
  G^m_2(0) & 0 & 0
  \\
  G^Z_2(0) & 1 & 0
  \\
  G^g_2(0) & 0 & 1
\end{pmatrix}
,
\ee
and so
\be
V =
\begin{pmatrix}
 1 - G^m_2(0){}^{-1} & 0 & 0
  \\
 G^m_2(0){}^{-1} G^Z_2(0) & 0 & 0
  \\
 G^m_2(0){}^{-1} G^g_2(0) & 0 & 0
\end{pmatrix}
.
\label{eq:4.64}
\ee
$G^m_2(0)$ is finite and $G^Z_2(0)$ and $G^g_2(0)$ have a canonical
logarithmic divergence, hence the divergences in $V$ are canonical too (and
exceptionally in the present case, also those of $W$). The renormalized
derivatives of the action follow from using \nec{4.b3}, and this
yields\footnote{ We have neglected the correlation functions with
  $n\in\{0,1\}$ throughout. Their inclusion adds terms to these relations
  that do not affect the sector $n\ge 2$.}
\bes
G^m_R[\phi] &= \frac{G^m[\phi]}{G^m_2(0)}
,
\\
G^Z_R[\phi] &= G^Z[\phi] - \frac{G^Z_2(0)}{G^m_2(0)}G^m[\phi]
,
\\
G^g_R[\phi] &= G^g[\phi] - \frac{G^g_2(0)}{G^m_2(0)}G^m[\phi]
.
\label{eq:4.b7}
\ees
These equation are, of course, identical to those in \nec{3.48} when the
latter are expressed in terms of $G^\alpha_n(p)$.

In Sec. \ref{sec:3} for $\phi^3_4$ and $\phi^4_3$ it was possible to rearrange
the integrals expressing $G^\alpha_{R,n}(p)$ in such a way that convergence
was explicit and an UV regulator was not needed. The divergences in the
renormalizable case are more severe than those in the super-renormalizable
case. For $\phi^3_6$ or $\phi^4_4$ all three parameters in the action require
renormalization, and in the massive case, the divergences are quadratic instead
of logarithmic. It can be conjectured that, in the renormalizable case, the
very same rearrangements of the integral made in Sec. \ref{sec:3} (but
increasing the dimension and keeping the regulator) while not completely
eliminating the divergences, will remove quadratic divergences and leave
logarithmic ones only. This is based on the observation that in the
super-renormalizable case, the matrix $W$ has the divergence structure
\be
V =
\begin{pmatrix}
 \mbox{fin} & \mbox{fin} & \mbox{fin}
  \\
 \log & \mbox{fin} & \mbox{fin}
  \\
 \log & \mbox{fin} & \mbox{fin}
\end{pmatrix}
\ee
while in the renormalizable case
\be
V =
\begin{pmatrix}
 \log & \log & \log
  \\
 \mbox{quad} & \log & \log
  \\
 \mbox{quad} & \log & \log
\end{pmatrix}
.
\ee

The relation \nec{4.b3} can be expressed more explicitly as
\be
G^\alpha[\phi;\Lambda] = W^\alpha{}_\beta(\Lambda;\muRC) \left(
  G^\beta_R[\phi;\muRC]
  + R^\beta[\phi;\Lambda;\muRC]
  \right)
,
\ee
or
\be
G^\alpha_n(p;\Lambda) = W^\alpha{}_\beta(\Lambda;\muRC) \left(
  G^\beta_{R,n}(p;\muRC)
  + R^\beta_n(p;\Lambda;\muRC)
  \right)
,
\ee
where $\muRC$ indicates the dependence on the RC and $R^\beta(\Lambda;\muRC)$
vanishes for large $\Lambda$. The information encoded here is that
asymptotically, in the large cutoff limit, $G^\alpha[\phi;\Lambda]$ is the
product of two factors, one that depends on $\Lambda$ but not on $\phi$, and
the other conversely. This is the necessary and sufficient condition for the
theory to be renormalizable in the linearized version.  The $\phi$-independent
factor $W$ can be extracted from $G^\alpha$ by applying the RC. The quantities
$G^\alpha_n(p)$, and hence $W$, depend only polynomially on the basic elements
$\fp(p)$ and $\Gamma_n(p)$, with regulated momentum integrals inserted in
various places. In the combination $G_R = W^{-1} G$ such dependence is no
longer polynomial but of the rational type. From this point of view, a
probably crucial simplification is that in the super-renormalizable case, the
determinant of $W$ (namely $G^m_2(0)$) is UV finite, and this guarantees that
the divergences in the matrix $W^{-1}$ have only a polynomial dependence on
the basic regulated integrals. This fact makes it easier to devise a
rearrangement of the terms to produce explicitly convergent blocks, as done in
Sec. \ref{sec:3}. Let us note that in the super-renormalizable case, the
divergence structure of $W$ is particularly simple and the equation of
$G^g_R[\phi]$ in \nec{4.b7} would suggest a simple subtraction scheme to
cancel the divergences, yet actually two subtractions were needed to achieve
manifestly convergent momentum integrals in \nec{3.108}.

The situation is much harder in the renormalizable case. While
renormalizability guarantees that the combination $W^{-1} G$ has a limit as
the regulator is removed, it is far from clear that the momentum integrals can
be explicitly rearranged as in the super-renormalizable case to yield
manifestly UV convergent blocks. Therefore, in the renormalizable case, one
arrives at expressions of $G$ and $W$ with an explicit regulator, and only
after the combination $W^{-1} G$ has been constructed the regulator can be
removed to yield $G_R$.  By way of illustration, one can imaging two
conditionally convergent series; in one case it has been possible to rearrange
the series into an absolutely convergent one with the same sum while for the
other series no rearrangement has been found or it may not exist.  In the
super-renormalizable case the theory is manifestly renormalized, while in the
renormalizable case, it is renormalized but not in an explicit manner.

To achieve an explicitly convergent expression of $G_R$ in the renormalizable
case (if that is at all possible), it might be of help the constraint that the
divergences in $W^{-1}$ are not arbitrary, but instead they follow a canonical
pattern. Thus $W$ enjoys two non-trivial properties: an {\em anti-canonical
  pattern} (meaning that $W^{-1}$ is canonical) and also its divergences are
{\em explicit} (meaning that they come from a single cutoff momentum integral
with a cutoff-independent integrand).

The divergences of $G^\alpha[\phi]$ are also explicit and anti-canonical,
which we define to mean that they can be canceled by multiplication with a
canonically divergent matrix, namely, $W^{-1}$. Another way to express the
anti-canonical divergence-pattern of $G^\alpha[\phi]$ is to take three
independent field configurations $\phi_A$, $A=1,2,3$, and form a $3\times 3$
matrix $G^\alpha{}_A$ with the three column-vectors
$G^\alpha[\phi_A;\Lambda]$, then the inverse matrix
$(G^{-1})^A{}_\alpha(\Lambda)$ has only canonical divergences,
\bes
&
(W^{-1})^\beta{}_\alpha G^\alpha{}_A =:  G^\beta{}_{R,A}
\\
&
(G^{-1})^A{}_\alpha = (G^{-1})^A{}_{R,\beta} (W^{-1})^\beta{}_\alpha
\ees
Instead of using three field configurations, one can use three sets of momenta
$p$, or in fact any three observables linearly extracted from the functional
$G^\alpha[\phi]$, the RC producing $W^\alpha{}_\beta$ being a particular case.
It is important to note that $(G^{-1})^A{}_\alpha$ is canonical only with
respect to its dependence on $\Lambda$. In general, however, its divergent
component will not be a polynomial on external momenta and masses (see
Appendix \ref{app:C}).

Another observation is that while the regulator cannot be fully removed in
the explicit non-perturbative expressions of $G^\beta_R[\phi]$, it can be
removed in a perturbative treatment. That is, if the equation for
$G^g_R[\phi]$ is solved perturbatively in powers of $g_R$, starting from
$m_R^2 O^m + Z_R O^Z$ at zeroth order, what is obtained are the usual
(regulated) momentum integrals from the standard Feynman rules of $S_R$ plus
systematic subtraction of divergences and subdivergences (with $T$ as operator
isolating the divergences) implementing Bogoliubov's $R$ operation on the
$S_R$-graphs. Then the regulator can be removed since the subtracted
integrals are already convergent. This is illustrated in Appendix \ref{app:C}.
In the super-renormalizable case $G_R$ is still obtained from $W^{-1} G$, as
explained around \nec{4.64}, hence the $R$ operation acts in the standard way.
The treatment in Sec. \ref{sec:3} looks different because it uses
$\Gamma$-graphs and the RC in \nec{4.62} automatically set to zero the action
of $\hT$ on most subgraphs.

A technical assumption has not yet been explicitly addressed.  We have
considered renormalizable theories, at least in a perturbative sense. This
means that for finite values $g_R$ (meaning here the various $g_{R,\alpha}$)
and some regularization with cutoff $\Lambda$, there are parameters
$g_0(\Lambda)$ in the action $S_0$ producing the functional
$\Gamma[\phi;\Lambda;g_R]$ which fulfills the RC of $g_R$ for all $\Lambda$,
and has a finite limit $\Gamma[\phi;g_R]$ as the regulator is removed.
Clearly, in this case, for $g_R+\delta g_R$ there will be
$g_0(\Lambda)+\delta g_0(\Lambda)$ such that
$\Gamma[\phi;\Lambda;g_R+\delta g_R]$ also has a finite limit. This can be
denoted the theory `$\Gamma(\Lambda)+\delta S_0(\Lambda)$'. The point here is
that $\Lambda$ appears both in the original (unperturbed) system and in the
perturbation, as the regulator is removed. However, our point of view has been
slightly different since we have been working with the theory
`$\Gamma+\delta S_0(\Lambda)$', i.e., the effective action of the
unperturbed system is already renormalized and only the perturbation is
regulated. Technically, the validity of such an approach is not guaranteed as
an immediate consequence of renormalizability. While the assumption worked
well in the super-renormalizable case, the divergences of just renormalizable
theories are more untamed. It is argued in Appendix \ref{app:D} that the
assumption is indeed valid: even if the parameters $\delta g_0(\Lambda)$
required to have a finite limit for `$\Gamma+\delta S_0(\Lambda)$' need not be
identical to those from `$\Gamma(\Lambda)+\delta S_0(\Lambda)$', they do exist.
This is not surprising since such final-result-independence also holds in
other aspects of renormalization. For instance, in the projective scheme $T$
acts according to the overlap of the graph with the space $\HS$, whether the
graph is actually divergent or not, while in the standard BPHZ approach
$\Tmin$ only acts if the graph is divergent; the two approaches produce the
same effective action when the cutoff is removed, although the detailed
parameters $g_{0,\alpha}(\Lambda)$ are different in both cases.

We conclude our discussion of the renormalizable case by considering the use
of DR as the UV regulator. In DR, the perturbative amplitudes are meromorphic
functions of $\epsilon$ with poles at most at integer values, hence finite for
$\epsilon$ close but different from $0$. The operators are still fully local
in $\bd=d-2\epsilon$ dimensions but (slightly) non-local from the point of
view of the $d$ dimensional degrees of freedom (i.e., after integrating out
the $(\bd-d)$-dimensional degrees of freedom), and in this way $\epsilon$ acts
as a regulator. A logarithmic mass scale $\nu$ is introduced as the measure
$\int d^d p/(2\pi)^d$ must be replaced by
$\nu^{2\epsilon}\int d^\bd p/(2\pi)^{\bd}$ \,to restore proper dimensions in
the Feynman graphs. All the amplitudes are then finite but depend on
$\epsilon$. By definition, the fact that a quantity $A(\epsilon)$ is UV-finite
means that the limit $\epsilon\to 0 $ exists and is finite, equivalently, the
residue at $\epsilon=0$ vanishes.

In our discussion a cutoff was applied as a regulator of the momentum
integrals. As it turns out, a possibility is open to regulate the UV
divergences by means of DR within the linearized approach. A way to do this
can be illustrated with $G^m_n(p)$ in \nec{3.32b},
\be
G_n^m(p) = \delta_{n,2} + \nu^{2\epsilon}
\int \frac{d^\bd q}{(2\pi)^\bd} \frac{1}{2}H^2_n(q,-q;p)
.
\label{eq:4.b2}
\ee
Here, the integrand is expressed in terms of the functions $\fp(k)$ and
$\Gamma_n(k)$. In the spirit of the linearized approach, these $\Gamma_n(k)$
are derived from $\Gamma[\phi]$ at $\epsilon=0$ (i.e., for $d$ spacetime
dimensions), but they are needed in $\bd=d-2\epsilon$ dimensions in
\nec{4.b2}. An obvious prescription is to exploit the Lorentz (or Euclidean)
invariance to express $\Gamma_n(k)$ in $d$ dimensions in terms of the
invariants $p_i p_j$, which are then identified with the corresponding
invariants in $\bd$ dimensions. The implementation of DR is then
straightforward if the vertices $\Gamma_n(k)$ are explicitly given as
functions of the invariants $p_i p_j$ (this would be the case, for instance,
within a perturbative calculation).  This prescription does not cover the
possibility of pseudoscalar invariants (involving the Levi-Civita symbol), but
this is rather an issue of DR itself, related to the presence of anomalies. A
more serious obstacle is that a Lorentz invariant function $\Gamma_n(k)$,
known only in $d$ dimensions, can be written in many inequivalent ways in
terms of the invariants $p_i p_j$ when $n-1 > d$, and therefore the extension
would not be unique.\footnote{For instance, $k_1^2 k_2^2 - (k_1 k_2)^2$
  vanishes identically in $d=1$, so a multiple of this function can be added
  to $\Gamma_3(k)$ in $d=1$ producing an ambiguity in its extension to
  $\bd$. Ultimately, this problem is also related to the Levi-Civita symbol;
  the scalar quantity $|k_1 \wedge k_2 \wedge \cdots \wedge k_{d+1}|^2$
  vanishes in $d$ dimensions.}  This ambiguity does not arise in the standard
application of DR in perturbation theory because in that case there are
explicit formulas in terms of $k_i k_j$ for the functions. In our case the
functions $\Gamma_n(k)$ are defined by their values rather than by explicit
formulas.  A prescription would be required for a ``natural'' or ``minimal''
extension of $\Gamma_n(k)$ when $n>d+1$. A compromise solution would be to use
only the $d$-dimensional components of the momenta in the vertex functions
with $n>d+1$ and the full $\bd$-dimensional momenta otherwise, or even to make
the extension only in the propagator function $\fp(k)$.  On the other hand,
the ambiguity is of order $\bd-d$, so its likely effect would be to redefine
the bare couplings without changing $\delta\Gamma[\phi]$ as the regulator is
removed. This point and the use of DR for $\delta \Gamma$ is illustrated in
Appendix \ref{app:D}. The possibility of using DR as a regulator is of
interest because the formulas of $G^\alpha_n(k)$ are non-perturbative, and a
non-perturbative formulation does not exist for DR at present.

\section{ \textsf{ Composite operators }
  \label{sec:5} }

\subsection{ \textsf{ General considerations }
  \label{sec:5.1} }

In the previous section we introduced the quantities
\be
G^\alpha[\phi] = \frac{\partial \Gamma[\phi]}{\partial g_{0,\alpha}}
 = \espp{O^\alpha}
.
\ee
They represent the (unrenormalized) amputated matrix elements of the operators
$O^\alpha$ (discussed after \nec{2.36} in Sec. \ref{sec:3.a}). The
renormalized ones can be obtained as
\be
G_R^\beta[\phi] = \frac{\partial \Gamma[\phi]}{\partial g_{R,\beta}}
 =: \espp{O^\beta}_R.
\ee
In terms of $S_R$-graphs, all divergences not touched by the composite
operator are subtracted; hence, they are already renormalized. In
$\espp{O^\alpha}$ the new divergences induced by the composite operator remain
unsubtracted, while in $\espp{O^\beta}_R$ they are removed through the $R$
operation. In view of the relation
%
$G_R^\beta[\phi] = (W^{-1})^\beta{}_\alpha G^\alpha[\phi]
,
$
%
one can define the {\em renormalized operators}
\be
O^\beta_R := (W^{-1})^\beta{}_\alpha O^\alpha
,
\ee
(hence $ \hT_\alpha O^\beta_R = (W^{-1})^\beta{}_\alpha $) such that
\be
\espp{O^\beta_R} = \espp{O^\beta}_R
.
\ee

The operators $O^\alpha[\phi]$ are functionals that are finite for regular
field configurations, and so finite at the tree or classical level; however,
they have divergent matrix elements when introduced in the dynamics described
by $\Gamma[\phi]$ due to the quantum fluctuations of the fields, i.e.,
radiative corrections. In turn, the renormalized operators are divergent as
functionals but have finite matrix elements.

A first-order variation in the relation $S_0 = g_{0,\alpha} O^\alpha$ implies
\be
\delta S_0 = \delta g_{0,\alpha} O^\alpha
= \delta g_{R,\beta} O^\beta_R
.
\ee
$\delta S_0$ diverges at the tree level but has finite matrix elements,
\be
\delta \Gamma[\phi] = \espp{ \delta S_0}
= \delta g_{0,\alpha} \espp{ O^\alpha }
= \delta g_{R,\beta} \espp{ O^\beta_R }
.
\ee

As indicated by these formulas, the amputated matrix elements of the operator
$O^\alpha$ can be obtained from the perturbation it produces on the effective
action when it is coupled to the action. By the same token, one can obtain the
matrix elements of other operators not necessarily present in the original
action. For instance\footnote{$\hat\phi(x)$ is the field regulated with a
  cutoff, as defined in \nec{3.11}. The operator is at $x=0$.}
\be
O^a[\phi] := \frac{1}{2} \hat\phi^2(0)
,
\ee
can be coupled as $\delta S_0 = \delta \lambda_{0,a} O^a$, and
\be
\delta \Gamma[\phi] = \delta \lambda_{0,a} \espp{O^a}
.
\ee
As a rule, $\espp{O^a}$ will present UV-divergences, and in general, it will not
be possible to cancel them (for all $\phi$) by a suitable choice of the cutoff
dependence in the bare parameter $\delta \lambda_{0,a}$. The cancellation will
require mixing with other composite operators of equal or smaller
dimension \cite{Collins:1984xc}. For the operator $O^a$ above in the
$\phi^3_6$ theory, the mixing requires the operators
\be
O^b := \hat\phi(0),
\qquad
O^c := -(\partial^2 \hat\phi)(0)
,
\qquad
O^d := 1
.
\ee
That is
\be
\delta S_0 = \delta \lambda_{0,a} O^a + \delta \lambda_{0,b} O^b
+ \delta \lambda_{0,c} O^c + \delta \lambda_{0,d} O^d
.
\ee
The bare parameters $\delta \lambda_{0,\mu}(\Lambda)$ can be chosen so that
\be
\delta \Gamma[\phi] = \delta \lambda_{0,\mu} \espp{O^\mu}
\quad
\mu \in \{ a,b,c,d \}
,
\ee
with implicit sum over the label $\mu \in \{ a,b,c,d \}$, is UV-finite. Once
again, here the unperturbed effective action $\Gamma[\phi]$ (and so its
propagator and all its correlations functions) is already renormalized and
UV-finite.

Note that also in the renormalization of the operators $O^\alpha$ in $S_0$
further operators $\propto 1$ and $\propto \phi$ are needed in the mixing to
produce a finite $\delta\Gamma_n(p)$ for $n=0,1$. Such operators were mostly
disregarded in the discussion as they do not affect the renormalization of
$\delta\Gamma_n(p)$ for $n\ge 2$. An important difference between the
operators in the action and the sporadic composite operators considered here,
is that the former act many times while the latter act only once (or at any
rate, act a finite number of times). As a consequence, the non-renormalizable
operators in the action (i.e., with negative mass-dimension coupling) require
the introduction new counterterm operators in the action of increasingly
larger dimensions, while the composite operators require only a finite number
of counterterms.

The new divergences introduced by the loops in the amputated matrix elements
of the composite operators $O^\mu$ are to be canceled by the bare parameters
$\delta \lambda_{0,\mu}(\Lambda)$. In order to do this, RC plus a projective
renormalization scheme can be employed.

Until now the effective action was parameterized by the three values
$g_{R,\alpha}$, so $\Gamma[\phi;g_{R,\alpha}]$, and these values are enforced
through RC. One can consider the enlarged action
$S_0 + \lambda_{0,\mu} O^\mu $, and correspondingly
$\Gamma[\phi;g_{R,\alpha},\lambda_{R,\mu}]$. The new parameters
$\lambda_{R,\mu}$ require additional RC and the following are suitable in the
present case,\footnote{Actually $\tilde\Gamma_0(p)$ has zero arguments $p$; we
  set $0$ for uniformity of the notation.}
\bes
\tilde\Gamma_2(0) &= \lambda_{R,a}
,
\\
\tilde\Gamma_1(0) &= \lambda_{R,b}
,
\\
\partial_{p^2}\tilde\Gamma_1(0) &= \lambda_{R,c}
,
\\
\tilde\Gamma_0(0) &= \lambda_{R,d}
.
\label{eq:5.11}
\ees
Here $\tilde\Gamma_n(p)$ denotes the momentum amplitudes in the expansion of
$\Gamma[\phi]$, as defined in \nec{3.4}. Because the operators $O^\mu$ are not
translationally invariant, the total momentum is not conserved and $\sum p$
needs not vanish in $\tilde\Gamma_n(p)$ in the theory with non-vanishing
$\lambda_{0,\mu}$.

If only the matrix elements of the operators $O^\mu$ are needed, and not more
general correlations functions, the conditions can be relaxed to
\bes
\delta \tilde\Gamma_2(0) &= \delta \lambda_{R,a}
,
\\
\delta \tilde\Gamma_1(0) &= \delta \lambda_{R,b}
,
\\
\partial_{p^2}\delta \tilde\Gamma_1(0) &= \delta \lambda_{R,c}
,
\\
\delta \tilde\Gamma_0(0) &= \delta \lambda_{R,d}
,
\label{eq:5.11a}
\ees
as a perturbation around the action $S_0$. i.e, the point
$\lambda_{R,\mu}=0$. We refer to this point as the point $\lambda=0$.

To include the operators $O^\mu$ it is necessary to extend the space of
functionals, as the space $\HC$ only includes the translationally invariant
ones. The total space will be
\be
\cH = \HC \oplus \HNC
\ee
(conserving and non-conserving momentum). A neat way to keep the two spaces
well separated and avoid problems, such as evaluating at $p=0$ when
$\delta(p)$ is present, is to introduce a projector $P_C$ so that for a
generic functional $F[\phi]$
\be
P_C \tilde{F}_n(p) := \lim_{\eta \to 0^+}
h\left((\textstyle{\sum} p)^2 /\eta\right) \tilde{F}_n(p)
.
\ee
Here $h(x)$, defined for $x\ge0$, is a decreasing smooth function with compact
support such that $h(0)=1$, and $h^{(k)}(0)=0 ~~\forall k\ge 1$. Then
\be
\HC := P_C \cH,
\qquad
\HNC := (1-P_C) \cH
.
\ee

The RC in  \nec{5.11} can be cast in the form
\be
\hT_\mu \Gamma[\phi] = \lambda_{R,\mu}
,
\ee
where the $\hT_\mu \in \cH^*$. These operators fulfill the duality
property\footnote{Indeed $\tilde{O}^a_n(p) = \delta_{n,2}$,
  $\tilde{O}^b_n(p) = \delta_{n,1}$,
  $\tilde{O}^c_n(p) = p^2 \delta_{n,1}$
  and $\tilde{O}^d_n(p) = \delta_{n,0}$,
  up to regulator-factors $F(p)$, which have no effect at $p=0$.}
\be
\hT_\nu O^\mu  = \delta^\mu_\nu
\quad
\mu,\nu \in \{ a,b,c,d \}
.
\ee
Furthermore, due to the separation between
$\HC$ and $\HNC$,
\bes
\hT_\mu \HC &= 0
\qquad
\mu \in \{ a,b,c,d \}
,
\\
\hT_\alpha \HNC &= 0
\qquad
\alpha \in \{ m,Z,g \}
.
\ees
Since $O^\alpha \in \HC$ and $O^\mu \in \HNC$,
\be
\hT_\mu O^\alpha = \hT_\alpha O^\mu = 0
.
\ee

The projector
\be
T = O^\alpha \hT_\alpha + O^\mu \hT_\mu
\ee
is such that
\be
T \HC = \HS,
\qquad
T \HNC = \HNS
,
\ee
where $\HNS\subseteq \HNC$ is the space spanned by the operators $O^\mu$. The
projector $T$ serves to construct the $R$ and $\bR$ operations to
renormalize both the action and the composite operators.

Specifically, following the previous steps, one can define the functionals
\be
G^\mu[\phi] := \espp{ O^\mu } =
\frac{\partial \Gamma[\phi]}{\partial\lambda_{0,\mu}}
.
\ee
These functions, or equivalently the functions $\tG^\mu_n(p)$, are needed only
at $S_0$ i.e. at the point $\lambda=0$. We assume that scenario in what
follows.

The functions $\tG^\mu_n(p)$ are expressed in terms of the vertex
$\tilde{O}^\mu_\ell(q)$ and the functions $H_n^\ell(q;p)$, similar to
\nec{4.b0a}, except that $O^\mu$ does not conserve momentum, namely,
\be
\tG^\mu_n(p) = \tilde{O}^\mu_n(p) + 
\sum_{\ell \ge 2} \frac{1}{\ell!} \int^\Lambda
\prod_i^{\ell-1}\frac{d^dq_i}{(2\pi)^d}
\tilde{O}^\mu_\ell(q) H^\ell_n(q;p)
.
\label{eq:4.b0b}
\ee
Here $\textstyle{\sum} p$ is arbitrary but the condition
$\sum q +\textstyle{\sum} p=0$ fixes one of the $q_i$. Likewise,
\be
G_R^\nu[\phi] := 
\espp{ O^\nu }_R
:= \frac{\partial \Gamma[\phi]}{\partial\lambda_{R,\nu}}
.
\ee
The relation
\be
W^\mu{}_\nu := \frac{\partial \lambda_{R,\nu}}{\partial \lambda_{0,\mu}}
\ee
then implies
\be
G_R^\nu[\phi] = (W^{-1})^\nu{}_\mu \, G^\mu[\phi]
.
\label{eq:4.b3b}
\ee
Since there is no mixing between the two sectors $\HC$ and $\HNC$ the would-be
matrix elements $W^\alpha{}_\nu= W^\mu{}_\beta$ are zero. The matrix $W$ in
the $\HNC$ sector follows from
\be
W^\mu{}_\nu = \hT_\nu G^\mu[\phi]
.
\label{eq:4.b4a}
\ee
A composite operator of a mass-dimension can only produce UV divergences of
that very dimension or less; therefore, in minimal schemes $W^\mu{}_\nu$ would
be a triangular matrix, as the matrix element would vanish when the dimension
of $O^\nu$ is larger than that of $O^\mu$. In the projective scheme, those
matrix elements are UV-finite but not necessarily zero. By assumption, the
space $\HNS$ spanned by the $O^\mu$ is such that all divergences generated by
those operators are also contained in the same space.

The renormalized composite operators are\footnote{If the set of composite
  operators is enlarged by adding new operators, the previous matrix elements
  of $W$ do not change, however $W^{-1}$, and so the combinations $O^\mu_R$ in
  terms of the $O^\nu$, may change in the unextended sector, unless the new
  operators are chosen so that the matrix $W$ is triangular.}
\be
O_R^\nu = (W^{-1})^\nu{}_\mu O^\mu
\ee
and they fulfill the relation
\be
\espp{ O^\nu_R} = \espp{ O^\nu }_R
.
\ee

For the operator $O^a$ in $\phi^3_6$ above, the renormalization yields
\be
O^a_R  = \frac{1}{\tG^q_2(0)} \left( O^a
- \tG^a_1(0) O^b - \partial^2_p\tG^a_1(0) O^c
- \tG^a_0(0) O^d
\right)
\ee
where $\tG^q_2(0)$ and $\partial^2_p\tG^a_1(0)$ are $O(L_\Lambda)$,
$\tG^a_1(0) = O(\Lambda^2)$ and $\tG^a_0(0) = O(\Lambda^4)$. In the present
case, the matrix $W$ is particularly easy to invert because, regardless of the
interaction, $\tG^A_n(p)= \tilde{O}^A_n(p)$ whenever $A$ contains at most one
field $\phi$, hence $\tG^b_n(p) = \delta_{n,1}$,
$\tG^c_n(p) = p^2 \delta_{n,1}$, and $\tG^d_n(p) = \delta_{n,0}$.

As already happened for the action, also for composite operators the
$\tG^\mu_n(p)$ and $W^\mu{}_\nu$ are explicit in terms of the cutoff momentum
integrals and present an anti-canonical divergence-pattern, while
\be
(W^{-1})^\nu{}_\mu = \delta^\nu_\mu - V^\nu{}_\mu
\ee
is canonical. Some related ideas are discussed in Appendix \ref{app:C}.

\subsection{ \textsf{ Renormalization of $\phi^3(0)$ in the theory $\phi^3_4$
  }
  \label{sec:5.2} }

In the super-renormalizable theory $\phi^3_4$, we consider the renormalization
of the composite operator
\be
O^a := \frac{1}{3!} \hat{\phi}^3(0)
.
\ee
Dimensionally, it mixes with the operator
\be
O^b := \frac{1}{2} \hat{\phi}^2(0)
,
\ee
and the operators $\partial^2\hat{\phi}(0)$, $\hat{\phi}(0)$, and $1$. We
disregard these because they only involve the renormalization of $G^\mu_n(p)$
for $n\le 1$.

Note that the relations
\be
O^g = \int d^d x \, O^a(x),
\qquad
O^m = \int d^d x \, O^b(x),
\ee
imply
\be\begin{split}
\tG^a_n(p) &= G^g_n(p)
\\
\tG^b_n(p) &= G^m_n(p)
\end{split}
\quad
\mbox{if~} \sum p=0
.
\ee

For the operators $O^a$ and $O^b$ we will adopt RC similar to those of $O^g$
and $O^m$ in \nec{4.62}, that is,
\bes
\lambda_{R,a}
=
\hT_a \Gamma &:= 
\tilde\Gamma_3(p)\Big|_{\muRC=\infty}
,
\\
\lambda_{R,b}
=
\hT_b \Gamma &:= 
\tilde\Gamma_2(p)\Big|_{\muRC=0}
.
\ees
Because the kinematical conditions \nec{4.12} enforce $\sum p=0$ for $n\ge2$,
our choice of RC in turn ensures that
\be\begin{split}
\tG^a_{R,n}(p) &= G^g_{R,n}(p)
\\
\tG^b_{R,n}(p) &= G^m_{R,n}(p)
\end{split}
\quad
\mbox{if~} \sum p=0
.
\ee
Furthermore, the matrix $W^\mu{}_\nu$ of the composite operators  (the basis
is ordered first $a$ then $b$) is
\bes
W^\mu{}_\nu &=
\PM{
  \tG^a_3(\infty) &  \tG^a_2(0) \\
  \tG^b_3(\infty) &  \tG^b_2(0) \\
}
=
\PM{
  G^g_3(\infty) &  G^g_2(0) \\
  G^m_3(\infty) &  G^m_2(0) \\
  }
\\ &=
\PM{
  1 &  G^g_2(0) \\
  0 &  G^m_2(0) \\
}
=
W^\alpha{}_\beta
,
\ees
That is, it coincides with the matrix $W^\alpha{}_\beta$ of the operators in
the action. As a consequence relations similar to \nec{4.b7} apply (for
$n\ge2$)
\bes
\tG^a_{R,n}(p) &= \tG^a_n(p) - \frac{G^g_2(0)}{G^m_2(0)} \tG^b_n(p)
,
\\
\tG^b_{R,n}(p) &= \frac{1}{G^m_2(0)} \tG^b_n(p)
,
\label{eq:4.b7a}
\ees
or equivalently
\bes
O^a_R  &= O^a - \frac{G^g_2(0)}{G^m_2(0)} O^b
,
\\
O^b_R &= \frac{1}{G^m_2(0)} O^b
,
\label{eq:4.b7b}
\ees
up to terms proportional to the operators $1$, $\phi(0)$ and
$\partial^2\phi(0)$.

It should be noted that not only the expressions for $\tG^a_{R,n}(p)$ and
$\tG^b_{R,n}(p)$ are explicit and finite, in fact, they can be arranged so that
no regulator is needed, as already happened for the functions
$G^\alpha_{R,n}(p)$ in Sec. \ref{sec:3.b}. This follows from the fact that the
expressions in both cases are very similar, namely (with implicit form
factors)
\bes
\tG^a_n(p) &=
\delta_{n,3}
\\\quad&+
\int^\Lambda \frac{d^d q_1}{(2\pi)^d} \frac{d^d q_2}{(2\pi)^d}
\frac{1}{3!} H^3_n(q_1,q_2,-q_1-q_2-\textstyle{\sum p} ; p)
\\
\tG^b_n(p) &=
\delta_{n,2}
+
\int^\Lambda \frac{d^d q_1}{(2\pi)^d} 
\frac{1}{2!} H^2_n(q_1,-q_1-\textstyle{\sum p} ; p)
,
\ees
only differ from $G^g_n(p)$ and $G^m_n(p)$ in \nec{3.32b} and \nec{3.32} by
the term $\textstyle{\sum p}$. Because $\textstyle{\sum p}$ is finite, the
asymptotic properties of the functions are identical and the same arrangements
carried out in Sec. \ref{sec:3.b} work here with straightforward
modifications. Entirely similar considerations apply to the renormalization of
the composite operator $\frac{1}{2}(\partial\phi)^2(0)$.  Moreover, they also
apply to the corresponding operators in the theory $\phi^4_3$.

\section{ \textsf{ Schwinger-Dyson equations}
  \label{sec:6} }

The relation%
\be
0 = \int D\varphi \partial_i \left( e^{S+J\cdot\varphi} A \right)
\ee
is valid for any sufficiently convergent observable $A[\varphi]$. It
immediately produces the Schwinger-Dyson equations (SDE)
\be
0 = \esp{ \partial_i A + A \partial_i S + A J_i }^J
.
\ee
More specifically, the choice $A=1$ yields ~$\esp{ \partial_i S}^J = -J_i$,
and hence
\be
\partial_i\Gamma[\phi]   = \espp{ \partial_i S}
.
\label{eq:3.3s}
\ee
The same result is also deduced directly from the identity
$\delta\Gamma[\phi] = \espp{\delta S}$ by noting that a shift
$S[\phi]\to S[\phi+\epsilon]$ is accompanied by
$\Gamma[\phi]\to \Gamma[\phi+\epsilon]$. In what follows by SDE we will refer
to the relations in \nec{3.3s}, or equivalent to them.

The SDE can be made more explicit by expanding both sides of \nec{3.3s} in
powers of $\phi$. This procedure yields
\bes
\sum_{n\ge 1}
&
\frac{1}{(n-1)!}
\fv_{i j_2 \ldots j_n} \phi^{j_2} \cdots \phi^{j_n}
\\ &=
\sum_{\ell\ge 1} \frac{1}{(\ell-1)!} g_{i i_2\ldots i_\ell}
\espp{ \varphi^{i_2} \cdots \varphi^{i_\ell} }
\ees
and so
\be
\fv_{i j_2 \ldots j_n}
=
g_{i j_2\ldots j_n}
+
\sum_{\ell\ge 3} \frac{1}{(\ell-1)!} g_{i i_2\ldots i_\ell}
H^{i_2 \ldots i_\ell}_{j_2 \ldots j_n}
\qquad n\ge 1
.
\label{eq:3.5a}
\ee
The LHS is completely symmetric, so the RHS is symmetric too, although not
manifestly so. These are the (unsymmetrized) SDE. The coefficients $H$ are
tree graphs of $\Gamma$, then after joining the $i$-legs in the vertex
$g_\ell$ of $S$, graphs with at most $\ell-2$ explicit loops are obtained.

Let us consider in more detail the particular case of a $\phi^3$ theory, i.e.,
$g_{i_1\ldots i_n}=0$ ~for $n\ge 4$
\bes
\fv_i
&=
h_i
+
\frac{1}{2} g_{i j k}
H^{j k}
\\
\fv_{i j_2 }
&=
m_{i j_2}
+
\frac{1}{2} g_{i j k}
H^{j k}_{j_2}
\\
\fv_{i j_2 j_3}
&=
g_{i j_2 j_3}
+
\frac{1}{2} g_{i j k}
H^{j k}_{j_2 j_3}
\\
\fv_{i j_2 \ldots j_n}
&=
\frac{1}{2} g_{i j k}
H^{j k}_{j_2 \ldots j_n}
\qquad n\ge 4
.
\ees

\begin{figure}[ht]
  \begin{center}
\includegraphics[height=37mm]{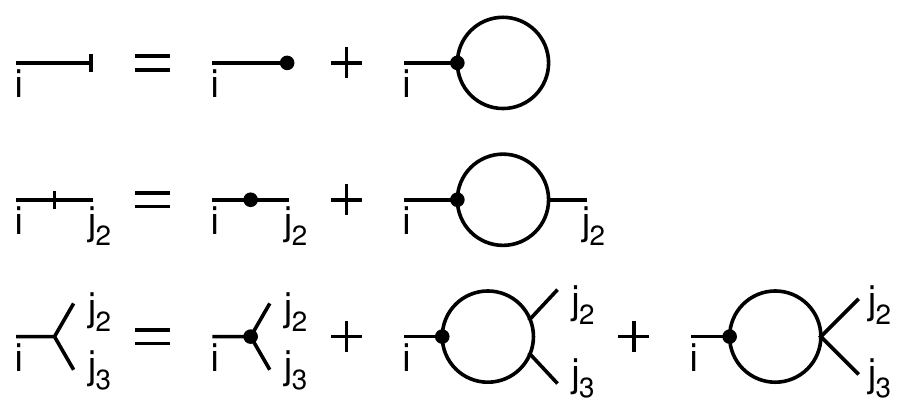}
\end{center}
\caption{Graph representation of SDE in \nec{6.7}. The dots represent the
  vertices of $S$, the other vertices and the lines are those of $\Gamma$.}
\label{fig:12}
\end{figure}

More explicitly, for $1 \le n \le 3$,
\bes
\fv_i
&=
h_i
+
\frac{1}{2} g_{i j k}
\fp^{j k}
\\
\fv_{i j_2 }
&=
m_{i j_2}
+
\frac{1}{2} g_{i j k}
\fp^{ja}\fp^{kb} \fv_{j_2 ab }
\\
\fv_{i j_2 j_3}
&=
g_{i j_2 j_3}
+
 g_{i j k}
\fp^{ja}\fp^{kb} \fp^{cd}
\fv_{j_2 ac } \fv_{j_3 bd }
\\ &\quad
+
\frac{1}{2} g_{i j k}
\fp^{ja}\fp^{kb} \fv_{j_2 j_3 ab }
.
\label{eq:6.7}
\ees

The diagrammatic representation of these equations is displayed in \fig{12}.
This is an infinite hierarchy of exact relations fulfilled by the theory
$\phi^3$. Furthermore, through systematic iteration (inserting the LHS in the
RHS) the hierarchy produces the perturbative series in powers of the coupling
$g$ in terms of Feynman graphs ($S$-graphs). In such a construction the
perturbative counting is assigned as follows:
\begin{itemize}
  \item[i)] $m_{ij}$,
    $\fv_{ij}$ and $\fp^{ij}$, as well as $h_i$ and $\fv_i$, are of $O(g^0)$,
  \item[ii)]
      $g_{ijk}$ and $\fv_{ijk}$ are of $O(g)$, and
  \item[iii)] the vertices
        $\fv_{i_1\ldots i_n}$ ($n \ge 4$) are of $O(g^n)$.
\end{itemize}
In each iteration $\fp^{ij}$ and $\fv_{i_1\ldots i_n}$ are expressed using
$s^{ij}$ and $g_{ijk}$ plus terms of higher order. Then, at each step, there
are contributions involving only $s^{ij}$ and $g_{ijk}$ plus a remainder that
contains the full propagator and vertices. The terms constructed solely with
$S$ no longer evolve and these are the standard Feynman graphs of the theory
$S$, while the remainder becomes of higher order at each iteration. Hence, the
procedure not only produces the perturbative series but also provides a closed
form for the exact remainder corresponding to the truncated series.

Since, as noted above, the RHS of \nec{3.5a} is not manifestly symmetric in
the indices $i,j_2,\ldots,j_n$, there is an ambiguity in how to expand
$\fp^{ij}$ and $\fv_{i_1\ldots i_n}$ present on the RHS using the SDE. A
simple prescription is to use a symmetrized version of the SDE, which will
also be useful later; namely, the RHS of \Eq{3.5a} is symmetrized by
hand. This gives rise to the {\em symmetrized} SDE:
\be
\fv_{j_1 \ldots j_n}
=
g_{j_1\ldots j_n}
+
\sum_{\ell\ge 3} \frac{1}{\ell!} g_{i_1\ldots i_\ell}
\Hs^{i_1 \ldots i_\ell}_{j_1 \ldots j_n}
\qquad n\ge 1
\label{eq:6.8}
\ee
where
\be
\Hs^{i_1 \ldots i_\ell}_{j_1 \ldots j_n}
:=
\frac{1}{n} \sum_{q=1}^\ell \sum_{p=1}^n \delta^{i_q}_{j_p}
H^{i_1 \ldots \widehat{i_q}\ldots i_\ell}_{j_1 \ldots \widehat{j_p}\ldots j_n}
\qquad n\ge 1, \quad \ell\ge 3
.
\label{eq:6.9}
\ee
By construction the coefficients $\Hs^{i_1 \ldots i_\ell}_{j_1 \ldots j_n}$
are completely symmetric and vanish for $\ell=1,2$.
Correspondingly, in the momentum space, there are functions
$H^\prime{}^\ell_n(q;p)$ in the subspace $\sum q+\sum p=0$.

In complete analogy with the relations
\bes
\delta \Gamma[\phi]
&=
\delta S_0[\phi]
+
\sum_{\ell \ge 2} \frac{1}{\ell!} \delta g_{i_1 \ldots i_\ell}
\hH^{ i_1 \ldots i_\ell } [\phi]
=
\delta g_{0,\alpha} G^\alpha[\phi]
,
\\
G^\alpha[\phi] &= O^\alpha[\phi] + H^\alpha[\phi]
,
\ees
the SDE in \nec{6.8} can be cast in the form
\bes
\Gamma[\phi]
&=:
S_0[\phi]
+
\sum_{\ell \ge 3} \frac{1}{\ell!} g_{i_1 \ldots i_\ell}
\hH^\prime{}^{ i_1 \ldots i_\ell } [\phi]
=:
g_{0,\alpha} G^\prime{}^\alpha[\phi]
,
\\
G^\prime{}^\alpha[\phi] &=: O^\alpha[\phi] + H^\prime{}^\alpha[\phi]
.
\label{eq:6.11}
\ees

Similar to the functionals $G^\alpha[\phi]$, the $G^\prime{}^\alpha[\phi]$
or the functions $H^\prime{}^\alpha_n(p)$ are expressed in terms
$\Gamma$-graphs, a finite number of graphs for each given number $n$ of
(amputated) $p$-legs. As in \nec{3.23}, $\ell$ $q$-legs are
attached to a vertex $O^\alpha$.

The integral over $q$ is, in general, UV divergent and {\em explicit} in the
sense that a regulator $\Lambda$ bounds the integration region, but otherwise
it is not present in the integrand. As was the case for $\delta \Gamma[\phi]$
through Schwinger's principle in previous Sections, here we assume that the
effective action, which determines the lines and vertices to construct the
$\Gamma$-graphs, is already renormalized. The regulator is needed only for the
explicit loops in the $\Gamma$-graphs and the divergences are compensated
through a suitable cutoff dependence in the bare couplings
$g_{0,\alpha}(\Lambda)$ explicit in the formula. The consistency of this
assumption is based on the analysis carried out in Appendix \ref{app:D}.

Applying the forms $\hT_\beta$ on both sides of \nec{6.11}, in order to
enforce the RC,
\be
\hT_\beta\Gamma[\phi]=
g_{R,\beta} =
g_{0,\alpha} W^\prime{}^\alpha{}_\beta 
\label{eq:6.12}
\ee
with
\be
W^\prime{}^\alpha{}_\beta := \hT_\beta G^\prime{}^\alpha[\phi]
.
\ee
Clearly, the matrix $V^\prime$, defined by
\be
(W^\prime{}^{-1})^\beta{}_\alpha  =: \delta^\beta_\alpha -
V^\prime{}^\beta{}_\alpha
\ee
fulfills the relation
\be
g_{R,\alpha}
=
g_{0,\alpha} 
+ g_{R,\beta} V^\prime{}^\beta{}_\alpha
.
\label{eq:6.15}
\ee
This equation conforms to the structure $S_R = S_0 + T \bR (\Gamma_R - S_R )$.
Therefore $V'$ has only canonical divergences (those in $g_{0,\alpha}$) and in
turn $W^\prime{}^\alpha{}_\beta$ and $G^\prime{}^\alpha[\phi]$ are
anti-canonical. Specifically, the functionals
\be
G^{\prime\beta}_R[\phi]
:=
(W^\prime{}^{-1})^\beta{}_\alpha G^\prime{}^\alpha[\phi]
\label{eq:6.16}
\ee
are UV-finite, and
\be
\Gamma[\phi] = g_{R,\beta} G^{\prime\beta}_R[\phi]
.
\label{eq:6.17}
\ee
However, note that the functionals $G^{\prime\beta}_R[\phi]$ are not
univocally determined by this equation, they are defined by the very hierarchy
of SDE.  When \nec{6.16} is expanded perturbatively the pattern
$\Gamma = S_R + R(\Gamma_R-S_R)$ is reproduced.  The equation paralleling
\nec{4.b5},
\be
G^\prime{}^\beta[\phi]
 =
G^{\prime\beta}_R[\phi]
+
V^\prime{}^\beta{}_\alpha G^\prime{}^\alpha[\phi]
,
\label{eq:4.b5a}
\ee
also holds.

Clearly, the relations fulfilled by the SDE and their renormalization are very
similar to those of the linearized renormalization in Sec. \ref{sec:4}.
Nevertheless, one noteworthy difference between the matrices $W$ and $W'$ (or
$V$ and $V'$) arises, namely, while the matrix elements of $W$ are all 
non-trivial and divergent in the renormalizable case, $W'$ and $V'$ have the
structures
\be
W' = \PM{ I & 0 \\ X & X }
,
\qquad
V' = \PM{ 0 & 0 \\ X & X }
,
\ee
where $I$ is the identity matrix and the first subset of indices refers to
operators $O^\alpha$ with $\ell \le 2$. This pattern follows from
$\hH^\prime{}^{i_1 \ldots i_\ell}[\phi]=0$ for $\ell\le 2$. Also, for a
$\phi^\kappa_d$ theory, the $\Gamma$-graphs of the SDE involve $\kappa-2$
explicit loops, for $\kappa-1$ loops in the case of $\delta\Gamma$.

The two sets of matrices $W$ and $W'$, or $V$ and $V'$, can be related. Using
\nec{4.b2.5} and \nec{6.12},
\be
W^\alpha{}_\beta = W^\prime{}^\alpha{}_\beta
+
g_{0,\gamma} \frac{ \partial W^\prime{}^\gamma{}_\beta }
{ \partial g_{0,\alpha} }
,
\ee
or equivalently
\be
V^\beta{}_\alpha = V^\prime{}^\beta{}_\alpha
+
g_{R,\gamma} \frac{ \partial V^\prime{}^\gamma{}_\alpha }
{ \partial g_{R,\beta} }
.
\label{eq:6.21}
\ee
Furthermore, from \nec{6.17}
\be
G^\beta_R[\phi] = G^{\prime\beta}_R[\phi]
+
g_{R,\gamma} \frac{ \partial G^{\prime\gamma}_R[\phi] }
{ \partial g_{R,\beta} }
.
\ee
The RHS automatically implements the consistency conditions \nec{4.59}.

In the renormalizable case, the $G^{\prime\beta}_R[\phi]$ are UV-finite but
not manifestly so: the regulator appears in $W^\prime{}^\alpha{}_\beta$ and
$G^\prime{}^\alpha[\phi]$ in such a way that the combination
$W^\prime{}^{-1}G^\prime[\phi]$ has a finite limit. This is exactly the same
situation as for $G_R[\phi] = W^{-1}G[\phi]$ in the linearized renormalization
of a renormalizable theory.  On the other hand, for super-renormalizable
theories, the renormalization of the SDE is manifest. For instance, for
$\phi^3_4$, using the RC in \nec{4.62},
\bes
W^{\prime g}{}_m &= \hT_m G^\prime{}^g = G^{\prime g}_2(0),
\\
W^\prime{}^g{}_Z &= \hT_Z G^\prime{}^g = \partial_{p^2}
G^{\prime g}_2(\infty) = 0,
\\
W^\prime{}^g{}_g &= \hT_g G^\prime{}^g = G^{\prime g}_3(\infty) = 1
.
\ees
Hence,
\be
W^\prime = \PM{ 1 & 0 & 0 \\ 0 & 1 & 0 \\  G^{\prime g}_2(0) & 0 & 1 }
,
\quad
V^\prime = \PM{ 0 & 0 & 0 \\ 0 & 0 & 0 \\  G^{\prime g}_2(0) & 0 & 0 }
.
\ee  
In view of the SDE for this theory, \fig{12},
\bes
G^{\prime \,m}_{R,n}(p) &= G^{\prime \,m}_n(p) = \delta_{n,2},
\\
G^{\prime \,Z}_{R,n}(p) &= G^{\prime \,Z}_n(p) = p^2\delta_{n,2},
\\
G^{\prime \,g}_{R,n}(p) &= G^{\prime \,g}_n(p)
- G^{\prime \,g}_2(0)\delta_{n,2}
.
\ees
Hence
\be
\Gamma_n(p) =
( m_R^2 + Z  p^2 ) \delta_{n,2}
+
g ( G^{\prime \,g}_n(p) - G^{\prime \,g}_2(0)\delta_{n,2} )
.
\label{eq:6.25}
\ee
Of course, the divergent component of the quantity
$\ds g G^{\prime \,g}_2(0) = \frac{1}{2}\Big(
\includegraphics[height=8mm]{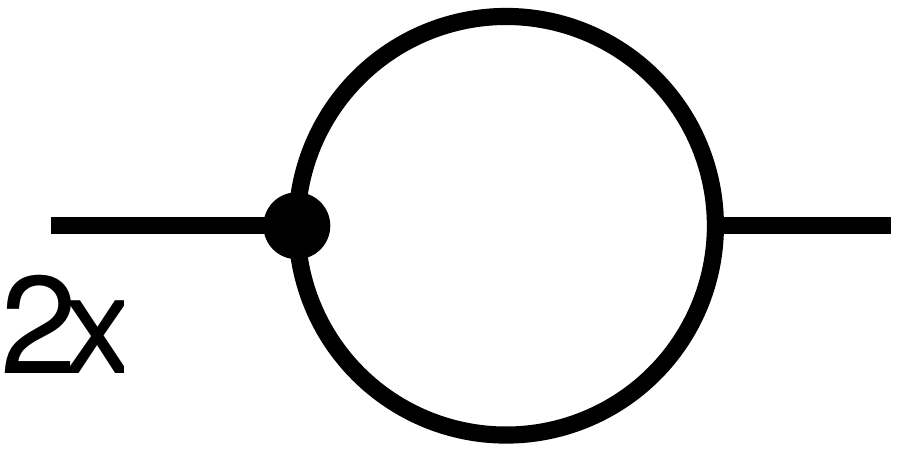} \Big)_{p=0}$ is just
$-m_\ct^2(\Lambda)$ in \nec{3.22}. If the iteration discussed after \nec{6.7},
eliminating $\Gamma[\phi]$ in favor of $S_R[\phi]$,
is systematically applied from \nec{6.25}, it produces the subtracted Feynman
graphs of the theory.

Using the expression for $V^\beta{}_\alpha$ in \nec{4.64}, the identities
\nec{6.21} imply the relations
\bes
1 &= \frac{1}{G^m_2(0)}
+ g \frac{ \partial G^{\prime\,g}_2(0) }{ \partial m_R^2 }
\\
\frac{ G^Z_2(0) }{ G^m_2(0) }
&=
g \frac{ \partial G^{\prime\,g}_2(0) }{ \partial Z }
\\
\frac{ G^g_2(0) }{ G^m_2(0) }
&=
\frac{\partial ( g G^{\prime\,g}_2(0) )}{\partial g} 
.
\ees
Because the divergent component of $G^{\prime \,g}_2(0) $ is independent of
$m_R^2$ the RHS of the first equation is UV-finite.


\acknowledgments This work has been partially supported by
MICIU/AEI/10.13039/501100011033 under grant PID2023-147072NB-I00 and by the
Junta de Andaluc{\'\i}a under grant No. FQM-225.  I thank the referee for
bringing Ref. \cite{Binosi:2005yk} to my attention, where a related approach
to renormalization is developed.


\appendix

\section{ \textsf{ Derivation of some formulas }
  \label{app:A} }

\subsection{ \textsf{ Proof of \Eq{2.1}  } }

Under the first-order variation $S\to S+\delta S$, it follows from \nec{2.7}
that $\delta W[J]=\esp{\delta S}^J$.  Hence, applying the variation $\delta$
to $W[J]= \Gamma[\phi]+J_i\phi^i$, with $J=J[\phi]$ and $\delta \phi = 0$, we
obtain
\be
\delta W[J] + \delta J_i \partial^i W[J]
= \delta \Gamma[\phi] + \delta J_i \phi^i
\,.
\ee
Since $\phi^i= \partial^i W[J]$, it follows that
$\esp{\delta S}^J = \delta \Gamma$ which proves \Eq{2.1}.

\subsection{ \textsf{ Proof of the Theorem around \Eq{2.19} } }

The action producing the expectation values
$\espp{\varphi^{i_1}\cdots\varphi^{i_n}}$ is just
$S'[\varphi] = S[\varphi]+J[\phi]\cdot\varphi$, and therefore the
corresponding generator of the connected Green functions is $W[J'+J]$. Here
the current is $J'$ while $\phi$, and so $J=J[\phi]$, are parameters. For
$n\ge2$ such generator may be changed to
\be
W'[J'] =  W[J+J'] - (J+J') \cdot \phi
\,.
\ee
Although this functional would give a vanishing value for $\esp{\varphi}^J$,
it correctly reproduces the expectation values for $n\ge 2$. The effective
action $\Gamma'[\phi']$ that at the tree level produces the same Green
functions (for $n\ge 2$) is then obtained as the Legendre transform of
$W'[J']$, as follows.

Using the Legendre-transform relations of the type
\be
\Gamma[\phi] =
\inf_{J} ( W[J] - J \cdot \phi  )
,
\qquad
 W[J] =  \sup_{\phi} (
\Gamma[\phi] + J \cdot \phi
),
\ee
we can write
\bes
\Gamma'[\phi'] &=
\inf_{J'} ( W'[J'] - J' \cdot \phi'  )
\\
&=\inf_{J'} ( W[J+J'] - (J+J') \cdot \phi - J' \cdot \phi'  )
\\
&=\inf_{J'} ( \sup_{\phi_1} (
\Gamma[\phi_1] + (J+J') \cdot \phi_1
)
- \phi \cdot (J+J') - J' \cdot \phi'  )
.
\\
\ees
The extremum with respect to $J'$ requires $\phi_1 = \phi + \phi'$, hence
\be
\Gamma'[\phi'] =
\Gamma[\phi + \phi'] + J[\phi] \cdot \phi'
\,.
\label{eq:A7}
\ee
Here $\phi'$ is the classical field and $\phi$ is a
parameter. This effective action has a vanishing one-point vertex.

The statement of the Theorem in \Eq{2.19} is that the derivatives of
$\Gamma[\phi]$ of order two or higher produce the correct propagator and
vertices to be used in the Feynman rules at tree level to reproduce
$\espp{\varphi^{i_1}\cdots \varphi^{i_n}}$, for $n \ge 2$. Those Feynman
rules obviously derive from the effective action in \Eq{A7}, hence the
statement is proven.

It can be noted that the Legendre transformation of the functional $W[J+J']$,
namely, $\Gamma[\phi']+J\cdot\phi'$, also produces the correct Green
functions, albeit with different propagator and vertices; however, it contains
a non-vanishing one-point vertex. A resummation of those graphs to eliminate
the one-point vertex is directly provided by the effective action in \Eq{A7}.

Alternatively, the Theorem can be obtained by recursively applying
derivatives, starting from the two-point function:
\be
\esp{\varphi^i\varphi^j}_c^J = \partial^i\partial^j W[J]
= \frac{\partial \phi^j} {\partial J_i}
.
\ee
By the symmetry $W \leftrightarrow -\Gamma$ of the Legendre transformation, we
also have
\be
-\partial_i\partial_j \Gamma[\phi] = \frac{\partial J_i} {\partial \phi^j}
.
\ee
This proves
$\espp{\varphi^i\varphi^j}_c = ((-\partial^2\Gamma)^{-1})^{ij}$. Then for
$n=3$
\be
\esp{\varphi^i\varphi^j\varphi^k}_c^J
=
\partial^i \esp{\varphi^j\varphi^k}_c^J
=
\partial^i \phi^a \partial_a ((-\partial^2\Gamma)^{-1})^{ij}
\ee
produces the third equation in \nec{20}. The rule is that each
$\partial^a= \partial/\partial J_a$ generates a new leg from either a vertex
or a (external or internal) line, while each
$\partial_a= \partial/\partial \phi^a$ generates an amputated leg.

\begin{widetext}
\section{ \textsf{ Renormalization as reparametrization }
  \label{app:B} }

In this Appendix, we want to illustrate that the relation \nec{3.31}
\be
\Gamma - S_0 = \bR(\Gamma_R - S_R)
\ee
expresses a change of variables from $S_0$ to $S_R$, where the latter is
defined by $S_R = T\Gamma$. $\Gamma$ and $\Gamma_R$ denote the effective
actions diagrammatically constructed with $S_0$ and $S_R$, respectively. As
previously noted, this relation already implies the other two:
$S_R - S_0 = T\bR(\Gamma_R - S_R)$ and $\Gamma - S_R = R(\Gamma_R - S_R)$. We
will work at low orders of the perturbative expansion in the $\Gamma_n$
sectors $n=2,3$ and assume that $S_0$ has only operators with $\ell=2,3$.  The
expansion of $\Gamma$ as one-particle irreducible amputated graphs of $S_0$,
yields
\bes
\Gamma_2 &=
\includegraphics[width=10mm]{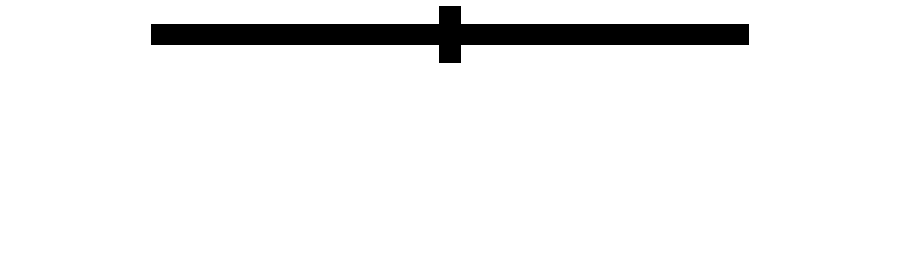}
+
\includegraphics[height=8mm]{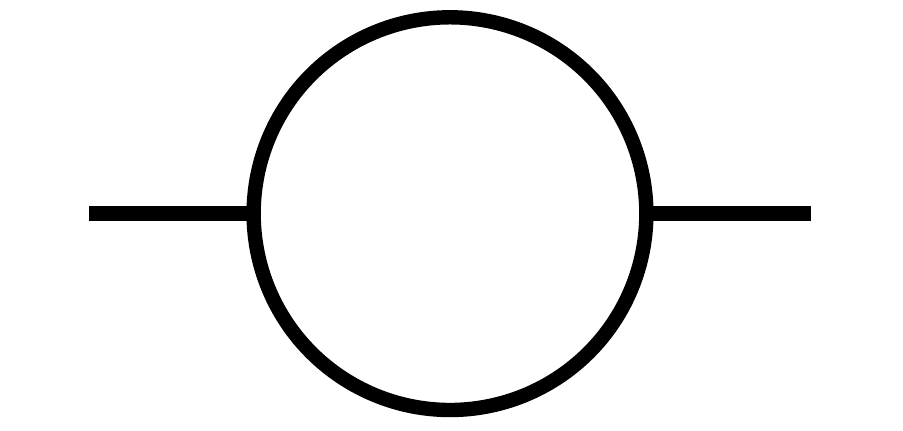}
+
\includegraphics[height=8mm]{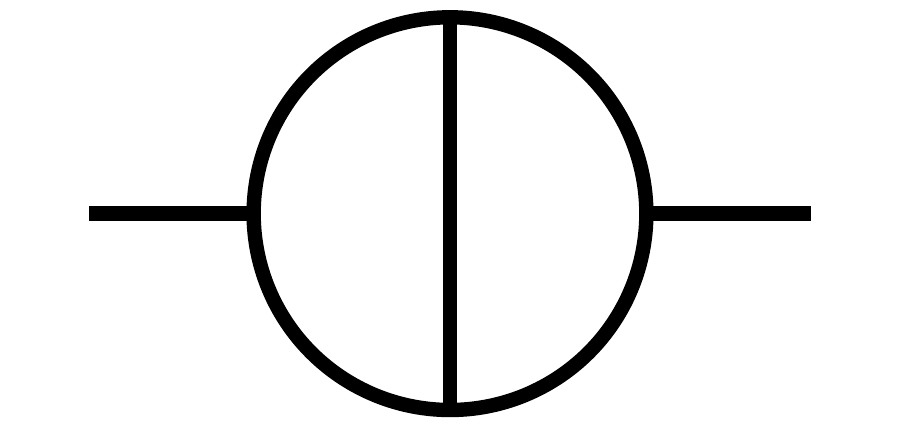}
+
\includegraphics[height=8mm]{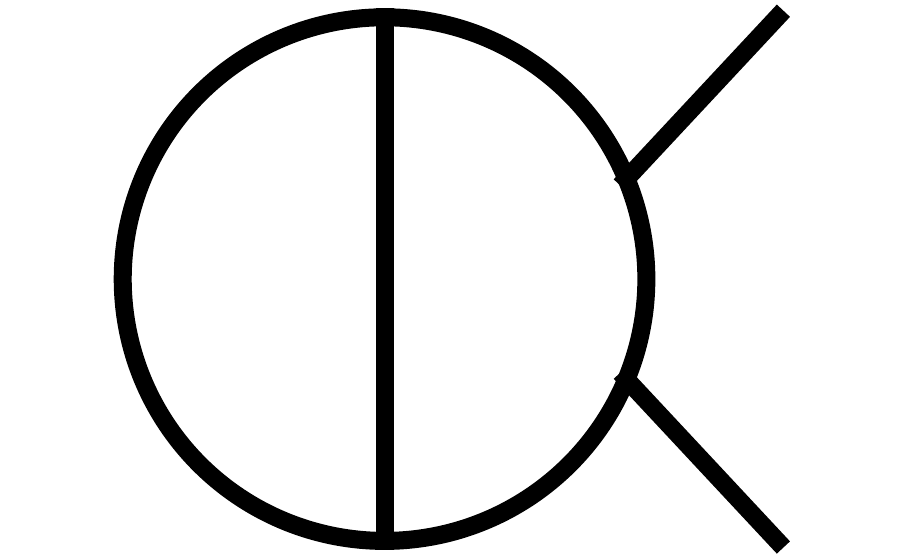}
+ O(\hbar^3)
\quad [S_0]
\\
\Gamma_3 &= 
\includegraphics[height=8mm]{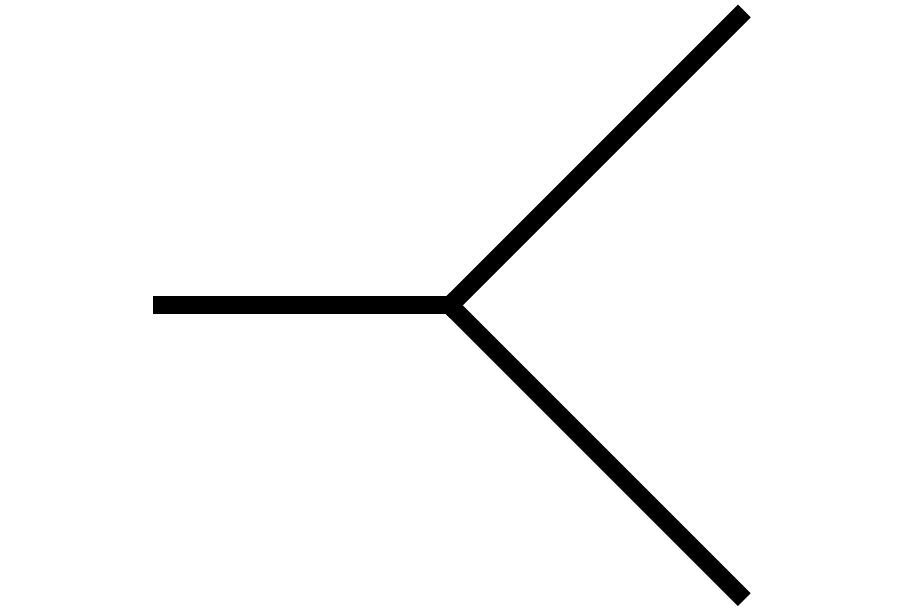}
+
\includegraphics[height=8mm]{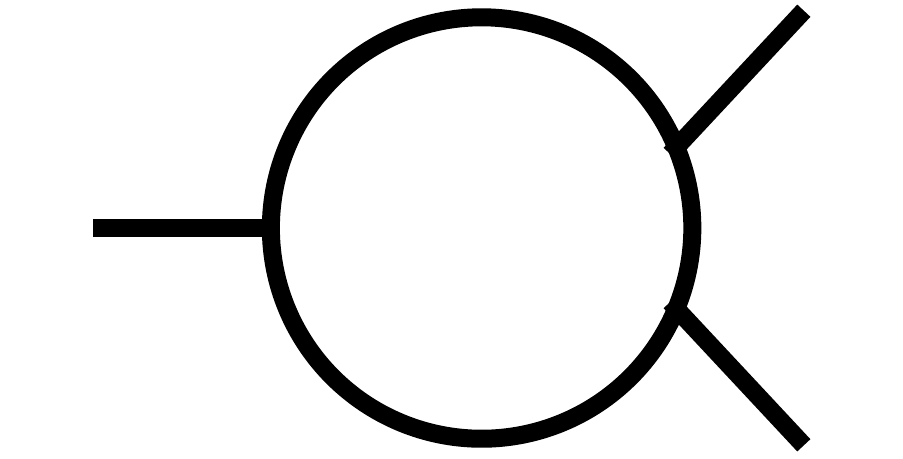}
+
3\times\hspace{-2mm}
\includegraphics[height=8mm]{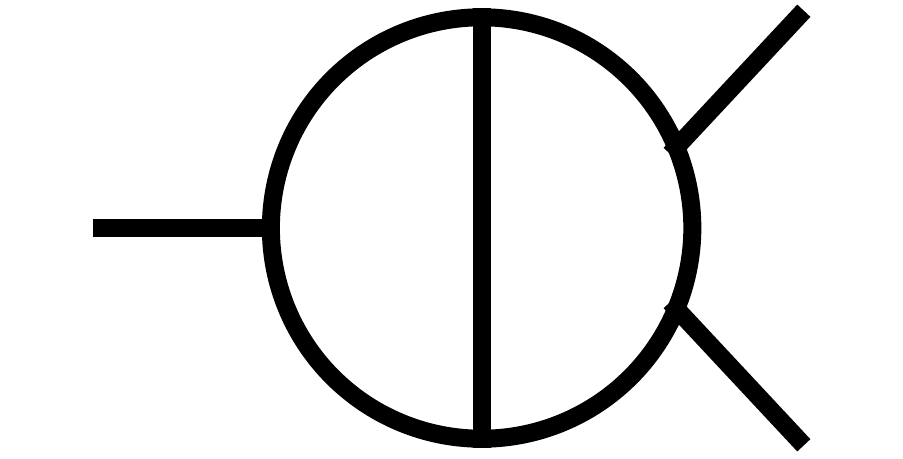}
+
3\times\hspace{-2mm}
\includegraphics[height=8mm]{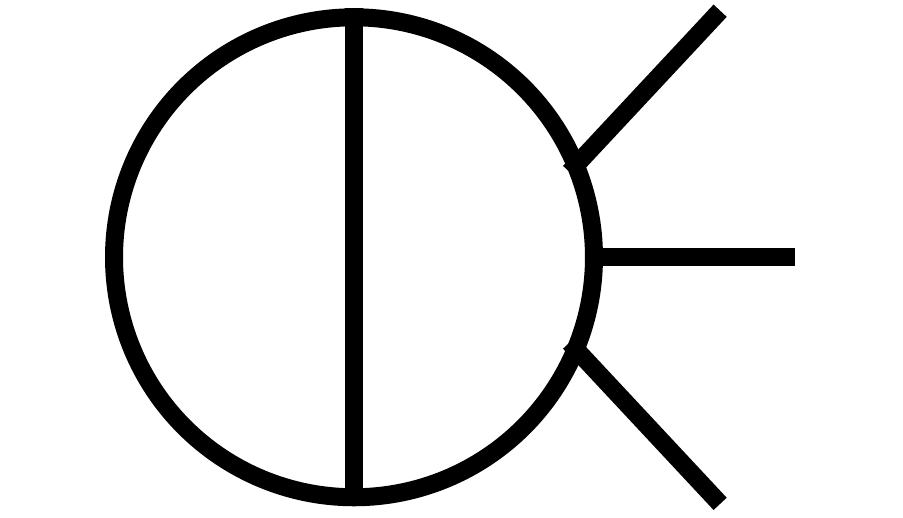}
+
\includegraphics[height=8mm]{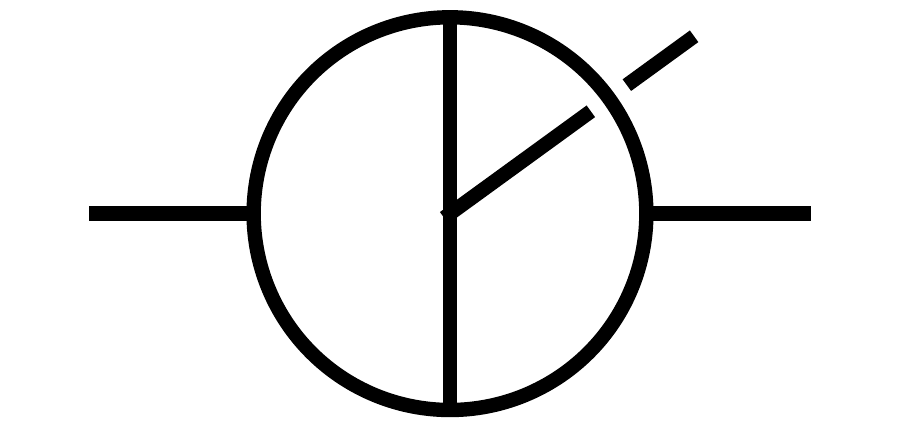}
+ O(\hbar^3)
\quad [S_0]
.
\label{eq:B2}
\ees
Here the external legs are amputated. The label $[S_0]$ on the RHS indicates
that the lines and vertices are those of $S_0$ unless otherwise indicated.
The prescription $S_R=T\Gamma$ then implies
\bes
\includegraphics[width=10mm]{AG1.pdf}\hspace{-1mm}{}_{[S_R]}
&=
\includegraphics[width=10mm]{AG1.pdf}
+
T  \hspace{-1mm}
\includegraphics[height=8mm]{AG2.pdf}
+
T  \hspace{-1mm}
\includegraphics[height=8mm]{AG3.pdf}
+
T  \hspace{-1mm}
\includegraphics[height=8mm]{AG4.pdf}
+ O(\hbar^3)
\quad [S_0]
\\
\includegraphics[height=8mm]{AG5.pdf}\hspace{-1mm}{}_{[S_R]}
&= 
\includegraphics[height=8mm]{AG5.pdf}
+
T \hspace{-1mm}
\includegraphics[height=8mm]{AG6.pdf}
+
3\times
T \hspace{-1mm}
\includegraphics[height=8mm]{AG7.pdf}
+
3\times
T \hspace{-1mm}
\includegraphics[height=8mm]{AG8.pdf}
+
T \hspace{-1mm}
\includegraphics[height=8mm]{AG9.pdf}
+ O(\hbar^3)
\quad [S_0]
.
\label{eq:B3}
\ees
We want to express $S_0$ in terms of $S_R$. The $n=3$ relation in \nec{B3}
yields
\bes
\includegraphics[height=8mm]{AG5.pdf}\hspace{-1mm}{}_{[S_0]}
&=
\includegraphics[height=8mm]{AG5.pdf}_{[S_R]}
-
T \hspace{-1mm}
\includegraphics[height=8mm]{AG6.pdf}
+ O(\hbar^2)
\quad [S_0]
\\
&=
\includegraphics[height=8mm]{AG5.pdf}
-
T \hspace{-1mm}
\includegraphics[height=8mm]{AG6.pdf}
+ O(\hbar^2)
\quad [S_R]
.
\label{eq:B4}
\ees
On the other hand, for $n=2$, the free propagator relations
$D_{0,0}=-S_{0,2}^{-1}$ and $D_{R,0}=-S_{R,2}^{-1}$ apply. Upon inversion, one
immediately finds, for the bare and renormalized free propagators,
\bes
\includegraphics[height=8mm]{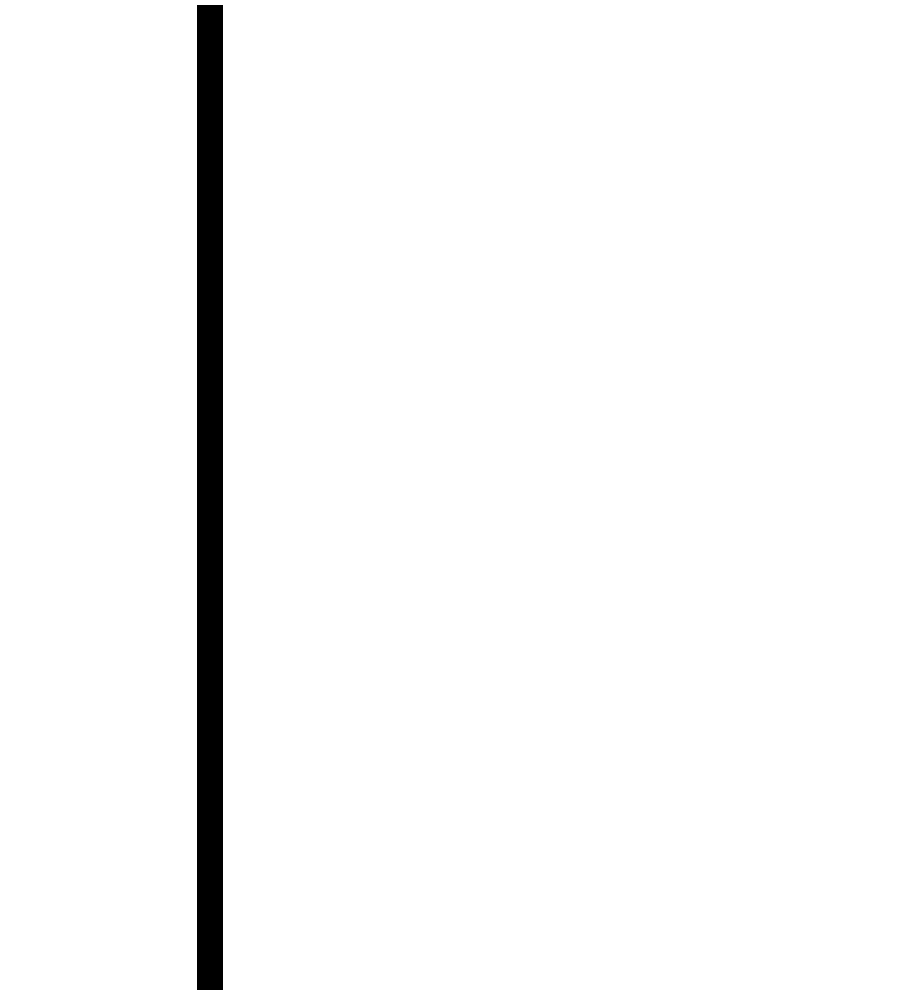}\hspace{-4mm}{}_{[S_0]} &=
-\left(
\includegraphics[width=10mm]{AG1.pdf}\hspace{-1mm}{}_{[S_0]}
\right)^{-1}
=
-\Big(
\includegraphics[width=10mm]{AG1.pdf}
-
T \hspace{-1mm}
\includegraphics[height=8mm]{AG2.pdf}
+ O(\hbar^2)
\Big)^{-1}_{[S_R]}
\\
&=
\includegraphics[height=8mm]{AG10.pdf}
\hspace{-4mm}
-
T \hspace{-1mm}
\includegraphics[height=8mm]{AG2.pdf}
+ O(\hbar^2)
\quad [S_R]
\label{eq:B5}
\ees
The lines are not amputated here.

Applying \nec{B4} and \nec{B5} one can proceed to the systematic elimination
of the line and vertices of $S_0$ in favor of those of $S_R$ for $\Gamma_n$ in
\nec{B2}. This produces, for $n=2$,
\bes
(\Gamma - S_0)_2 &=
\Big(
  \includegraphics[height=8mm]{AG2.pdf}
  -
  T_\gamma\hspace{-1mm}
  \includegraphics[height=11mm]{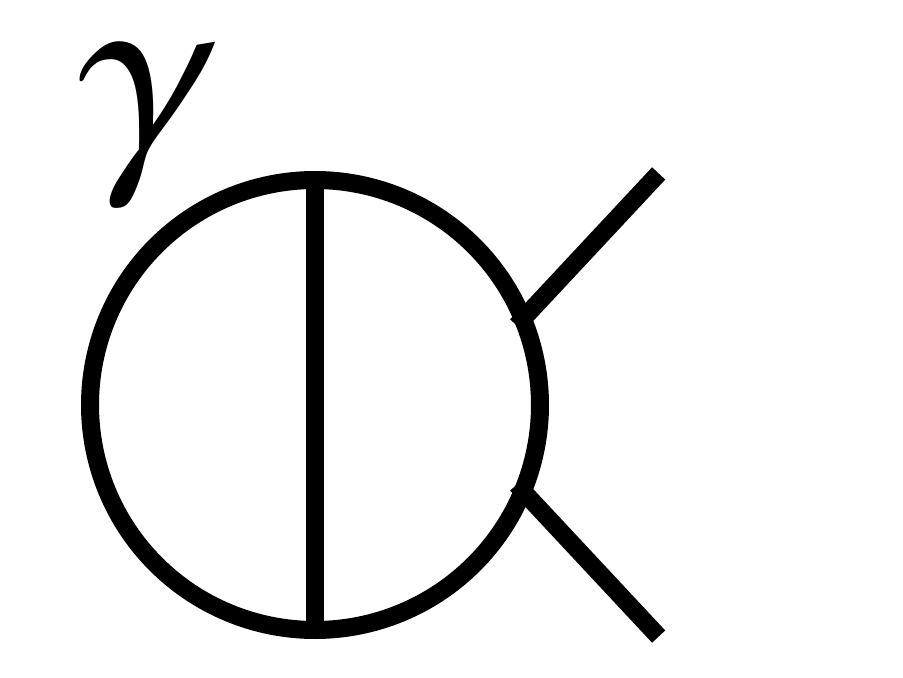}
  -
  2\times
  T_\gamma\hspace{-1mm}
  \includegraphics[height=11mm]{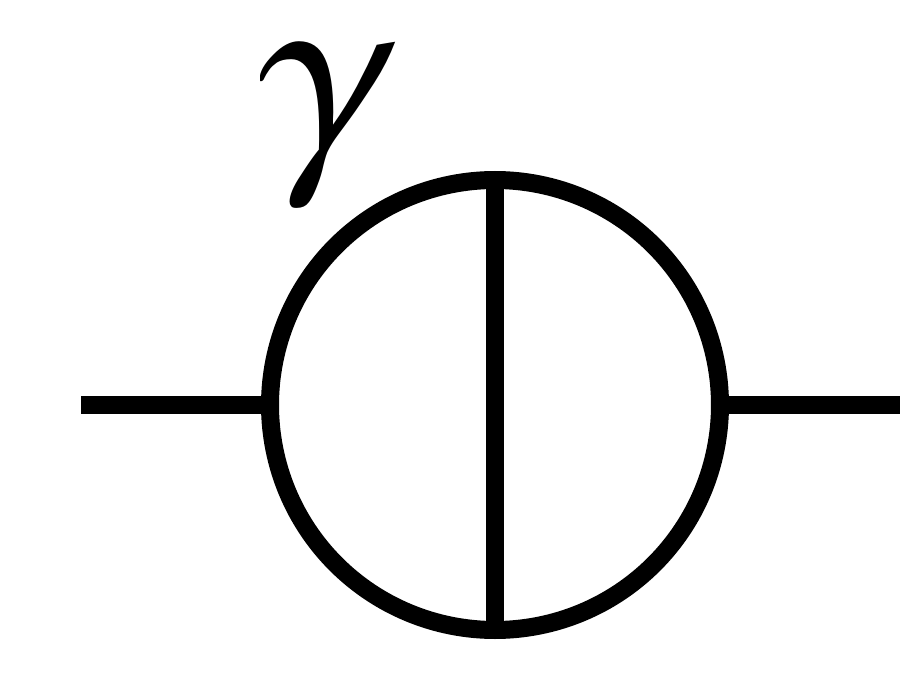}
~\Big)
+
\includegraphics[height=8mm]{AG3.pdf}
+
\includegraphics[height=8mm]{AG4.pdf}
+ O(\hbar^3)
\quad [S_R]
\\ &=
  \includegraphics[height=8mm]{AG2.pdf}
  +
 (1 - T_\gamma)\hspace{-1mm}
  \includegraphics[height=11mm]{AG4g.pdf}
  +
  2\times
  ( 1 - T_\gamma)\hspace{-1mm}
  \includegraphics[height=11mm]{AG3g.pdf}
  + O(\hbar^3)
\quad [S_R]
  \\ &=
  \bR \Big(
    \includegraphics[height=8mm]{AG2.pdf}
  +
  \includegraphics[height=8mm]{AG4.pdf}
  +
  \includegraphics[height=8mm]{AG3.pdf}
  + O(\hbar^3)
\Big)
\quad [S_R]
  \\ &=
  \bR ( \Gamma_R - S_R )_2
  .
\ees

Likewise, for $n=3$,
\bes
(\Gamma-S_0)_3 &=
\Big(
  \includegraphics[height=8mm]{AG6.pdf}
  -
  3\times
  T_\gamma\hspace{-1mm}
  \includegraphics[height=11mm]{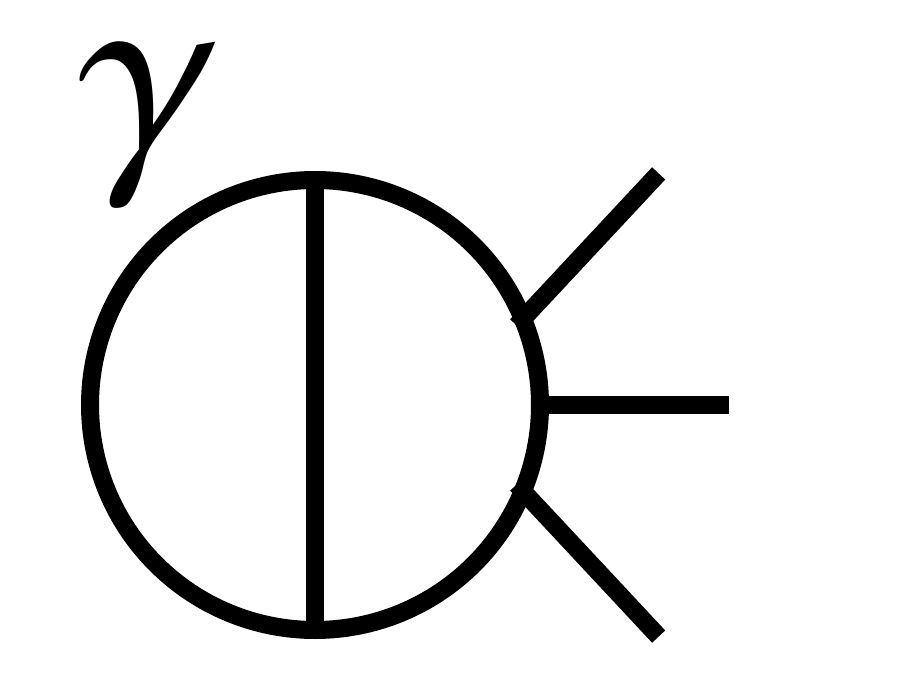}
  -
  3\times
  T_\gamma\hspace{-1mm}
  \includegraphics[height=11mm]{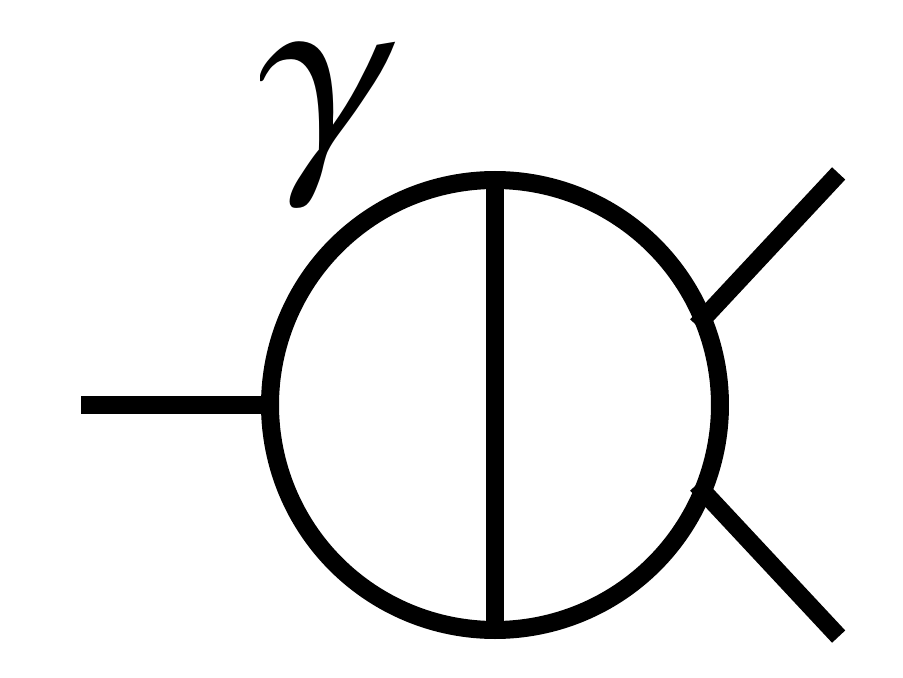}
\Big)
+
  3\times\hspace{-2mm}
\includegraphics[height=8mm]{AG7.pdf}
+
  3\times\hspace{-2mm}
\includegraphics[height=8mm]{AG8.pdf}
+
\includegraphics[height=8mm]{AG9.pdf}
+ O(\hbar^3)
\quad [S_R]
\\ &=
  \includegraphics[height=8mm]{AG6.pdf}
  +
  3\times
  ( 1 - T_\gamma)\hspace{-1mm}
  \includegraphics[height=11mm]{AG8g.pdf}
  +
  3\times
  ( 1 - T_\gamma)\hspace{-1mm}
  \includegraphics[height=11mm]{AG7g.pdf}
+
\includegraphics[height=8mm]{AG9.pdf}
+ O(\hbar^3)
\quad [S_R]
\\ &=
\bR \Big(
  \includegraphics[height=8mm]{AG6.pdf}
  +
  3\times\hspace{-2mm}
  \includegraphics[height=8mm]{AG8.pdf}
  +
  3\times\hspace{-2mm}
  \includegraphics[height=8mm]{AG7.pdf}
+
\includegraphics[height=8mm]{AG9.pdf}
+ O(\hbar^3)
\Big)
\quad [S_R]
\\ &=
R( \Gamma_R - S_R )_3
.
\ees
Hence, the statement is verified to this order.

\end{widetext}

  \section{ \textsf{
Cancellation of divergences and canonical and anti-canonical patterns
      }
\label{app:C} }

As stated in Sec. \ref{sec:5.1}, the matrix $W^\mu{}_\nu$ has an anti-canonical
pattern of divergences, while $V^\nu{}_\mu$ is canonical, and similarly for
$W^\alpha{}_\beta$ and $V^\beta{}_\alpha$ in  Sec. \ref{sec:4b}. In addition, the
construction $(W^{-1})^\mu{}_\nu G^\nu[\phi] =G^\mu_R[\phi]$ eliminates the
divergences.

These statements can be explicitly verified through a perturbative calculation
to any desired order. Nevertheless, to illustrate them in a simple setting, we
consider the renormalization of the composite operator
$O = \frac{1}{2} \phi^2(0)$ in the theory $\phi^3_6$, but including only a
restricted set of graphs of $S_R$. To be more explicit, we consider the class
\be
\includegraphics[height=15mm]{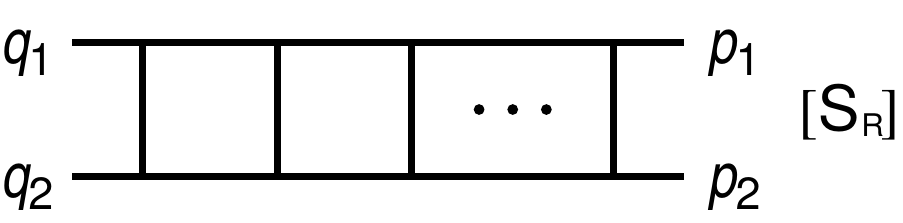}
\label{eq:C1}
\ee
($q_1+q_2+p_1+p_2=0$) with $k=0,1,2,\ldots$ exchanges in the $t$-channel, such
$S_R$-graphs are of perturbative order $g_R^{2k}$. Within this approximation,
only the term $\tG_2(p)$ (i.e., $n=2$) can be described, so we denote it as
$\tG(p)$:
\be
\includegraphics[height=25mm]{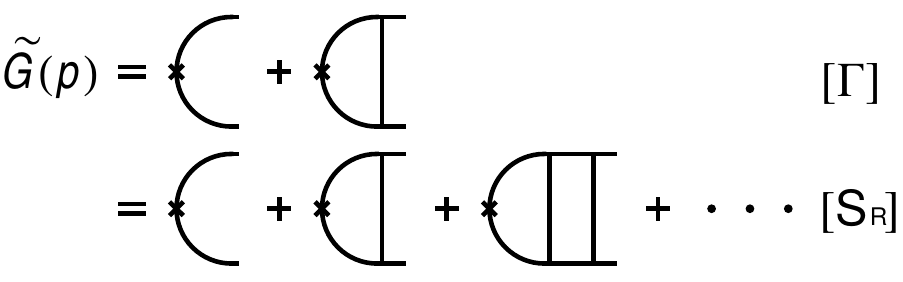}
\label{eq:C2}
\ee
Analytically,
\be
\tG(p) = 1 + \frac{1}{2} \int^\Lambda \frac{d^6q}{(2\pi)^6}
\, D_q^2 \, H(q;p)
.
\label{eq:C2a}
\ee
In our approximation, there is no mixing with other operators. As the
renormalization condition we adopt $\hT G = \tG(0)$, hence
\be
W = \tG(0) = 1 + \frac{1}{2} \int^\Lambda \frac{d^6q}{(2\pi)^6}
\, D_q^2 \, H(q;0)
,
\ee
and $G^\mu_R[\phi] = (W^{-1})^\mu{}_\nu G^\nu[\phi] $ reduces to
\be
\tG_R(p) = \frac{ \tG(p)}{ \tG(0)}
.
\ee

In the theory $\phi^3_6$ all the box subgraphs in the $S_R$-graph in \nec{C1}
are UV-finite. On the other hand, the composite operator induces divergences
with up to $k$-loops in the graph of order $g_R^{2k}$.

In \nec{C2}, the $S_R$-graph of order $g_R^0$,
\includegraphics[height=8mm]{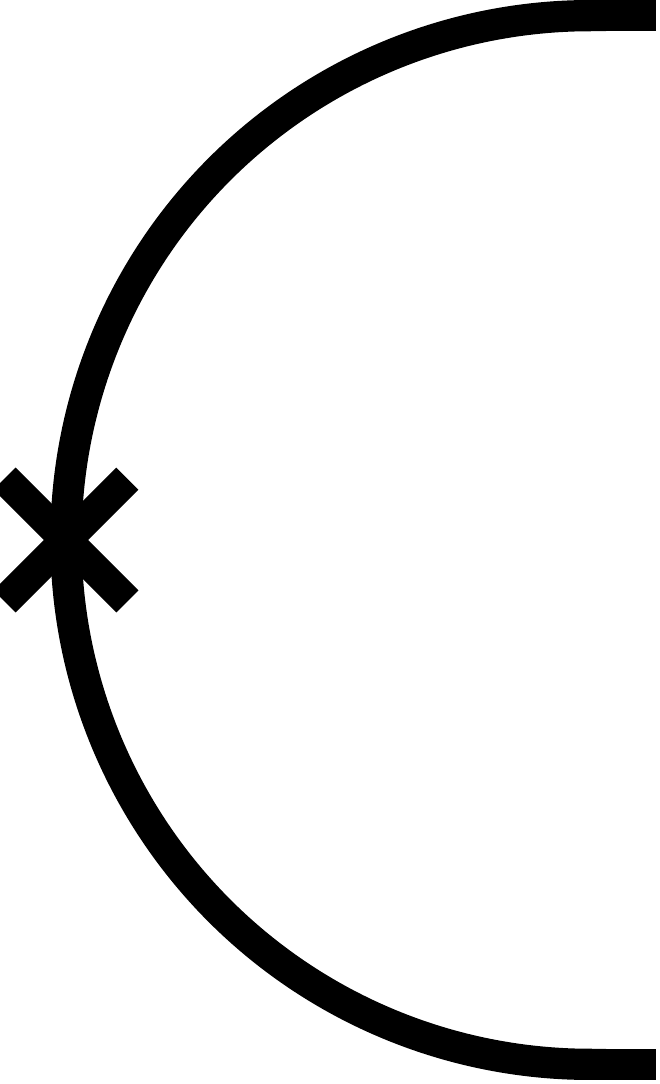}
, is a tree graph with the value $1$.

The one-loop $S_R$-graph of order $g_R^2$ in \nec{C2},
\includegraphics[height=8mm]{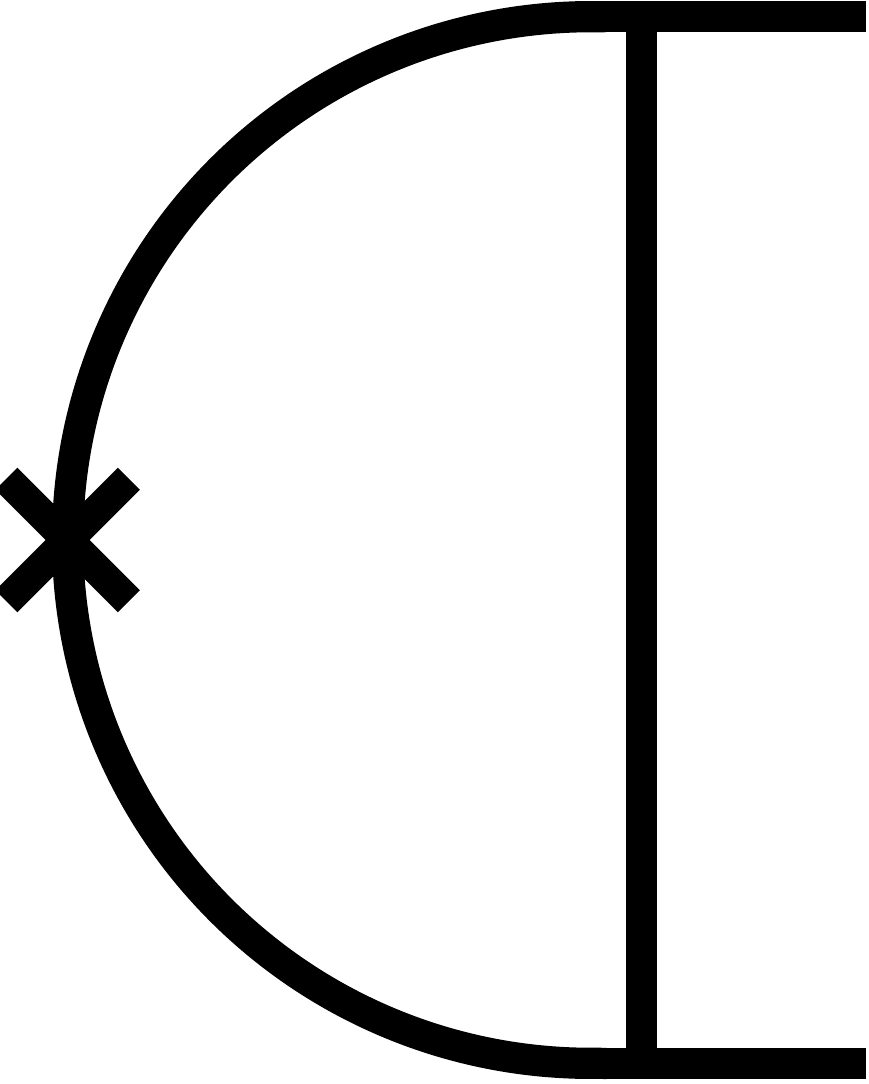} , has only a logarithmic
superficial divergence, without subdivergences.  Applying the identity
$1 = (1-T) + T$, the term $1-T$ gives the subtracted graphs and $T=O\hT$
isolates the divergence in the sector $O$, hence:
\be
\includegraphics[height=15mm]{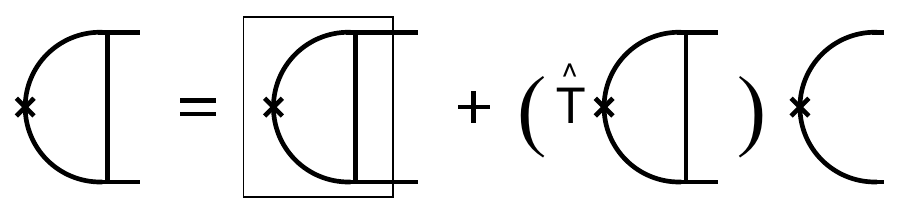}
\label{eq:C3}
\ee

The two-loop $S_R$-graph, as well as the higher-order graphs, can be treated
similarly in diagrammatic form; however, we turn to a more convenient algebraic
notation. Let $\gamma_k$, $k=0,1,2,\ldots$, denote the $S_R$-graph as in
\nec{C2} of order $g_R^{2k}$, which also appears as a subgraph of
$\gamma_\ell$ for $\ell \ge k$.  Then \nec{C2} becomes
\be
\tG(p) = \gamma_0 + \gamma_1 + \gamma_2 + O(g^6)
\label{eq:C7}
\ee
and \nec{C3} becomes
\be
\gamma_1 =
T \gamma_1 + (1-T)\gamma_1 =
(\hT \gamma_1) \gamma_0 + R \gamma_1
.
\label{eq:C4}
\ee
As always, $R$ denotes the operation of subtracting all sub- and overall
divergences, while $\bR$ subtracts only subdivergences.

Likewise, denoting $T_{\gamma_k}$ the projection acting on the (sub)graph
$\gamma_k$,
\be
\gamma_2 =
T_{\gamma_1} \gamma_2 + (1-T_{\gamma_1})\gamma_2
\label{eq:C5}
\ee
which can be worked out so that any divergence is either subtracted or is a
projected overall divergence:
\bes
(1-T_{\gamma_1})\gamma_2
&=
(1-T_{\gamma_2})(1-T_{\gamma_1})\gamma_2
+
T_{\gamma_2}(1-T_{\gamma_1})\gamma_2
\\ &=
R\gamma_2 + T \bR \gamma_2
=
R\gamma_2 + (\hT \bR \gamma_2) \gamma_0
,
\\
T_{\gamma_1} \gamma_2 &=
( \hT \gamma_1 ) \gamma_1
=
( \hT \gamma_1 )
(
(\hT \gamma_1) \gamma_0 + R \gamma_1
)
,
\ees
that is,
\bes
\gamma_2 =
(
\hT \bR \gamma_2 + (\hT\gamma_1)^2
) \gamma_0
+
(\hT\gamma_1) R \gamma_1
+
R \gamma_2
.
\ees

Collecting the various terms
\bes
\tG(p) &=
(
1 + \hT \gamma_1 +( \hT \gamma_1 )^2 + \hT \bR \gamma_2
) \gamma_0
\\
&\quad +
( 1 + \hT \gamma_1 ) R \gamma_1
+
R \gamma_2 + O(g^6)
.
\label{eq:C12}
\ees
Using $\hT \gamma_0 = 1$ and $\hT R = 0$,
\be
W = \hT \tG(p) =
1 + \hT \gamma_1 +( \hT \gamma_1 )^2 + \hT \bR \gamma_2 + O(g^6)
\ee
and
\be
\tG(p) = W \tG_R(p)
\label{eq:C14}
\ee
with
\be
\tG_R(p) = 
\gamma_0 + R ( \gamma_1 + \gamma_2 + O(g^6) )
.
\ee
Thus $\tG_R(p)$ is UV-finite and conforms to the expected structure
$\Gamma = S_R + R( \Gamma_R - S_R)$. Effectively, in \nec{C12} all
divergences and subdivergences present in $\tG(p)$ have been systematically
transferred to a factor $W$ with no dependence on $p$, leaving a factor
$\tG_R(p)$ with dependence on $p$ but no divergences.

On the other hand, $W$ has an anti-canonical pattern of divergences, because
\bes
V &= 1-W^{-1} =\hT \gamma_1 + \hT \bR \gamma_2 + O(g^6)
\\&=
\hT \bR ( \gamma_1 + \gamma_2 + O(g^6) )
\ees
is canonical, i.e., $\hT$ acts only after all subdivergences have been
subtracted. $V$ provides the counterterms and also conforms to the expected
structure $S_0 = S_R - T \bR ( \Gamma_R - S_R)$. Let us point out that
$\tG(p)$ does not know about $T$ (i.e., about the RC chosen for the composite
operator $O$) and $T$ appears just once in $W$; nevertheless, the construction
$V= 1- W^{-1}$ arranges the insertions of $T$ to produce the correct ordered
subtractions to select superficial divergences only. Not only that,
$\tG_R(p)=W^{-1} \tG(p)$ implies that $W^{-1}$ introduces the projectors and
subtractions in the graphs in the correct form to subtract all divergences.

In what follows, we present some speculation about the possible use of the
anti-canonical pattern to extract useful information. The fact that the
expressions are only guaranteed to exist perturbatively is disregarded here.
As discussed in Sec. \ref{sec:4b} $\tG(p)$ also has an anti-canonical
dependence on the regulator. In fact since $W^{-1}$ diverges logarithmically,
\Eq{C14} implies a $\Lambda$-dependence in $\tG(p)$ of the form
\be
\tG(p;\Lambda) = ( f(p;\Lambda) - h(p) L_\Lambda)^{-1}
\label{eq:C17}
\ee
for some UV-finite functions $f(p;\Lambda)$ and $h(p)$. The anti-canonical
divergence-pattern of $\tG(p;\Lambda)$ does not mean that $1/\tG(p;\Lambda)$
is also canonical in its dependence on $p$; a finite number of derivatives
with respect to $p$ does not render this quantity UV-finite.\footnote{Since
  the RC centered at $p=0$ are arbitrary, and $W^{-1}$ can be expressed using
  superficially divergent integrals only, a suitable RC centered at $p$
  (instead of $p=0$) should also express $1/\tG(p;\Lambda)$ in terms of
  superficially divergent integrals; however, as the very RC depend on $p$, it
  would not follow that the divergences are polynomial in $p$.}

According to the expression in \nec{C17}, $\tG(p)$ has a finite limit as the
cutoff is removed; however, it still needs renormalization since the limit is
zero. The renormalized value is
\be
\tG_R(p) = \lim_{\Lambda\to\infty} \frac{\tG(p;\Lambda)}{\tG(0;\Lambda)}
=\frac{ h(0) } { h(p) }
.
\ee
The limit is approached at a rate $O(1/L_\Lambda)$. Knowledge of the
anti-canonical pattern can be exploited to accelerate the rate of
convergence. For instance, a single subtraction produces
\be \frac{ \tG^{-1}(p;\Lambda_1) - \tG^{-1}(p;\Lambda_2) }{ L_{\Lambda_1} -
  L_{\Lambda_2} } = - h(p) + \frac{ f(p;\Lambda_1) - f(p;\Lambda_2) }{
  L_{\Lambda_1} - L_{\Lambda_2} }
.
\ee
The second term would vanish if the cutoff dependence in $f$ were
negligible. Further acceleration should be achieved by applying more
subtractions.

In the super-renormalizable case (Secs. \ref{sec:3} or \ref{sec:5.2}) it was
possible to arrange the momentum integrals to avoid the need for a regulator
altogether, and using a finite number of $\Gamma$-graphs in each case. Here,
such a rearrangement does not seem possible. In fact the bare $\tG(p)$ in
\nec{C7} comes from a series of unsubtracted graphs, and each new loop should
produce a new factor $L_q = \log(q^2/\mu^2)/2$ in the integrand, that in turn
produces a new factor $L_\Lambda$ in $\tG(p;\Lambda)$. This is reflected in the
expansion
\be
\tG(p) = f^{-1}_p + f^{-2}_p h_p L_\Lambda + f^{-3}_p h^2_p L_\Lambda^2
+ \cdots
\ee
based on the anti-canonical pattern of $\tG(p)$ in \nec{C17}.

The information encoded in the anti-canonical pattern can be exploited as
follows. Let $\bar{H}(q;p)$ be the angular-averaged value of $H(q;p)$ in
\nec{C2a} over the directions of $q$. Then
\bes
\tG(p;\Lambda)
&= 1 + \int^\Lambda dq \, Q(q;p)
\\
Q(q;p) &:=
 \frac{1}{2} \Omega_6 q^5 D_q^2 \, \bar{H}(q;p)
.
\label{eq:C21}
\ees
Neglecting the dependence of $f_p$ on $\Lambda$, which should be a valid
assumption for a large cutoff,
\be
Q(\Lambda;p) = \frac{\partial \tG(p;\Lambda)}{\partial \Lambda} =
\frac{1}{(f_p-h_p L_\Lambda)^2} \frac{h_p}{ \Lambda }
.
\ee
This relation implies an asymptotic behavior for large $q$
\be
\bar{H}(q;p) =
\frac{Z^2}{\Omega_6}
\frac{1}{h_p}
\frac{1}{q^2 L_q^2}
,
\ee
which allows us to extract the renormalized value of $\tG(p)$ directly from
$\bar{H}(q;p)$, namely,
\be
\tG_R(p) = \lim_{q\to\infty} \frac{\bar{H}(q;p) }{\bar{H}(q;0) }
.
\ee
To arrive at this result, we have exploited not only the anti-canonical
pattern of $\tG(p)$, but also its ``explicit'' property, namely, the
dependence on $\Lambda$ comes only from the boundary of the momentum integral
and not from the integrand itself.

\section{ \textsf{ `$\Gamma+\delta S_0(\Lambda)$' vs
`$\Gamma(\Lambda)+\delta S_0(\Lambda)$' }
\label{app:D} }

In this Appendix, we discuss the technical point raised near the end of
Sec. \ref{sec:4b}: we argue that using an already renormalized effective
action $\Gamma$ (cutoff removed) to make the perturbation (the
`$\Gamma+\delta S_0(\Lambda)$' approach) does not modify the final result
$\Gamma+\delta\Gamma$ as compared to the standard approach of using the same
regulator everywhere when doing the perturbation and then removing the
regulator (the `$\Gamma(\Lambda)+\delta S_0(\Lambda)$' approach).

\begin{figure}[ht]
  \begin{center}
\includegraphics[height=30mm]{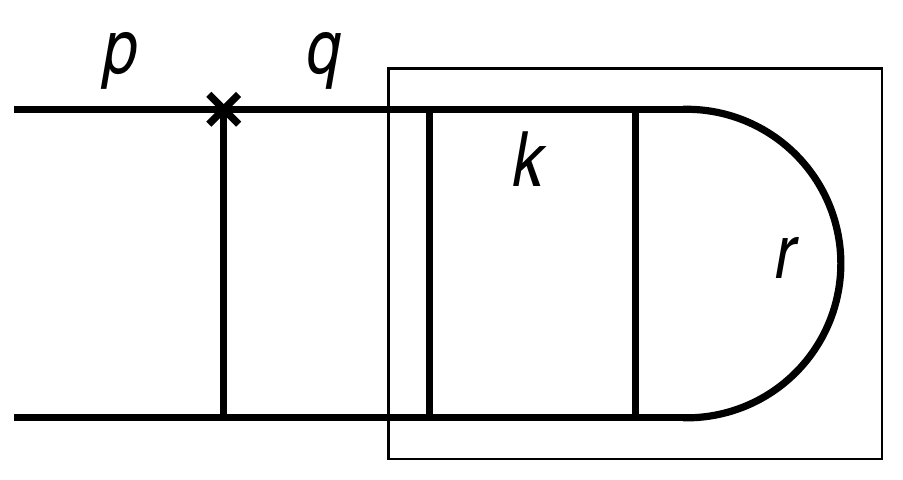}
\end{center}
\caption{An $S_R$-graph of the massless $\phi^3_6$ theory contributing to
  $G^g_{R,2}(p)$. The graph belongs to the $\Gamma$-graph class
  \includegraphics[height=8mm]{GF9.pdf}. The two internal loops are part
  of $\Gamma[\phi]$ and they are already subtracted (indicated by the
  box). The outer loop is induced by the perturbation and requires a further
  subtraction.}
\label{fig:8}
\end{figure}

To show this we will consider the $S_R$-graph of the $\phi_6^3$ theory shown
in \fig{8}. Let us denote $\gamma_1$ the one-loop subgraph defined by the
inner-loop, with internal momentum $r$, $\gamma_2$ the two-loop subgraph with
internal momenta $r$ and $k$, and $\gamma_3$ the full three-loop graph, with
internal momenta $r$, $k$ and $q$ and external momentum $p$. Also,
$\gamma_1(k^2;\epsilon)$, $\gamma_2(q^2;\epsilon)$ and
$\gamma_3(p^2;\epsilon)$ denote their values. For simplicity, we work in the
massless theory with the RC in \nec{3.6} and apply DR.

Let us first consider the standard calculation where a non-vanishing $\epsilon$
is held until the graph is computed and then set to zero. For the first
subgraph (omitting trivial factors $g_R$, $-Z_R$ and $\delta g_R$) 
\be
\gamma_1(k^2;\epsilon) =
\frac{1}{2}\nu^{2\epsilon} \int \frac{d^\bd r}{(2\pi)^\bd}
\frac{1}{r^2}\frac{1}{(k-r)^2}
,
\label{eq:D1}
\ee
with $\bd= 6-2\epsilon$. This and all other required integrals can be
explicitly obtained from the basic
expression \cite{Pascual:1984zb}
\be
\int \frac{d^dq}{(2\pi)^d} \frac{1}{(q^2)^\alpha}
\frac{1}{((q-p)^2)^\beta}
=
(p^2)^{d/2-\alpha-\beta}
C_2(\alpha,\beta;d)
\ee
with
\be
C_2(\alpha,\beta;d) := \frac{\Gamma(\alpha+\beta-d/2)}{\Gamma(\alpha+\beta)}
\frac{B(d/2-\alpha,d/2-\beta)}{B(\alpha,\beta)}
\ee
and $B (a,b)\equiv\Gamma (a)\,\Gamma (b)/\Gamma (a+b)$. For instance,
\be
\gamma_1(k^2;\epsilon) =
\frac{1}{2}\nu^{2\epsilon}
(k^2)^{1-\epsilon}C_2(1,1;6-2\epsilon).
\ee

The graph $\gamma_1$ is quadratically divergent and requires subtractions.
The canonical divergences in the $\phi^3_6$ massless theory are proportional
to $k^2$.  It is clear from the integral in \nec{D1}, that three derivatives
with respect to $k^\mu$ make the integral convergent; thus, the divergence in
$\gamma_1$ is canonical and is removed by $1-T$,
\bes
\bar{\gamma}_1(k^2;\epsilon) &:= (1-T) \gamma_1(k^2;\epsilon)
\\ &=
\gamma_1(k^2;\epsilon) - \frac{k^2}{\muRC^2}\gamma_1(\muRC^2;\epsilon)
.
\ees
The subtracted integral $\bar{\gamma}_1(k^2;\epsilon)$ has only a regular
power series in $\epsilon$. A further feature is that the zeroth-order term,
\be
\bar{\gamma}_1(k^2;0) =
k^2 \frac{\log(k^2/\muRC^2)}{768 \pi^3}
,
\ee
is independent of the scale $\nu$. The renormalization condition erases the
dependence on $\nu$ in the $\epsilon$-independent term.  The second subgraph
(already using the subtracted $\gamma_1$) is
\be
\gamma_2(q^2;\epsilon) :=
\nu^{2\epsilon} \int \frac{d^\bd k}{(2\pi)^\bd}
\frac{1}{(k^2)^2}\frac{1}{(q-k)^2} \bar\gamma_1(k^2;\epsilon)
.
\ee
Again, it has only a superficial divergence that is canonical and is canceled
by the subtraction
\bes
\bar{\gamma}_2(q^2;\epsilon) &:= (1-T)\gamma_2(q^2;\epsilon)
\\ &=
\gamma_2(q^2;\epsilon) - \frac{q^2}{\muRC^2} \gamma_2(\muRC^2;\epsilon)
,
\ees
and
\be
\bar{\gamma}_2(q^2;0) =
q^2 \frac{\log(q^2/\muRC^2)
  (
3 \log(q^2/\muRC^2) - 11
  )}{1769472 \pi^6}
.
\ee

Likewise
\be
\gamma_3(p^2;\epsilon) :=
\nu^{2\epsilon} \int \frac{d^\bd q}{(2\pi)^\bd}
\frac{1}{(q^2)^2}\frac{1}{(p-q)^2} \bar\gamma_2(q^2;\epsilon)
,
\label{eq:D9}
\ee
and
\bes
\bar{\gamma}_3(p^2;\epsilon) &:= (1-T)\gamma_3(p^2;\epsilon)
\\ &=
\gamma_3(p^2;\epsilon) - \frac{q^2}{\muRC^2} \gamma_3(\muRC^2;\epsilon)
.
\label{eq:D10}
\ees
The final result of the standard calculation is then
$\bar{\gamma}_3(p^2;0)$,
\bes
\bar{\gamma}_3(p^2;0) &= p^2 \log(p^2/\muRC^2)
\\ &\quad \times
\frac{  
  3 \log^2(p^2/\muRC^2)
- 33 \log(p^2/\muRC^2) + 103
}
{
  2038431744 \pi^9
}
.
\ees

In the alternative computation, the two inner loops are contributions to
$\Gamma[\phi]$, which is already renormalized, hence $\epsilon$ is set to zero
in $\bar\gamma_2(q^2;\epsilon)$. The outer loop comes from the perturbation
and there $\epsilon$ is non-vanishing when performing the integration over $q$,
\be
\gamma_3^{\,\prime}(p^2;\epsilon) :=
\nu^{2\epsilon} \int \frac{d^\bd q}{(2\pi)^\bd}
\frac{1}{(q^2)^2}\frac{1}{(p-q)^2} \bar\gamma_2(q^2;0)
.
\ee
Upon subtraction
\be
\bar{\gamma}_3^{\,\prime}(p^2;\epsilon) = 
\gamma_3^{\,\prime}(p^2;\epsilon)
- \frac{q^2}{\muRC^2} \gamma_3^{\,\prime}(\muRC^2;\epsilon)
,
\ee
and the final result of the alternative calculation is
$\bar{\gamma}_3^{\,\prime}(p^2;0)$.  By direct calculation of the integrals it
is easy to verify that the two calculations yield the same result,
\be
\bar{\gamma}_3^{\,\prime}(p^2;0) = \bar{\gamma}_3(p^2;0)
.
\ee

As already said $\bar\gamma_2(q^2;\epsilon)$ has a regular power series in
$\epsilon$,
\be
\bar\gamma_2(q^2;\epsilon) = \bar\gamma_2(q^2;0) + \epsilon q^2 R(q^2;\epsilon)
\ee
and the function $R(q^2;\epsilon)$ is also regular. The argument why
neglecting such remainder $\epsilon q^2 R$ has no effect in
$\bar\gamma_3(p^2;0)$ is actually simple: when the extra term
$q^2 R(q^2;\epsilon)$ is introduced in $q$-integral \nec{D9} it will produce
both finite and divergent contributions. The finite contributions are
irrelevant due to the extra factor $\epsilon$. On the other hand, the
divergent contributions can compensate for the factor $\epsilon$ and yield a
net result, however such divergent terms are necessarily canonical and so they
are removed by the subtraction $1-T$ in \nec{D10}. That the divergences
induced by $q^2 R(q^2;\epsilon)$ are necessarily canonical follows from the
fact that the function $R$ has only a soft dependence on $q^2$, namely, of the
type $(q^2)^{n\epsilon}$ (for a few small values $n$). Upon expansion in
$\epsilon$, only terms of the type $\log^m(q^2)$ are produced. Such soft
dependence cannot overturn the dominant factor $1/(q^2(p-q)^2)$: still taking
three derivatives with respect to $p^\mu$ makes the integral convergent;
hence, the divergence is proportional to $p^2$ and so is canonical.

In the graph considered above, no subdivergences were induced by adding the
composite operator (the crossed vertex), only a superficial
divergence. However, this is not relevant to the argument. One can consider
instead the graph in \fig{9}. There, the subgraph $(345)$ is already
subtracted, but the composite operator induces both a subdivergence and an
overall divergence. Again, the subtracted subgraph $(345)$ has only a soft
momentum dependence. The remainder after subtracting its value at $\epsilon=0$
is $O(\epsilon)$ so, after performing the last momentum integration, the
finite contributions vanish when $\epsilon$ is set to zero, while the
divergent contributions are canonical due to the soft momentum
dependence. They are then removed by the last subtraction $1-T$.


\end{document}